\documentclass{nature}
\usepackage{graphicx}
\usepackage{epstopdf}
\usepackage[position=t,singlelinecheck=off]{subfig}
\usepackage{caption}
\usepackage{multicol}
\usepackage{float}
\usepackage{CJK}
\usepackage{amsmath}
\usepackage{lineno}
\usepackage[colorlinks,linkcolor=blue]{hyperref}
\usepackage{amsfonts}
\usepackage{appendix}

\title{Investigating and Modeling the Dynamics of Long Ties}

\author{Ding Lyu$^{1}$, Yuan Yuan$^{2,3*}$, Lin Wang$^{1}$, Xiaofan Wang$^{1,4}$, Alex Pentland$^{2,5}$}

\begin{document}
\maketitle
\begin{affiliations}
    \item Department of Automation, Shanghai Jiao Tong University, Shanghai, Shanghai, China
    \item Krannert School of Management, Purdue University, West Lafayette, IN, USA
    \item Connection Science, Massachusetts Institute of Technology, Cambridge, MA, USA
    \item Department of Automation, Shanghai University, Shanghai, Shanghai, China
    \item Media Lab, Massachusetts Institute of Technology, Cambridge, MA, USA
\end{affiliations}
\captionsetup[figure]{labelfont={bf},name={Fig.}}
\captionsetup[table]{labelfont={bf},name={Tab.}}

\begin{abstract}
Long ties, the social ties that bridge different communities, are widely believed to play crucial roles in spreading novel information in social networks. However, some existing network theories and prediction models indicate that long ties might dissolve quickly or eventually become redundant, thus putting into question the long-term value of long ties. Our empirical analysis of real-world dynamic networks shows that contrary to such reasoning, long ties are more likely to persist than other social ties, and that many of them constantly function as social bridges without being embedded in local networks.
Using a cost-benefit analysis model combined with machine learning, we show that long ties are highly beneficial, which instinctively motivates people to expend extra effort to maintain them. This partly explains why long ties are more persistent than what has been suggested by many existing theories and models. Overall, our study suggests the need for social interventions that can promote the formation of long ties, such as mixing people with diverse backgrounds.

\end{abstract}

\section*{Introduction}

Social network analysis provides a powerful instrument to investigate the structure of society by aggregating interpersonal relationships among individuals~\cite{watts1998collective,barabasi1999emergence,jackson2010social,barabasi2013network,broido2019scale}. In the social network literature, a large body of research centers on how tightly clustered social ties and groups are formed, as well as how they evolve, spread information and behaviors, and promote group solidarity~\cite{mcpherson1992social,jackson1996strategic,clauset2004finding,liben2007link,christakis2007spread,entwisle2007networks,flache2013weakness}. Meanwhile, a smaller but increasing number of studies focus on weak ties, which may function as ``bridges'' between different communities because of the unique roles they play in global network structures and information diffusion~\cite{burt1992structural,granovetter1973strength,watts1998collective,levin2004strength,onnela2007structure,zhao2010weak,ghasemiesfeh2013complex,larson2017weakness,gee2017paradox}. 

\begin{figure*}
    \centering
    \includegraphics[width=1\linewidth]{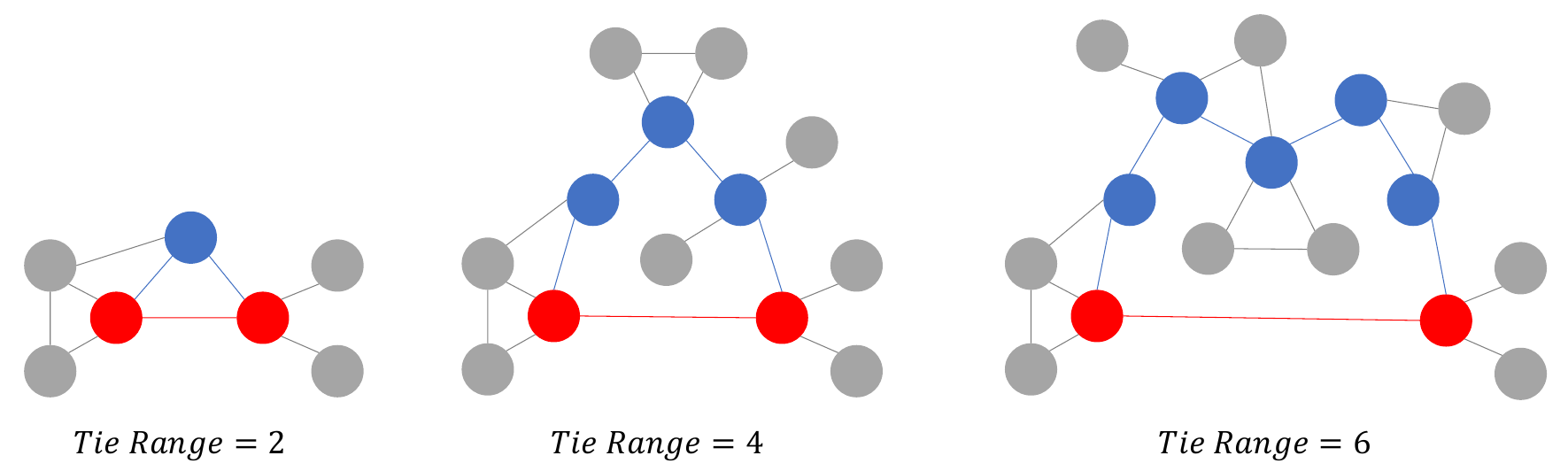}
    \caption{\textbf{Definition of tie range.} Tie range characterizes the length of the second shortest path between two connected nodes. The blue nodes are the nodes on the second shortest path between the two red nodes.}
    \label{fig:definitionTR}
\end{figure*}

One recent development in the literature is the concept of ``long ties.'' These are social ties that have a large tie range, which is measured by the length of the second shortest path between two connected nodes (see Fig.~\ref{fig:definitionTR}). Long ties -- social ties with a large tie range -- work as important social network bridges between different communities~\cite{montgomery1994weak,centola2007complex,centola2010spread,romero2011differences,park2018strength,trieu2019likes}. Structurally, long ties may be considered to be weak ties, as they are not positioned in a ``cohesive embedded network'' where individuals can easily contact or spend time with common neighbors~\cite{granovetter1973strength,centola2007complex,aral2011diversity}. Yet, despite the seeming weakness (in terms of low frequency or intensity of contact) of long ties, many studies have shown that long ties are crucial for the widespread dispersion of novel information and contagious behaviors~\cite{granovetter1973strength,watts1998collective,ghasemiesfeh2013complex,todo2016strength,park2018strength,eckles2019long,jahani2022origins}. Relatedly, these bridges may have other special characteristics such as exhibiting a higher level of direct reciprocity~\cite{block2015reciprocity}.

Still, one crucial perspective lacking in the literature of long ties is the dynamics. Evidence from static social networks may not be generalizable to dynamic networks~\cite{li2017fundamental}. In particular, existing social network theories and prediction models may indirectly imply that long ties should dissolve quickly or eventually become redundant, thus putting into question the long-term value of long ties. 

The critical role of long ties would be challenged if empirical evidence from dynamic networks suggests that long ties tend to dissolve or become short ties. Firstly, it is possible that long ties may dissolve rapidly. According to various theories~\cite{granovetter1973strength,aral2011diversity} and prediction models~\cite{liben2007link,easley2010networks}, social ties are likely to dissolve quickly when they lack sufficient common neighbors to reinforce their relationships or when they have few interactions (i.e., interactions with weak tie strength). Long ties likely satisfy this condition, and thus their role in bridging different communities might be limited~\cite{onnela2007structure}. Secondly, long ties may evolve to become redundant ``short ties.'' By triadic closure~\cite{easley2010networks,benson2018simplicial}, a person may introduce other friends to their long ties, thereby forming common neighbors and switching the long tie to a short tie. Therefore, two people who had a long tie may become increasingly similar, for example, regarding the information they digest or the opinions they hold~\cite{asikainen2020cumulative}. Eventually, the previously long tie becomes largely redundant, as there now exist other paths where the same piece of novel information can flow between the two individuals~\cite{aral2011diversity,brashears2018weakness}.

Our study combines empirical analysis and computational modeling to provide a dynamic perspective of long ties. First, using two-year social network data, we find that contrary to what is implied by existing theories and models, not only are long ties more likely to persist than shorter-range ties but also that many of them continue to be long ties. To explain this finding, we propose three possible hypotheses: degree heterogeneity, survival bias, and valuable long ties~\cite{santos2006cooperation,weng2018attention}. Investigating these hypotheses, we empirically show that the first two mechanisms might not fully explain our main results. 

Next, we propose a cost-benefit analysis model to support our last hypothesis -- that individuals spend extra effort to maintain relationships with long ties because they are highly beneficial, since they provide novel information or different expertise. The model combines strategic network formation models from the game theory literature~\cite{jackson1996strategic,jackson2010social} and node embedding techniques in machine learning~\cite{perozzi2014deepwalk,grover2016node2vec,velickovic2018graph} to simulate the dynamics of social networks. This interdisciplinary approach has been shown effective in trading off the model’s power to explain mechanisms versus to predict~\cite{yuan2018interpretable}. Our model describes the social tie formation process as a result of a meeting procedure and a subsequent rational decision procedure. We verify the model by utilizing real-world data. Ultimately, we find that our model partly explains the persistency of long ties, which is the main conclusion of our empirical analysis.

\section*{Results}

\subsection{Long ties last longer}
\
\newline
In this work, we employ tie range to characterize the local network structure of a social tie. As the length of our data is two years, we partition the data into eight phases; our results are robust to other ways of partitioning, as well (see Supplementary Note 2). To begin our analysis, we classify all social ties by tie range in the first phase, and then, we observe the evolution of those ties in the subsequent phases. 

\begin{figure*}
    \centering
	\includegraphics[width=0.66\linewidth]{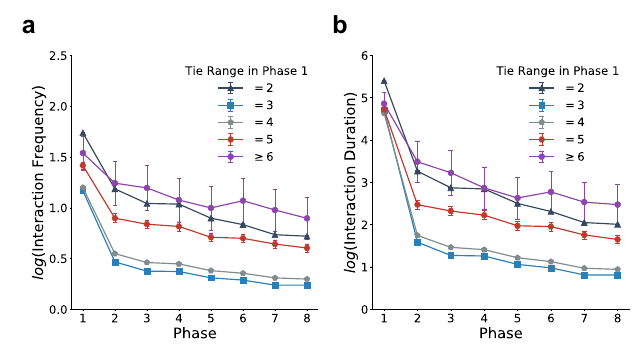}
    \caption{\textbf{Dynamics of tie strength given initial tie range.} Tie strength is measured by interaction duration (the total call volume in seconds) and interaction frequency (the number of calls or texts). Each phase represents a season (three months). We take logarithms ($\log$) for both interaction duration and frequency. All ties are classified according to their tie range in the first phase. The curves represent (\textbf{a}) the average ($\log$) interaction frequency and (\textbf{b}) the average ($\log$) interaction duration conditional on a tie existing in phase 1 with the given tie range. Error bars are 95\% confidence intervals for the mean $\log$ interaction duration and frequency (assuming normal distribution). Note that error bars are sometimes smaller than the data point markers.}
\label{fig:Fig.2}
\end{figure*}

First, we examine the dynamics of tie strength, which is measured by interaction frequency (the number of calls or texts) and interaction duration (the total duration of the calls). We define $y_t$ as the interaction frequency or duration in phase $t$. We present $E[y_t|y_1>0]$ in the eight phases, as shown in Fig.~\ref{fig:Fig.2}. This conditional expectation indicates that we focus our analysis on ties that already exist in phase 1 (see Supplementary Note 4). Observing the magnitudes in just the first phase, we find a ``U-shape'' in the data that is consistent with the results of the prior work~\cite{park2018strength}. Our result shows that interaction frequency and duration initially decrease with the tie range, but later increase with the tie range. In particular, long ties (tie range $\geq 6$) appear to be as intimate as those short ties with tie range $=2$ in that the average interaction frequency or duration for these two types of ties are close in the first phase.

By comparing the dynamics of short ties and long ties in Fig.~\ref{fig:Fig.2}, we find that long ties continue to be stronger. For example, in the long run, the average interaction duration and frequency of social ties with a tie range $\geq 6$ appear to be even slightly larger than those with a tie range of $2$. Furthermore, social ties with a tie range of $5$ also appear to be stronger than ties with a tie range of $3$ or $4$. In Supplementary Note 2, we discuss the robustness of our findings by adjusting the time window that determines the length of each phase.

To understand what mechanisms drive the patterns above, we decompose the dynamics of interaction frequency or duration into persistence probability and interaction increments. We let the difference in the interaction frequency or duration between phase $t$ and $1$ be $\Delta y_t = y_t - y_1$. Then, we define the persistence probability and interaction increments as follows:

\begin{equation}
    \begin{split}
        \mathbb{E}[y_{t} | y_{1} > 0] & = \mathbb{E}[y_{1} + \Delta y_t | y_t > 0, y_1 > 0] \mathbb{P}[y_{t} > 0 | y_1 > 0]  \\
        & = \big( \mathbb{E}[y_{1} | y_t > 0, y_1 > 0] + \underbrace{\mathbb{E}[\Delta y_t | y_t > 0, y_1 > 0]}_{\text{interaction increments}} \big) \times \underbrace{\mathbb{P}[y_{t} > 0 | y_1 > 0]}_{\text{persistence probability}}.
    \end{split}
\end{equation}

\begin{figure*}
    \centering
    \centering
    \includegraphics[width=\linewidth]{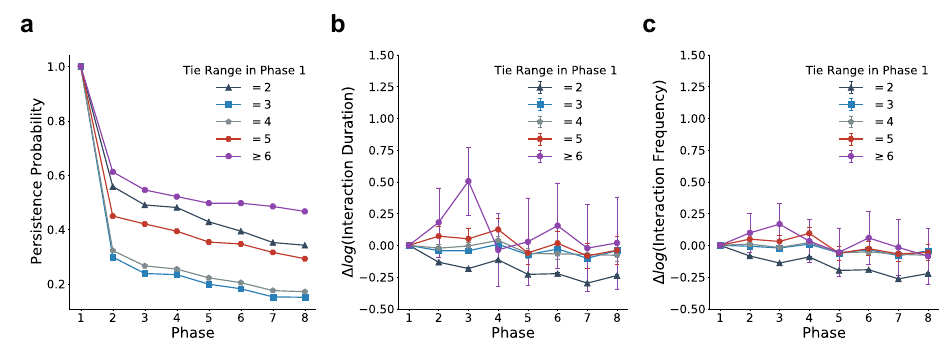}
    \caption{\textbf{Dynamics of persistence probability and interaction increments given initial tie range.} Each phase represents a season (three months). All ties are classified according to their tie range in the first phase. The curves represent (\textbf{a}) the probability of persisting, (\textbf{b}) the average ($\Delta \log$) interaction duration, and (\textbf{c})  and the average ($\Delta \log$) interaction frequency (\textbf{c}) conditional on a tie existing in phase 1 with the given tie range. Error bars are 95\% confidence intervals for the means (assuming normal distribution). Note that error bars are sometimes smaller than the data point markers.}
    \label{fig:persistence_increments}
\end{figure*}

The dynamics of the persistence probability and interaction increments are presented in Fig.~\ref{fig:persistence_increments}. As illustrated in the left panel of this figure, we find that social ties with a tie range $\geq 6$ have the largest persistence probability in all subsequent phases, followed by closely embedded ties with a tie range of $2$. Meanwhile, we find that social ties with a mid-sized tie range (i.e., $3$ or $4$) dissolve the fastest. This pattern is consistent with the overall effect presented in Fig.~\ref{fig:Fig.2}. In Supplementary Note 5, our additional analysis show that in general, long ties have longer lifespans. Note that when defining the lifespan, we explore two choices: (1) the social tie has to have interactions for every phase within the lifespan; and (2) a social tie has interactions in the first and the last phases no matter whether they have interactions in the phases in between. The latter considers the ties being re-established after termination. The conclusion does not change with the choice of the definition of lifespan (see Supplementary Note 5). These results also show that long ties tend to be persistent longer overtime.

Regarding the interaction increments, we find that they generally increase with tie range. This indicates that conditional on a persistent social tie, the interaction frequency and duration appear to be larger when there is a long tie. By contrast, social ties with a tie range of $2$ have the smallest interaction increments. From this, we conjecture that persistent short ties typically require less effort to maintain, as they can be indirectly maintained through their common friends; by contrast, we speculate that long ties require a lot of time investment in order to be maintained.

\subsection{Many long ties are persistently long}
\begin{figure*}
    \centering
    \includegraphics[width=\linewidth]{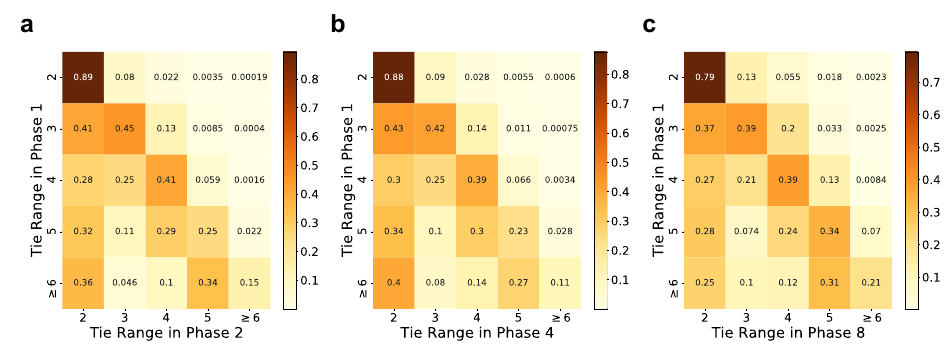}
    \caption{\textbf{Transition probability matrix of tie range from this to the next phase.} The y-axis and x-axis represent tie range of social ties in phase 1 and in a subsequent phase, respectively. Social ties that dissolved in the corresponding phase are disregarded in the analysis. The numbers on the cells indicate the corresponding transition probabilities from phase 1 to (\textbf{a}) phase 2, (\textbf{b}) phase 4, and (\textbf{c}) phase 8.}
    \label{fig:matrix}
\end{figure*}

\
\newline
Next, we investigate the dynamics of tie range. We first examine the dynamic trends of tie range in the first two phases by analyzing the social ties that exist in both phases. We present the transition probability matrix between tie ranges in the left panel of Fig.~\ref{fig:matrix}. As shown in the figure, all social ties have a large likelihood of evolving into short ties. In particular, for longer ties, i.e. those with a tie range of $=5$ or $\geq 6$, their probability of evolving into a tie range equal to $2$ is the largest: 32\% or 36\%, respectively. Few short ties become long ties, since such an evolution requires that all their common neighbors dissolve with either of them. In addition, long ties appear to be a stable status. For example, a social tie range $\geq 6$ in phase 1 has a probability of 34\% or 15\% to have a tie range of $5$ or $\geq 6$ in phase 2, respectively.

We further analyze the tie range dynamics in phase 4 and phase 8, which are presented in the middle and right panels of Fig.~\ref{fig:matrix}. We find the patterns in phases 4 and 8 are largely consistent with the pattern in phase 2. In particular, for those with a tie range $=5$ or $\geq 6$ in phase 1, they have a probability of 26\% or 38\%, respectively, to persist with a tie range $\geq 5$ in phase 4; they also have a probability of 41\% or 52\%, respectively, to persist with a tie range $\geq 5$ in phase 8.
These results indicate that although long ties have a high probability of becoming short ties, they can also persist as long ties. This finding suggests that it is not necessary for a social tie to become a short-range tie to be long-lasting. 
 
\begin{figure*}
    \centering
    \includegraphics[width=\linewidth]{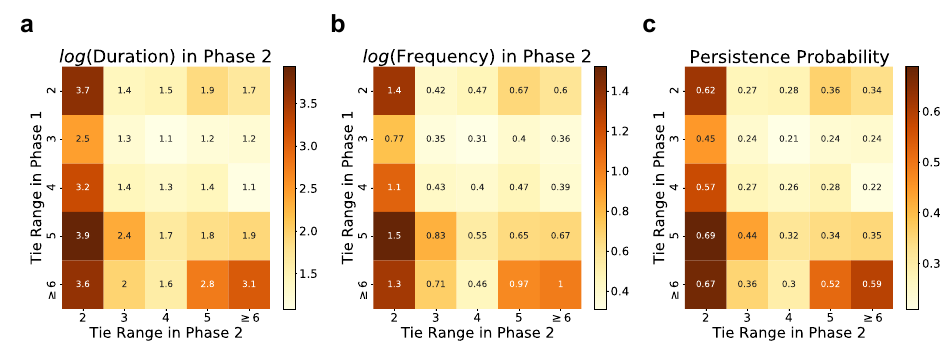}
    \caption{\textbf{Interaction duration, frequency, and persistent probability in the subsequent phase when tie range evolves.} The y-axis and x-axis represent tie range of social ties in phase 1 and in phase 2, respectively. Interaction duration is measured by call volume in seconds. Interaction frequency is the number of calls or texts. Persistence probability is defined as the probability of social ties persisting from phase 1 to phase 2. The numbers on the cells indicate (\textbf{a}) the $\log$ means of interaction duration, (\textbf{b}) the $\log$ means of interaction frequency, and  (\textbf{c}) the probability of persisting  in the next phase.}
    \label{fig:tie_range}
\end{figure*}

Next, we proceed to jointly investigate tie range and tie strength (i.e., the frequency and the total duration of interactions). As shown in Fig.~\ref{fig:tie_range}, in general, those ties that become short-range (e.g., tie range $=2$) are those with more interactions; for social ties that have an arbitrary initial tie range but later change to a tie range of $2$, the interaction frequency and duration are always the greatest. For the persistence probability, the same trend generally holds. The one exception here is for those with a tie range $\geq 6$: if they continue to be social ties with a tie range $\geq 6$, their tie strength remains strong. Note that although we are only discussing phase 1 and phase 2, our results are equally robust when we examine any phase $t$ and its first subsequent phase, $t+1$ (see Supplementary Fig.~S9).

\subsection{Explaining the results: Three hypotheses}
\
\newline
In the previous sections, we show that long ties are not only stronger but also last longer. Moreover, quite a few strong long ties continue to be long ties. To discuss the plausible explanations for the observed patterns, We next propose and discuss three hypotheses pertaining to degree heterogeneity, survival bias, and valuable long ties below. 

\noindent Degree heterogeneity.
\
\newline
First, one plausible explanation for the observed patterns is degree heterogeneity. As shown in Supplementary Fig.~S10, we find that individuals who have fewer friends are more likely to have long ties. Thus, they tend to retain relationships with a small number of friends, but with greater tie strength. 

To reduce the impact of degree heterogeneity, we plot the results conditional on the degree subgroup (see Supplementary Note 6). Specifically, we separate individuals by their degree and obtain multiple degree subgroups. We then plot the main results for each degree subgroup in Supplementary Fig.~S11. We find that the patterns observed in our main text are found in all degree subgroups. 
This finding shows that although degree heterogeneity may provide an explanation for the observed patterns, it does not fully explain our main results.

\noindent Survival bias. 
\
\newline
The second plausible explanation is survival bias -- that only very valuable long ties survived -- even though newly formed long ties are likely to be weaker than newly-formed short ties. Therefore, surviving long ties tend to continue to persist, or perhaps even become stronger, while others dissolve rapidly. To test this hypothesis, we need to examine (1) whether newly formed long ties are weaker than newly formed short ties in the beginning and (2) whether newly formed long ties have a smaller persistence probability, such that only very strong long ties survive. We find that while (1) is supported, (2) is not supported; thus, survival bias cannot fully explain our results.

To investigate these two ideas, we divide social ties into one of two categories: existing ties, and new ties. An existing tie is one that has had any interactions in the previous phase, while a new tie has had no such interactions. After separating all ties into existing or new ones, we perform the same analysis as that found in the previous sections. We use the tie range in phase 2 as the reference, and we investigate whether there was non-zero interaction frequency or duration in order to determine if it is a new or existing tie. 

\begin{figure*}
    \centering
    \includegraphics[width=0.66\linewidth]{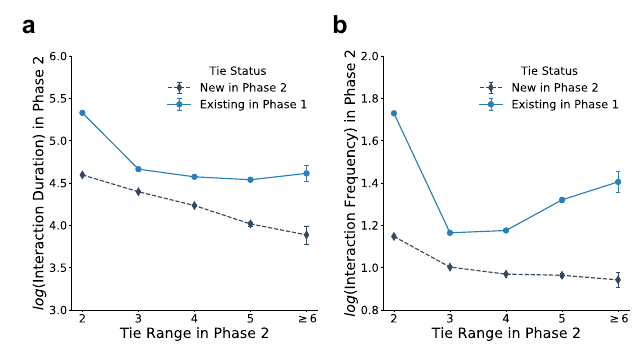}
    \caption{\textbf{Interaction duration and frequency for newly formed ties and existing ties conditional on an existing tie in phase 2.} An existing tie is one that has had any interactions in the previous phase, while a new tie has had no such interactions. All ties are classified according to their tie range in phase 2. The curves represent (\textbf{a}) the average ($\log$) interaction duration  and (\textbf{b}) the average ($\log$) interaction frequency  in phase 2 of newly formed ties and existing ties with the given tie range, respectively.  Error bars are 95\% confidence intervals for the means (assuming normal distribution). Note that error bars are sometimes smaller than the data point markers.}
    \label{fig:Fig.new_old}
\end{figure*}

We first examine whether newly formed long ties are weaker initially than newly formed short ties. In Fig.~\ref{fig:Fig.new_old}, we show that while existing ties present a ``U-shape'' in the relationship between interaction frequency (duration) and tie range in phase 2, this ``U-shape'' pattern does not hold for new ties. Instead, as indicated by Fig.~\ref{fig:Fig.new_old}, for new ties, the longer the new tie is, the fewer interactions the two people have in phase 2. This result supports our conjecture that newly formed long ties are likely to be weaker than newly formed short ties. 

Next, we investigate whether newly formed long ties have a smaller persistence probability. However, we observe that for newly formed ties, there exists a ``U-shape'' between tie range and persistence probability; newly formed long ties have the highest persistence probability (see Supplementary Note 7). This finding contradicts our conjecture that the persistence probability of newly formed long ties would be the smallest. Thus, for the two notions we examined, we find that (1) is supported while (2) is not supported. Therefore, the survival bias hypothesis does not fully explain our main results.

\noindent Valuable long ties. 
\
\newline
Our last hypothesis is that long ties tend to be more valuable. 
This hypothesis is consistent with weak tie theory and the roles of long ties, as conjectured in previous studies~\cite{granovetter1973strength,watts1998collective}.
However, while most computational models that simulate real-world networks highlight homophily~\cite{mcpherson2001birds} --
the phenomenon that individuals with similar attributes tend to be friends --
previous models do not typically consider the benefits of social exchange between people with different skills or information sets~\cite{yuan2018interpretable}. 
Recent work~\cite{yuan2018interpretable}, provides an example of how one can consider homophily and social exchange jointly, but this work is restricted to static social networks. Below, we propose a computational model that combines game theory and machine learning in order to examine long tie dynamics. This model helps support our hypothesis on valuable long ties, while also incorporating the first two hypotheses.

\subsection{The model explaining long ties' persistency}
\
\newline
Here, we propose a game-theoretical computational model that simulates the dynamics of social networks. Specifically, the model combines the embedding techniques in machine learning~\cite{perozzi2014deepwalk,grover2016node2vec,kipf2016semi,velickovic2018graph} and the strategic network formation in economics~\cite{jackson1996strategic,christakis2020empirical}. Compared to the common network formation game models in the economics literature, our model stresses the high-dimensional heterogeneity, as well as the values of social exchange. Compared to network embedding techniques, our model helps understand the social network formation mechanisms. Ultimately, our model integrates the strategic network formation approach to explain the mechanisms, while the embedding techniques improve the predictability of the computational model. Our study echoes Hofman's (2021) recent paper that discusses the trade-off between explanation and prediction in computational social science~\cite{hofman2021integrating}.

Our model considers two procedures during the formation of social ties: the meeting procedure, and the choice procedure. This two-step model takes into account the dynamics of social ties -- that people first meet others randomly, and then make their rational decisions about the choice of friends. The meeting procedure models reality, wherein people meet each other at random. There may exist many potential neighbor candidates who are mutually beneficial (e.g., some potentially valuable long ties), but the extremely low meeting probability can prevent the social tie from being formed. Moreover, when first meeting a new neighbor, a person may lack sufficient information to assess the person, and they are unable to make a rational decision about the social tie. After getting to know a new friend over a period of time (one phase in our study), the individual can then start to make a rational decision about that person. The choice procedure assumes that individuals are rational when choosing their network neighbors and that each individual maximizes their utility function.

Formally, let $\mathcal{I}$ be the set of individuals and let $i$ (or $j$, $\ell$) be their index. Additionally, let $t$ index the discrete time steps (or phases), and thus, $t \in \mathbb{N}^+$. Also, let $\mathbf{A}^{(t)}$ denote the adjacency matrix in phase $t$. $\mathbf{A}^{(t)}_{ij} = 1$ indicates that $i$ and $j$ are connected in phase $t$. $\mathbf{A}^{(t)}_{ij} = 0$ indicates that $i$ and $j$ are disconnected in phase $t$. For simplicity, we only consider an undirected network, i.e., $\mathbf{A}^{(t)}_{ij} = \mathbf{A}^{(t)}_{ji}$ for all $i, j \in \mathcal{I}$, and for all, $t \in \mathbb{N}^+$. To account for the heterogeneity of individual attributes, we use the ``endowment vector'' $\mathbf{w}_i$, which is a $K$-dimensional vector as in the embedding techniques~\cite{perozzi2014deepwalk,grover2016node2vec}. As embedding techniques do, each dimension measures a certain latent attribute of an individual, such as a type of skill or useful information. A larger $w_{ik}$ indicates that the individual retains a high endowment of the $k^{\text{th}}$ dimension.

In each phase, the neighbor's set of $i$ consists of two components: the new friend set $\mathcal{M}^{(t)}_i$, and the existing friend set $\mathcal{N}^\textit{(t)}_i$; which echoes our analysis newly formed ties and existing ties. The new friend set is formed in the random meeting procedure. We assume each pair of individuals has a different meeting probability. The concept of a ``meeting probability'' is found widely in several econometric studies that aim to model social network formation~\cite{mele2017structural,overgoor2019choosing,christakis2020empirical}. Specifically, for each pair of individuals, $i$ and $j$, they have a probability of $p_{ij}^{(t)}$ to ``meet'' each other in phase $t$. If $\mathbf{A}^{(t-1)}_{ij} = 1$, that is, the two individuals were connected in phase $t-1$, then the $p_{ij}^{(t)}$ is a large probability. Otherwise, $p_{ij}^{(t)}$ is a small probability, dependent on the network topology between $i$ and $j$. Inspired by our previous comparison between newly formed ties and existing ties, we can imagine that if this is a long tie, the probability would be much smaller. Formally, we parametrize $p_{ij}^{(t)}$ as follows:

\begin{equation}
   p_{ij}^{(t)} =  \begin{cases} 
      d_{t-1}(i, j) & \mathbf{A}^{(t-1)}_{ij} = 0\\
      q & \mathbf{A}^{(t-1)}_{ij} = 1
   \end{cases} 
\end{equation}

The distance metric $d_{t-1}(i,j)$ depends on the network topology between individual $i$ and individual $j$ in phase $t-1$. We define the distance metric to be proportional to the probability of random walks from $i$ to $j$. Here, $q$ is set to describe the probability of maintaining the meeting procedure in phase $t$. 

The second component is the existing friend set $\mathcal{N}^{(t)}_i$, which is determined by the rational choice procedure.  It is a subset of all friends in phase $t-1$, i.e., $\mathcal{N}^{(t)}_i \in \mathcal{M}^{(t-1)}_i \cup \mathcal{N}^{(t-1)}_i$. This means that individuals make rational decisions after maintaining their friendships for a period of one phase. The rationale behind this notion is that individuals need a significant amount of time to assess the value of an existing friend, so the rational choice procedure happens in the phase immediately following the meeting procedure. For a connected social tie in phase $t-1$, the friendship must survive both the meeting procedure (a random draw from Bern($q$)) and the rational choice procedure. The choice procedure is modeled using the following utility function:

\begin{equation}
    \begin{split}
        U_i^{(t)}(\mathbf{c}_{i}^{(t)}) = \sum_{j\in \mathcal{M}^{(t-1)}_i  \cup \mathcal{N}^{(t-1)}_i }  \left( c_{ij}^{(t)}\sum_k \left(\sigma \left(w_{jk} - w_{ik} \right) + \sum_{\ell\in  \mathcal{M}^{(t-1)}_j \cup \mathcal{N}^{(t-1)}_j } \delta \sigma \left(w_{\ell k} - w_{ik} \right) \right) - \left(c_{ij}^{(t)}\right)^2 \right), \\ 
        \text{ where } \sum_j \left( {c}_{ij}^{(t)} \right) ^2 = 1.
    \end{split}
    \label{eq:utility}
\end{equation}

Here, $U_i^{(t)}$ is the utility function of individual $i$ in phase $t$. $\mathbf{c}_{i}^{(t)} \in [0, 1]^{ \mathcal{M}^{(t-1)}_i \cup \mathcal{N}^{(t-1)}_i}$, which can be understood as a function that maps any $j$ in the neighbor set in phase $t-1$, i.e., each element in $\mathcal{M}^{(t-1)}_i \cup \mathcal{N}^{(t-1)}_i$, to a real number in $[0, 1]$. The utility function sums over all $i$'s neighbors in phase $t-1$. $\sigma$ is the ReLU function: if $w_{jk} - w_{ik} > 0$, the output is $w_{jk} - w_{ik}$; otherwise, 0. $\ell$ enumerates over all $j$'s neighbors in phase $t-1$, which are also $i$'s ``friends' friends.'' The depreciation factor $\delta$, which ranges in $(0, 1)$, measures how the value of a potential friend depreciates as the distance on the network increases. We refer to $ \sigma(w_{jk}-w_{ik}) + \sum_{\ell \in \mathcal{M}_j^{(t-1)} \cup \mathcal{N}_j^{(t-1)} }{\delta \sigma(w_{\ell k} - w_{ik})}$ as the benefit that $j$ brings to $i$. In addition, we separate the benefit into two: the direct benefit, $ \sigma(w_{jk}-w_{ik}) $, and the indirect benefit, $\sum_{\ell \in \mathcal{M}_j^{(t-1)} \cup \mathcal{N}_j^{(t-1)} }{\delta \sigma(w_{\ell k} - w_{ik})}$. The design of these benefit terms was intended for our valuable long tie hypothesis -- we hope to observe that long ties have, on average, larger values in the direct benefit term.

$c_{ij}^{(t)}$ measures the time investment of $i$ in $j$. A non-zero value of $c_{ij}^{(t)}$ indicates that $j$ belongs to $\mathcal{N}^t_i$. The restriction of the sum of squared $c_{ij}^{(t)}$ reflects that people have limited time or energy to invest in their neighbors. The benefit of each neighbor is proportional to the time or energy investment in each neighbor $j$; this is why we multiply the benefit term by $c_{ij}^{(t)}$. At the same time, the squared term $\left( c_{ij}^{(t)} \right) ^2$  is used to measure the cost of time or energy. The design of $c_{ij}^{(t)}$ echoes our degree heterogeneity hypothesis -- that those with many ties may have less investment in any one individual neighbor.

By the Cauchy-Schwarz inequality, Equation~(\ref{eq:utility}) can be solved by 

\begin{equation}
     (c_{ij}^{(t)})^* \propto  \sum_k \left(\sigma \left(w_{jk} - w_{ik} \right) + \sum_{\ell\in  \mathcal{M}^{(t-1)}_j \cup \mathcal{N}^{(t-1)}_j } \delta \sigma \left(w_{\ell k} - w_{ik} \right) \right), \text{ and }\sum_j \left((c_{ij}^{(t)})^*\right) ^2 = 1.
    \label{eq:attention}
\end{equation}
\noindent In particular, 
\begin{equation}
     j \in \mathcal{N}^{(t)}_i \text{ iff } \left({c}_{ij}^{(t)}\right)^* > 0; \\ 
     j \notin \mathcal{N}^{(t)}_i \text{ iff } \left({c}_{ij}^{(t)} \right)^*= 0. \\ 
\end{equation}

In other words, if the optimal solution informs $\left( c^{(t)}_{ij} \right)^* = 0$, then this indicates that $i$ and $j$ are no longer connected. Otherwise, $\left( c^{(t)}_{ij} \right)^*$ is the fraction of the call duration during which $i$ interacts with $j$ at time $t$ among $i$’s total call duration at time $t$.

This model provides major improvements based on the framework proposed in prior work~\cite{yuan2018interpretable}. First, different from their paper, we establish a model for network dynamics. In particular, we incorporate a meeting procedure; this addresses the phenomenon that, in reality, there are many neighbor candidates who do not form links purely because they have no opportunity to meet. Second, our model also takes into account the ``weight'' (i.e., the interaction frequency or duration) of the links. This is different from Yuan et al.~\cite{yuan2018interpretable}, where the weights between the links are binary. Third, Yuan et al.~\cite{yuan2018interpretable} assumes that the marginal utility of additional neighbors is not dependent on other existing neighbors; by contrast, our model does not incorporate this assumption, and it also accounts for the network externality (i.e., the benefits of friends of friends)~\cite{jackson1996strategic}. 
We provide additional analyses to verify our modeling fitting capacity in Supplementary Note 8.

Figure~\ref{fig:model} provides the main implications derived from the learning results of our model. We first present the average benefit, i.e., $ \sigma(w_{jk}-w_{ik}) + \sum_{\ell \in \mathcal{M}_j^{(t-1)} \cup \mathcal{N}_j^{(t-1)} }{\delta \sigma(w_{\ell k} - w_{ik})} $, given the different tie range in Panel (a) of Fig.~\ref{fig:model}. The average is taken over all candidate neighbors in $\mathcal{M}_j^{(t-1)} \cup \mathcal{N}_j^{(t-1)}$ given the tie range in phase $t-1$. From this, we find a ``U-shape'', i.e., the average benefit decreases with the tie range at the beginning, but later increases with the tie range. This is consistent with our previous findings regarding the ``U-shape'' between tie range and tie strength. 

\begin{figure*}
    \centering
    \includegraphics[width=\linewidth]{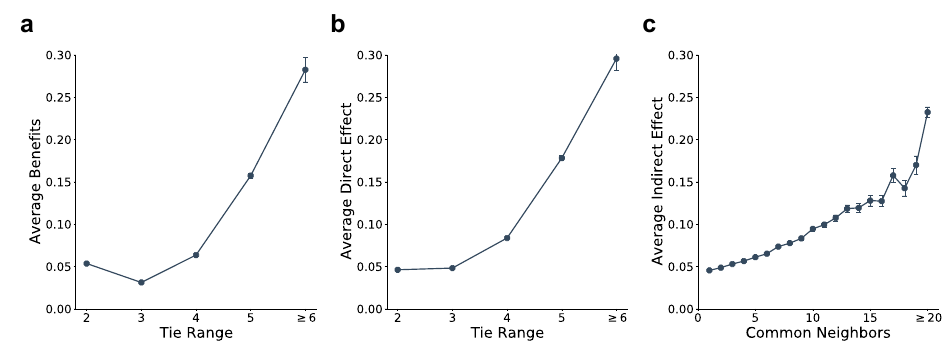}
    \caption{\textbf{Average benefits, average direct effects, and average indirect effects learned by our model.} 
    The curves represent \textbf{(a)} the average benefits, \textbf{(b)} average direct effects, and \textbf{(c)} average indirect effects for each tie range learned from our model, respectively. 
    Error bars are 95\% confidence intervals for the benefits (assuming normal distribution). Note that error bars are sometimes smaller than the data point markers.}
    \label{fig:model}
\end{figure*}

Next, we separate the benefits in Equation~(\ref{eq:utility}) into the direct effect and the indirect effect. We present the average direct effect, which is $\sigma(w_{jk}-w_{ik})$ in Panel (b) of Fig.~\ref{fig:model}. We observe an increasing pattern with the tie range, indicating that as the tie range increases, the average benefit that a tie brings also increases. This result supports our hypothesis that long ties tend to be more valuable, which also explains the results in the previous sections. We also compute the average indirect effect, i.e., $\sum_{\ell \in \mathcal{M}_j^{(t-1)} \cup \mathcal{N}_j^{(t-1)} }{\delta \sigma(w_{\ell k} - w_{ik})} $. In our model, only social ties with common friends, i.e., those with a tie range of $2$, have indirect effects. We plot the relationship between the number of common neighbors and the average indirect effect. The indirect effect echoes our previous discussion on patterns of social ties with a tie range of $2$. As observed in Panel (c) of Fig.~\ref{fig:model}, we find an increasing pattern. In particular, by examining the first several data points in the plot, we observe a seemingly convex pattern, indicating the increasing marginal utility of common neighbors. 

Overall, the results from our learning model suggest that long ties are generally more valuable (with greater direct effects). This model also takes into account degree heterogeneity and survival bias hypotheses, although they are probably not the primary drivers. We also compare our model with other baseline models in Supplementary Note 9, but they cannot provide the implications as we plot in Fig.~\ref{fig:model}.

\section*{Conclusion}

In this study, we combine empirical analysis and an interdisciplinary computational model to investigate the dynamics of long ties. We find that long ties persist longer than shorter-range ties and that many long ties are persistently long. These results are contrary to what is suggested by several prior theories and prediction models. To better understand our results, we propose three hypotheses -- degree heterogeneity, survival bias, and valuable long ties -- and then go on to discuss the limitations of both the degree heterogeneity hypothesis and the survival bias hypothesis. Finally, we discuss an interdisciplinary model that combines game theory and machine learning to support our valuable long-tie hypothesis. Verified by real-world data, our model partly explains why long ties are more persistent than what has previously been suggested by existing theories and models.

Our results also signal the importance of social interventions that promote the formation of long ties, such as mixing diverse people with diverse backgrounds. For example, both our empirical analysis and modeling results indicate that people who are dissimilar in certain attributes or who are distant in a social network may have significant mutual benefits to one another. However, as indicated by our model, the small likelihood of those people meeting can hinder the formation of their future interactions.

Based on this study, there are several interesting research directions that could be investigated. First, although we examine a large-scale social network with very few missing nodes, the generalizability of our results should be interrupted cautiously. On the one hand, our study replicates the U-shape in Park et al.~\cite{park2018strength} which examines multiple static phone communication and Twitter networks. The successful replication provides confidence in the potential generalizability of our additional dynamic analyses to these networks. On the other hand, there are many other types of social ties rather than phone communications, such as social media, offline interactions, or collaboration networks. We appeal for more studies on this important topic to verify the external validity of our conclusions. Second, although most existing studies on long ties, including ours, use the aggregate data to measure the tie range, it is interesting to investigate how to leverage advanced methods of analyzing temporal networks to further understand the mechanisms of dynamics of long ties, which can examine events occurring on network paths on a more fine-grained level~\cite{holme2012temporal,holme2015modern,sekara2016fundamental}. Finally, there may be intriguing variants of our model. For example, our model only reflects the absolute advantages that other people bring, but it would be interesting to incorporate comparative advantages in our model, as well.

\begin{methods}

\subsection{Data description}
\ 
\newline
In our study, we use a nationwide call detail record dataset. Users' private information has been anonymized and thus we are unable to identify them. This data provider is a company that functions as the main service provider for most of the mobile phone users in a European region. The time period covered by the data starts from Jan. 2015 to Dec. 2016. In the dataset, we retrieve the total number of calls, texts, as well as the duration of calls between any two people in each month. See Supplementary Note 1 for more details.

We establish a temporal social network with the dataset. We consider discrete time steps (or phases): for each phase, we construct a ``snapshot'' of the network, where the node indicates a user and the edge represents the interaction between two users. A key question is how we determine the length of the time window of each phase. In our main results, we treat every three months as a phase. In Supplementary Note 2, we also use one month or six months to verify the robustness of our results. 

To maintain a temporal network where the node set is stable and the global network structure does not change dramatically with the dynamics of a few nodes, we only consider the interactions among users who have at least one call or text in every phase. 
We construct a temporal directed network with 45,192 nodes and 385,533 edges on average for each phase.

In terms of the weight of the directed network, we consider two variables as mentioned in the main text: interaction frequency and duration. Interaction frequency is the total number of calls or text that node $i$ sends to $j$; there are a few calls with zero-second duration and we filter those calls out.  Interaction duration is the total time length that $i$ calls $j$  in each phase, and does not account for texting.

\subsection{Tie range and long ties}
\ 
\newline
Tie range~\cite{granovetter1973strength,park2018strength} is defined as the length of the second shortest path between two connected nodes (Fig.~\ref{fig:definitionTR}). It indirectly reflects the network distance of the connection. Consistent with previous long tie studies~\cite{centola2007complex,park2018strength}, there is no clear cutoff of tie range that decides whether a tie is short or long. A good reference is the Milgram experiment, which suggested that the average network distance between every two people is approximately $6$. In our study, we treat social ties with a tie range of $2$ as short ties, and ties with $5$ or $\geq 6$ as long ties. Besides, we do a sensitive check of our results by randomly dropping a proportion (5\%) of nodes or edges (see Supplementary Note 3). Our main results are verified not sensitive to a few nodes or edges happening to exist on the network. 

\subsection{Details in learning}
\ 
\newline
Based on Equation~(\ref{eq:attention}), we construct the loss function to minimize the MSE Loss between $c_{ij}$ and its right hand side. We use stochastic gradient descent to optimize the loss function. For each epoch, we construct our loss function as below: 

\begin{equation}
    \mathcal{L} = \mathcal{L}_{pos} + \mathcal{L}_{neg},
\end{equation}

\noindent The loss function is composed of the loss functions of positive (connected pairs), and negative samples (disconnected pairs).

\begin{equation}
    \mathcal{L}_{pos} = \sum_{i\in \text{sampled}} \left( \frac{\sum_{j\in \left( \mathcal{N}_i^{(t-1)} \cup \mathcal{M}_i^{(t-1)} \right) \cap \mathcal{N}_i^{(t)}} |\hat c^{(t)}_{ij} - c^{(t)}_{ij}|}{\sum_{j\in \left( \mathcal{N}_i^{(t-1)} \cup \mathcal{M}_i^{(t-1)} \right) \cap \mathcal{N}_i^{(t)}} 1} \right);
\end{equation}

\begin{equation}
    \mathcal{L}_{neg} = \sum_{i\in \text{sampled}} \left( \frac{\sum_{j\in \left( \mathcal{N}_i^{(t-1)}\cup \mathcal{M}_i^{(t-1)} \right) \setminus \mathcal{N}_i^{(t)}} \hat c^{(t)}_{ij}}{\sum_{j\in \left( \mathcal{N}_i^{(t-1)}\cup \mathcal{M}_i^{(t-1)} \right) \setminus \mathcal{N}_i^{(t)}} 1} \right).
\end{equation}

The set ``\text{sampled}'' denotes the set of sampled nodes in each epoch. For positive samples, we minimize the difference between  $c^{(t)}_{ij}$, the time investment of $i$ on $j$, and the predicted time investment denoted by $\hat c^{(t)}_{ij}$.

\begin{equation}
    c^{(t)}_{ij} = \frac{\log\left ( D^{(t)}_{ij}+1 \right )}{\sum_{j\in  \mathcal{M}^{(t-1)}_i \cup \mathcal{N}^{(t-1)}_i } \log\left ( D^{(t)}_{ij}+1 \right )},
\end{equation}

\noindent where $D_{ij}^{(t)}$ is the interaction duration between $i$ and $j$ in phase $t$. To reduce the impact of extreme values, we take the logarithm of $D_{ij}^{(t)}$. Since $D_{ij}^{(t)} \geq 0$, $c^{(t)}_{ij} \geq 0$.

\begin{equation}
     \hat c^{(t)}_{ij} = \frac{\exp \left\{\sum_k \left(\sigma \left (w_{jk} - w_{ik} \right) + \sum_{\ell\in  \mathcal{M}^{(t-1)}_j \cup \mathcal{N}^{(t-1)}_j } \delta \sigma \left(w_{\ell k} - w_{ik} \right) \right) \right\}}{\sum_{j'\in  \mathcal{M}^{(t-1)}_i \cup \mathcal{N}^{(t-1)}_i } \exp \left\{\sum_k \left(\sigma \left(w_{j'k} - w_{ik} \right) + \sum_{\ell\in  \mathcal{M}^{(t-1)}_{j'} \cup \mathcal{N}^{(t-1)}_{j'} } \delta \sigma \left(w_{\ell k} - w_{ik} \right) \right) \right\}}.
\end{equation}

When minimizing the loss function, we treat the time investment of $i$ in $j$, which is calculated by the interaction duration or frequency, as the input and endowment vectors in this loss function as the variables to be inferred. Note that the existence of the $\delta$ may result in an uncontrollable gradient issue. We thus use grid search for this variable and check the robustness of our results in Supplementary Note 8. Moreover, we also discuss the selection of the number of dimensions of the endowment vectors in Supplementary Note 8.

To facilitate the learning process, we apply mini-batch stochastic gradient descent with Adam optimizer~\cite{kingma2014adam}. Consistent with conventional network embedding algorithms, node sampling probability is proportional to node degree ($d^{\frac{3}{4}}$)~\cite{mikolov2013distributed}. In this case, the endowment vectors of both these sampled nodes and their neighbors will be updated in each epoch in the gradient descent. In Supplementary Note 8, we show that our learning converges under this setting. Details in the machine learning implementation are also discussed in Supplementary Note 8.

\subsection{Ethical declaration}
\ 
\newline
Our study has been determined to be exempt by MIT IRB (COUHES). Exempt ID: E-3442.

\end{methods}

\clearpage
\section*{Reference}
\bibliographystyle{naturemag}
\bibliography{IMDLT}
\clearpage
\begin{addendum}
    \item [Data Availability] Data is available at \href{https://github.com/DingLyu/Investigating-and-Modeling-the-Dynamics-of-Long-Ties}{https://github.com/DingLyu/Investigating-and-Modeling-the-Dynamics-of-Long-Ties}. Differential privacy is applied to protect the privacy of users.
    \item [Code Availability] Code is available at \href{https://github.com/DingLyu/Investigating-and-Modeling-the-Dynamics-of-Long-Ties}{https://github.com/DingLyu/Investigating-and-Modeling-the-Dynamics-of-Long-Ties}.
    \item [Author contributions] D.L. and Y.Y. conceived the present idea. Y.Y. collected and processed the data. D.L. and Y.Y. analyzed the results. D.L., Y.Y., L.W., X.W. and A.P discussed the analytical approach and furthered the results. D.L. and Y.Y. wrote the paper with input from L.W., X.W. and A.P. All authors have reviewed and commented on the manuscript.
    \item [Competing interests] The authors declare no competing interests.
    \item[Correspondence] Correspondence and requests for materials should be addressed to Yuan Yuan (email: yuanyuan@purdue.edu).
\end{addendum}

\clearpage
\section*{Supplementary Information}
\appendix
\setcounter{figure}{0} 
\setcounter{table}{0} 
\renewcommand\thefigure{S\arabic{figure}}    
\renewcommand\thetable{S\arabic{table}}

\section*{Supplementary Note 1: Data processing and summary statistics}
In our study, we use a nationwide mobile phone call dataset involving about 45 thousand (45192) people's phone call logs in 2 years from Jan. 2015 to Dec. 2016. This is a European region with more than 50 thousand but fewer than 100 thousand citizens. We aggregate the monthly phone call and texting log for each pair of users. Then we take a series of snapshots by aggregating all activities happening in a time window. We have flexibility in the choice of the time window. We establish a directed graph including all phone call logs in the time window. As mentioned in the main text, we primarily consider two types of edge weights -- interaction frequency and interaction duration. Interaction frequency is the phone call counts between two people, and interaction duration is the sum of call volumes of all phone calls in an interval. 

We next discuss how to select the time window. Note that the selection of the time window affects the proportion of each possible tie range. A too narrow time window may result in each snapshot being so sparse that many short-range ties might be treated as long-range ties. A too wide time window may result in too few snapshots for us to analyze the network dynamics. Eventually, we choose a season (three months) as the time window for the main text. Each season or three months is regarded as a ``phase.''

As the length of our data is two years, we partition the data into eight phases. As the definition of tie range, we classify all connections with respect to tie range in each phase. Due to the small magnitude of ties over range $6$, we merge them as $\geq 6$. In addition, some ties with an infinite tie range cannot be ignored. As illustrated in Tab.~\ref{tab:Table.S1}, social ties with a tie range of $5$ or $\geq 6$ only take a small proportion of all connections.

We also present the statistics for interaction duration, interaction frequency, degree, and tie range for each phase in Tab.~\ref{tab:Table.S2}. As shown in the Table, the average interaction frequency and duration do not change over time. We do see the average degree decreases and the average tie range increases over time, which is a result of the network getting sparser. However, this would not affect our main results (Fig.~2 and Fig.~3), as we examine the dynamics of interaction frequency or duration conditional on the tie range in the first phase only (see 
Supplementary Note 4 for more details).

\section*{Supplementary Note 2: Robustness of the choice of time windows}

\begin{figure*}
\captionsetup[subfigure]{labelformat=simple, font={small}}
\centering
\subfloat[]{
	\includegraphics[width=0.3\linewidth]{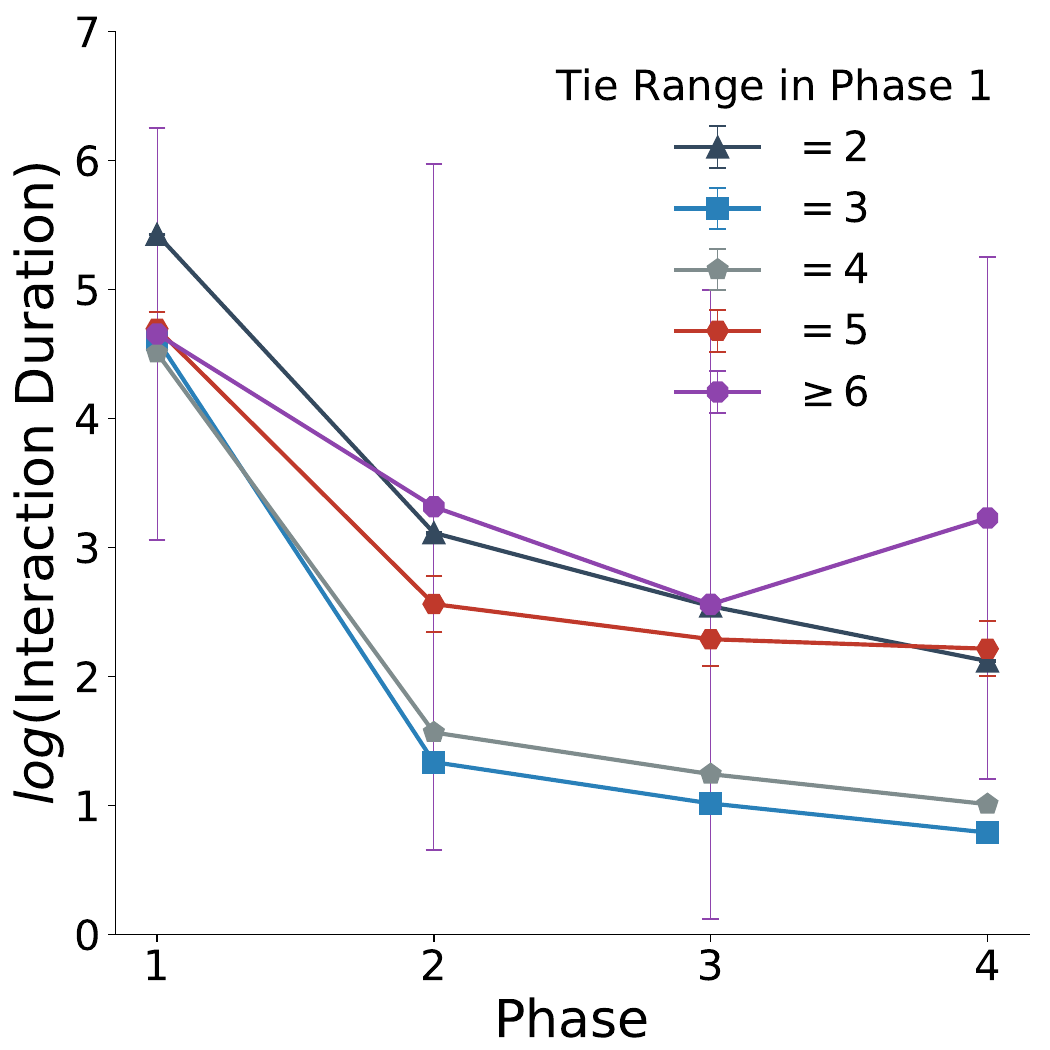}}
\quad
\subfloat[]{
	\includegraphics[width=0.6\linewidth]{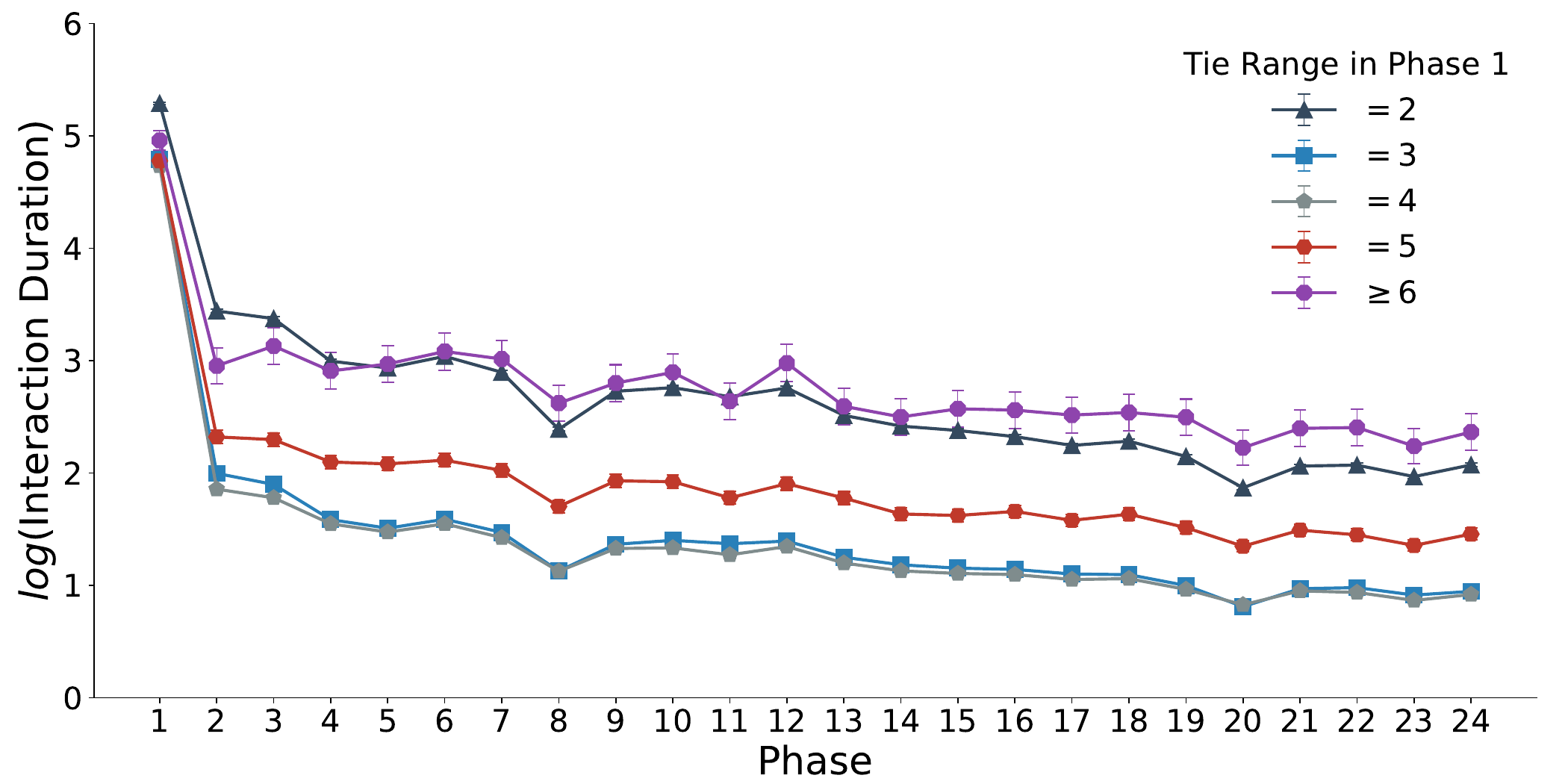}}
\vfill
\subfloat[]{
	\includegraphics[width=0.3\linewidth]{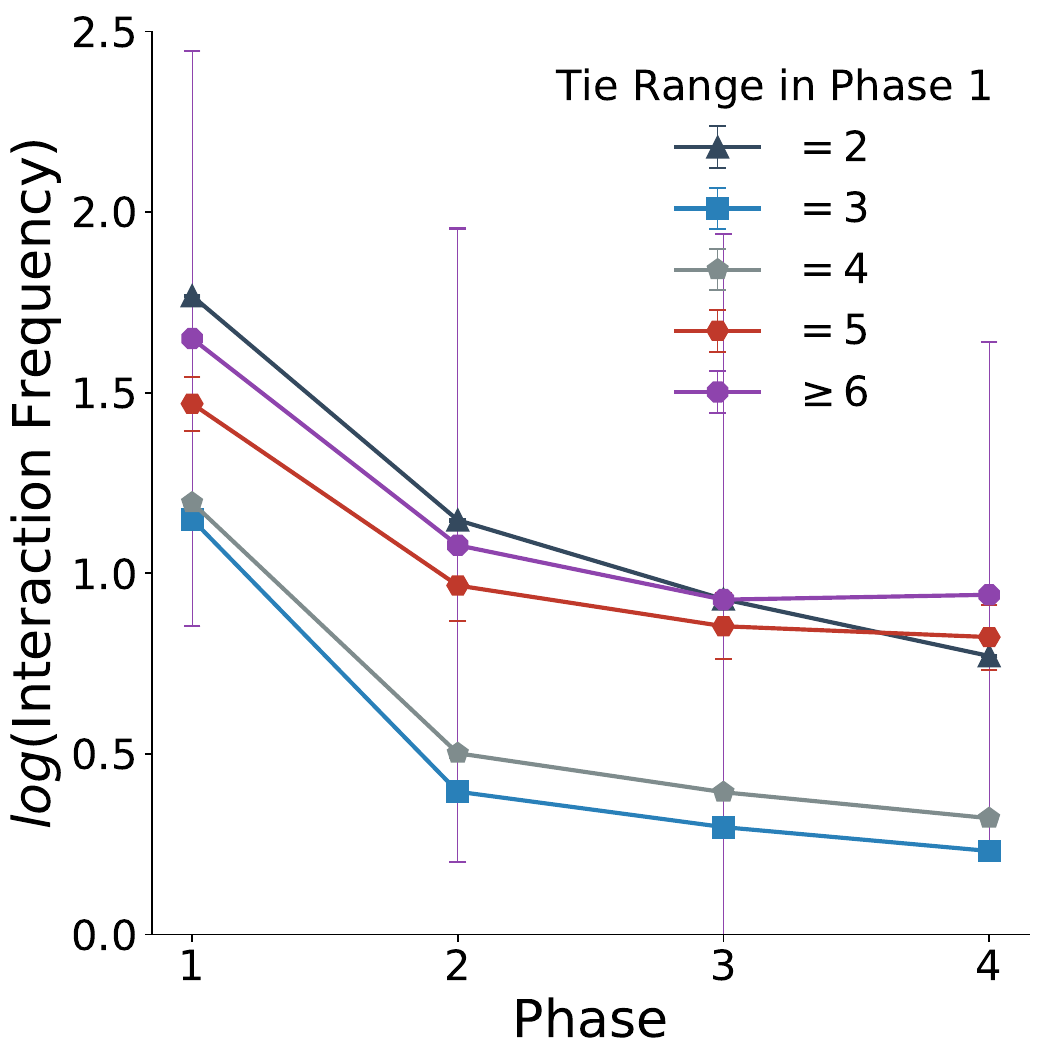}}
\quad
\subfloat[]{
	\includegraphics[width=0.6\linewidth]{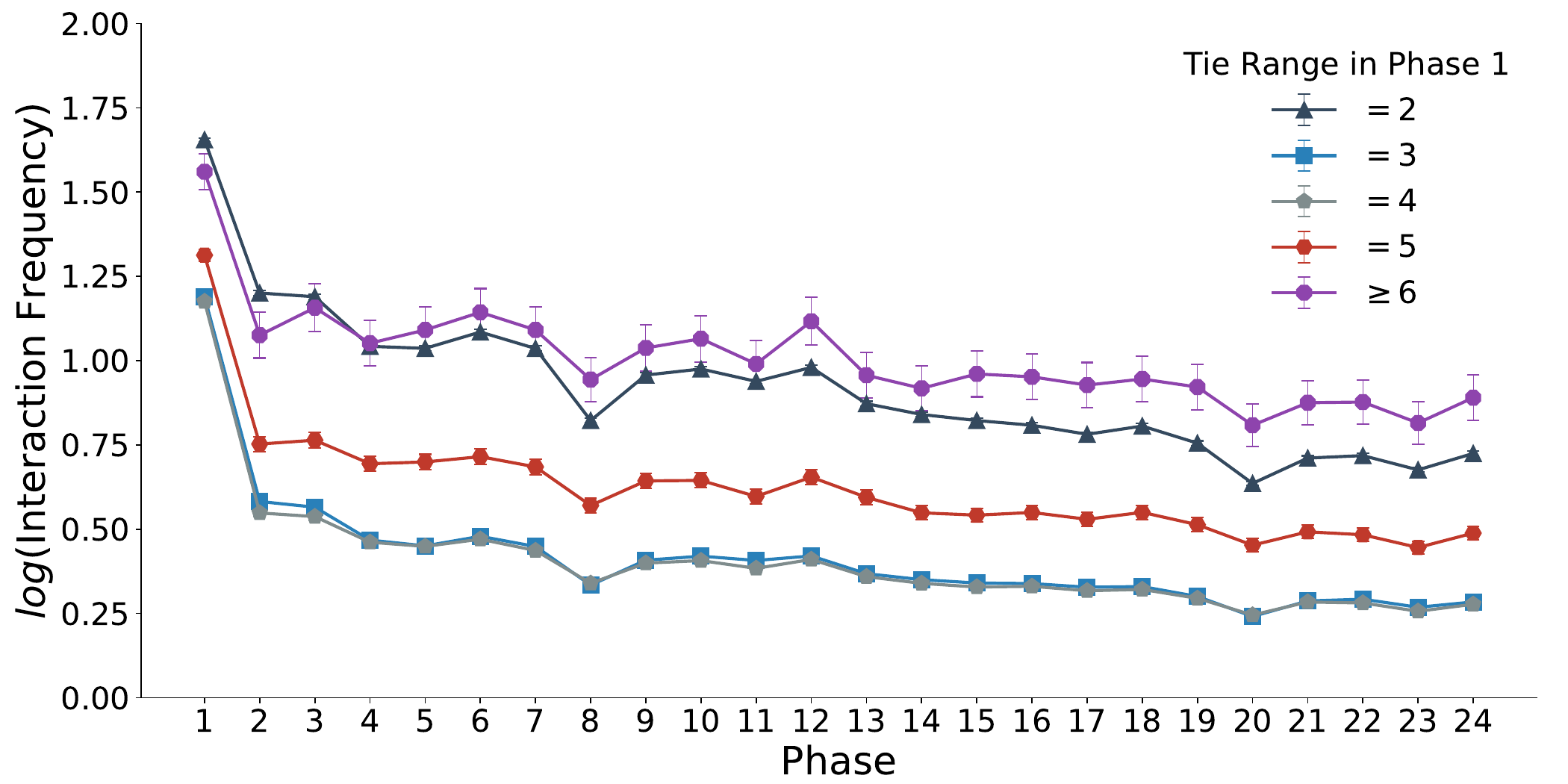}}
    \caption{\textbf{Dynamics of tie strength throughout semi-yearly snapshots and monthly snapshots.} Tie strength is measured by interaction duration (\textbf{a\&b}; the total call volume in seconds) and interaction frequency (\textbf{c\&d}; the number of calls or texts). Either a semi-year (\textbf{a\&c}; six months) or a month (\textbf{b\&d}) is set as the time window. We take logarithms ($\log$) for both interaction duration and frequency. All ties are classified according to their tie range in the first phase. The curves represent the average ($\log$) interaction duration or frequency conditional on that a tie exists in phase 1 with the given tie range. Error bars are 95\% confidence intervals for the mean $\log$ interaction duration and frequency (assuming normal distribution). Note that error bars are sometimes smaller than the data point markers.}
\label{fig:Fig.S1}
\end{figure*}

To test for the robustness of the choice of the time window, we further adjust the time windows. When the time interval is set as a month, we obtain 24 monthly snapshots. We respectively calculate the tie range of each edge in every snapshot. Consistent with the main text, we use the logarithm value of interaction frequency and duration so a few extreme values would not unreasonably affect the averages. Fig.~\ref{fig:Fig.S1}(a\&b), (c\&d) present our main results after adjusting the time window. We observe a very similar trend with the results when the time window is three months. 

The result from weekly aggregation is presented in Fig.~\ref{fig:S3}. We find that again, long range ties (especially those with tie range $\geq 7$) have greater interactions than short range ties (tie range $=2$) in the long term. Note that when we aggregate to small time windows, the distribution of tie range is shifted to have a fatter tail larger (Fig.~\ref{fig:S4}); we thus need to change the cutoff to 7 to maintain relatively the same proportion of ties as ``long range ties.''

We also examine the results from data aggregation. Since the snapshot of one-day interactions may miss a great number of persistent social ties that happen not to interact on one specific day, we use a sliding window with a length of seven days, but we move the window day by day. In this way, our resolution is still on the day level. The results are presented in Fig.~\ref{fig:S5} which are consistent with our main text.

\section*{Supplementary Note 3: Sensitivity check}

Since the tie range of an edge is easily impacted by another node or edge that is distant on the network, we need to conduct examine how our results are sensitive to the existence of a few nodes or edges. We examine the sensitivity of our results to the impacts of certain nodes or edges. We randomly drop a proportion (5\%) of nodes or edges and then replicate our main result. As shown in Fig.~\ref{fig:Fig.S6}, dropping either nodes or edges would not affect our main conclusions. This indicates that our results are not sensitive to a few nodes or edges happening to exist on the network.

\section*{Supplementary Note 4: Explanation of the decreasing pattern}

Note that Fig.~2 in the main text exhibit decreasing trends for all curves. This is because our analyses are the average interaction frequency and interaction given that a tie exists at phase 1 ($E[y_t|y_1>0]$). Therefore as $t$ ($>1$) increases, we expect a proportion of social ties to terminate, which drives the decreasing pattern.

We hope to clarify that the result is not driven by a decaying trend in activity ($E[y_t]$). In Fig.~\ref{fig:Fig.S7}, we plot the conditional ($E[y_t|y_1>0]$) and unconditional ($E[y_t]$). From the figure, we do not observe that the average activity ($E[y_t]$) changes over time. Therefore, our main result is not driven by a decaying trend in activity. 

\section*{Supplementary Note 5: Lifespan of social ties}

In the main text, we use the persistence probability and interaction increments to investigate the dynamics of social ties. Here we use the ``lifespan'' as the other dimension to measure the dynamics. When defining the lifespan, we explore two choices: (1) the social tie has to have interactions for every phase within the lifespan; and (2) a social tie has interactions in the first and the last phases no matter whether they have interactions in the phases in between. As shown in Fig.~\ref{fig:Fig.S8}, there are also U-shapes regarding the relationship between tie range and lifespans, regardless of the choice of the definition of the lifespan. This result further verifies our statement in the main text, i.e., ``long ties persist longer.''

\section*{Supplementary Note 6: Degree heterogeneity hypothesis}

Here we discuss our ``degree heterogeneity'' hypothesis. First, as shown in Fig.~\ref{fig:Fig.S10}, individuals with fewer neighbors, i.e., a lower degree, tend to have more long ties. We then categorize social ties by degree and plot the trends for each subgroup in Fig.~\ref{fig:Fig.S11}. We find that our main results persist in all degree subgroups. Therefore, the degree heterogeneity hypothesis cannot fully explain our main results.

\section*{Supplementary Note 7: Survival bias hypothesis}

To test for this hypothesis, we need to examine whether (1) newly formed long ties are weaker than newly formed short ties in the beginning; and (2) newly formed long ties have a smaller persistence probability such that only very strong long ties survive. 

The plot is presented in the main text.
We find that for new ties, the tie strength is weakest for those with tie range $\geq 6$. By contrast, for existing ties, the trend appears to be a ``U-shape.'' Thus, we support ``newly formed long ties are weaker than newly formed short ties in the beginning''. For hypothesis (2), we re-conduct the analysis by decomposing the outcome into persistence probability and interaction increments. However, we find that newly formed long ties still have the largest persistence probability. Thus (2) is not supported. We therefore believe that the survival bias hypothesis cannot fully explain our main results.

\section*{Supplementary Note 8: Details in learning}

Here we provide more technical details regarding the learning process of our proposed model. In our proposed model, we need to learn both hyper-parameter $\delta$ and endowments. However, simultaneously training $\delta$ and endowment vectors may cause an uncontrollable gradient issue. Therefore, we first try to find the optimal $\delta$ and then train endowment vectors by minimizing the loss. From the data, we observe there is a positive indirect effect from common friends, and thus $\delta$ should be a small positive value. As shown in Fig.~\ref{fig:Fig.S13}, we find that the model performs better when we set $\delta$ as 0.2 than other options -- the fit result $\hat{c}_{ij}$ is closest to the real-world data ${c}_{ij}$. 

After determining the value of $\delta$, we next infer the endowment vectors. To speed up the learning rate of the model, we adopt a sampling strategy. We set the maximum number of epochs as 500 and randomly sample 1000 nodes in each epoch. According to the loss function, sampled nodes and their neighbors will receive a gradient descent, and endowment vectors of them will be updated in each epoch. We set a testing set of 1000 nodes to track the learning curve of the model. As shown in Fig.~\ref{fig:Fig.S14}, the loss appears to converge to stable after 100 epochs. 

As to the dimension selection of endowment vectors, we investigate how different selections of the dimensions impact our main results. We test it from 2-dimensional to 5-dimensional endowment vectors. Note that a too large dimensionality may raise the issue of computational complexity. We present the results corresponding to Fig.~7(a) in the main text in Fig.~\ref{fig:Fig.S15}. As shown in the figure, the conclusions from different dimensions are largely similar. We therefore choose the dimensionality of four as an illustration in the main text. 

We implemented our algorithm in PyTorch. The endowment vectors are implemented as embeddings in PyTorch, and we use Adam optimizer with regularization for the optimization.

\section*{Supplementary Note 9: Baseline model comparisons}

We compare our model with two baselines: the classic connections model~\cite{jackson1996strategic} and a simplified version of our model.

The classic connections model also established a utility function describing the benefits and costs of forming additional links. Using the notations in our study, the utility function can be written down as:

\begin{equation}
    U_i^{(t)}(\mathbf{c}_{i}^{(t)}) = \sum_{j\in \mathcal{M}^{(t-1)}_i \cup \mathcal{N}^{(t-1)}_i } {\left( \delta+a_{ij}^{(t-1)}\delta^2-c_{ij}^{(t)} \right)},
\label{eq:baseline1}
\end{equation}

where $\delta$ is the direct benefit from the connection between node $i$ and node $j$. $a_{ij}^{(t-1)}$ denotes the number of common neighbors of node $i$ and node $j$ at phase $t-1$, thus $a_{ij}^{(t-1)}\delta^2$ is the indirect benefit from common neighbors. Again, $c_{ij}^{(t)}$ is the time investment of node $i$ in node $j$ at phase $t$. We can further consider a higher order of indirect benefits, such as the benefits of three-hop neighbors (neighbors' neighbors' neighbors).\footnote{That is, $U_i^{(t)}(c_{ij}^{(t)})=\delta+a_{ij}^{(t-1)}\delta^2+b_{ij}^{(t-1)}\delta^3-c_{ij}^{(t)}$, where $b_{ij}^{(t-1)}$ is the number of paths with a length of three between node $i$ and node $j$ at phase $t-1$. Note that considering even higher order indirect effects (e.g., four-hop neighbors) gives rise to the issue of high computational complexity.}

Furthermore, we introduce another baseline, which is a simplified version of our model with indirect effects removed, defined as below:

\begin{equation}
    \begin{split}
        U_i^{(t)}(\mathbf{c}_{i}^{(t)}) = \sum_{j\in \mathcal{M}^{(t-1)}_i \cup \mathcal{N}^{(t-1)}_i } c_{ij}^{(t)}\sum_k \left(\sigma \left(w_{jk} - w_{ik} \right) - \left(c_{ij}^{(t)}\right)^2 \right), \\ 
        \text{ where } \sum_j \left( {c}_{ij}^{(t)} \right) ^2 = 1.
    \end{split}
    \label{eq:baseline3}
\end{equation}

Compared to the version in the main text (i.e., Eq.~3), we remove the indirect effects; that is $\sum_{\ell\in \mathcal{M}^{(t-1)}_j \cup \mathcal{N}^{(t-1)}_j } \delta \sigma (w_{\ell k} - w_{ik})$. 

We present the results in Fig.~\ref{fig:Fig.S16} which tries to generate Fig.~7 -- like plots using different baseline models. The $\delta$ in the first two models are learned from empirical data. The (a), (b), and (c) panels correspond to the connections model of second-degree indirect effects, the connections model of three-degree indirect effects, and the simplified version of our model, respectively. We observe that these baselines cannot fully explain some of our key findings. The panels (a) and (b) do not reveal the specialty of long range ties -- the curves display the non-increasing patterns in tie range and the benefits would be a constant ($\delta$) after a certain cutoff ($3$ and $4$ respectively). The curves in these two panels do not exhibit a ``U-shape'' reflected in the main text anymore. Panel (c) is the result of the simplified version of our model. Although it reflects that longer-range ties have greater benefits, it also tends to consider short range ties (those with a tie range of $2$) the least beneficial. Thus, it neither presents a ``U-shape'' which we anticipated. Taken together, none of these baseline models reflects the ``U-shape'' observed in the previous empirical results. 

\clearpage

\begin{table}
    \centering
    \scalebox{0.75}{
    \begin{tabular}{c|cccccccc}
    \hline
    \textbf{Tie Range} & \textbf{Phase 1} & \textbf{Phase 2} & \textbf{Phase 3} & \textbf{Phase 4} & \textbf{Phase 5} & \textbf{Phase 6} & \textbf{Phase 7} & \textbf{Phase 8}\\
    \hline
         2 & 373,270 & 338,689 & 306,481 & 311,417 & 253,648 & 243,471 & 206,858 & 204,401\\
           & 71.2\% & 69.2\% & 68.2\% & 67.7\% & 64.3\% & 63.0\% & 61.0\% & 59.7\%\\
         3 & 105,438 & 102,713 & 93,316 & 100,617 & 88,051 & 91,261 & 77485 & 83,864\\
           &  20.1\% & 21.0\% & 20.8\% & 21.9\% & 22.3\% & 23.6\% & 22.9\% & 24.5\%\\
         4 & 40,729 & 42,366 & 43,115 & 41,968 & 44,102 & 43,540 & 43757 & 43,405\\
           & 7.77\% & 8.66\% & 9.59\% & 9.12\% & 11.2\% & 11.3\% & 12.9\% & 12.7\%\\
         5 & 4738 & 5264 & 6097 & 5,561 & 7,971 & 7636 & 9973 & 9727\\
           & 0.90\% & 1.08\% & 1.36\% & 1.21\% & 2.02\% & 1.98\% & 2.94\% & 2.84\%\\
         $\geq$ 6 & 284 & 255 & 433 & 409 & 686 & 663 & 1004 & 944\\
           & 0.05\% & 0.05\% & 0.10\% & 0.09\% & 0.17\% & 0.17\% & 0.30\% & 0.28\%\\
         \hline
    \end{tabular}}
    \caption{\textbf{Statistics of ties with different range throughout eight phases.} Each phase represents a season (three months).}
    \label{tab:Table.S1}
\end{table}

\begin{table}
    \centering
    \scalebox{0.75}{
    \begin{tabular}{c|cccccccc}
    \hline
    \textbf{Phase} & \textbf{1} & \textbf{2} & \textbf{3} & \textbf{4} & \textbf{5} & \textbf{6} & \textbf{7} & \textbf{8}\\
    \hline
    $Log(ID)_{min}$&0.69&0.69&0.69&0.69&0.69&0.69&0.69&0.69\\ 
    $Log(ID)_{0.25}$&4.04&4.01&4.04&4.04&4.03&4.03&4.03&4.04\\ 
    $Log(ID)_{avg}$&5.15&5.1&5.12&5.14&5.13&5.12&5.12&5.14\\ 
    $Log(ID)_{med}$&4.04&4.01&4.04&4.04&4.03&4.03&4.03&4.04\\ 
    $Log(ID)_{0.75}$&6.17&6.1&6.1&6.13&6.13&6.12&6.1&6.12\\ 
    $Log(ID)_{max}$&12.01&12.02&12.21&12.5&12.85&13.07&13.1&12.41\\ 
    \hline
    $Log(IF)_{min}$&0.69&0.69&0.69&0.69&0.69&0.69&0.69&0.69\\ 
    $Log(IF)_{0.25}$&0.69&0.69&0.69&0.69&0.69&0.69&0.69&0.69\\ 
    $Log(IF)_{avg}$&1.57&1.55&1.56&1.55&1.54&1.53&1.53&1.52\\ 
    $Log(IF)_{med}$&0.69&0.69&0.69&0.69&0.69&0.69&0.69&0.69\\ 
    $Log(IF)_{0.75}$&2.08&2.08&2.08&1.95&1.95&1.95&1.95&1.95\\ 
    $Log(IF)_{max}$&8.36&7.36&7.22&7.65&7.42&7.41&7.28&7.33\\ 
    \hline
    $d_{min}$&1&1&1&1&1&1&1&1\\ 
    $d_{0.25}$&6&6&5&5&5&5&4&4\\ 
    $d_{avg}$&18.13&17.51&16.57&17.08&15.09&15.11&13.53&13.75\\ 
    $d_{med}$&6&6&5&5&5&5&4&4\\ 
    $d_{0.75}$ &23.0&22.0&21.0&21.0&18.0&18.0&17.0&16.0\\
    $d_{max}$&337&351&325&301&296&295&254&301\\ 
    \hline
    $TR_{min}$&2&2&2&2&2&2&2&2\\ 
    $TR_{0.25}$&2&2&2&2&2&2&2&2\\ 
    $TR_{avg}$&2.39&2.42&2.44&2.44&2.51&2.53&2.59&2.59\\ 
    $TR_{med}$&2&2&2&2&2&2&2&2\\
    $TR_{0.75}$&3&3&3&3&3&3&3&3\\ 
    $TR_{max}$&7&7&7&8&7&7&8&7\\   
    \hline
    \end{tabular}}
    \caption{\textbf{Statistics on interaction duration (ID), interaction frequency (IF), degree (d), and tie range (TR) in eight snapshots at the interval of a season (three months).}}
    \label{tab:Table.S2}
\end{table}

\clearpage

\begin{figure*}
\captionsetup[subfigure]{labelformat=simple, font={small}}
\centering
\subfloat[]{
	\includegraphics[width=0.24\linewidth]{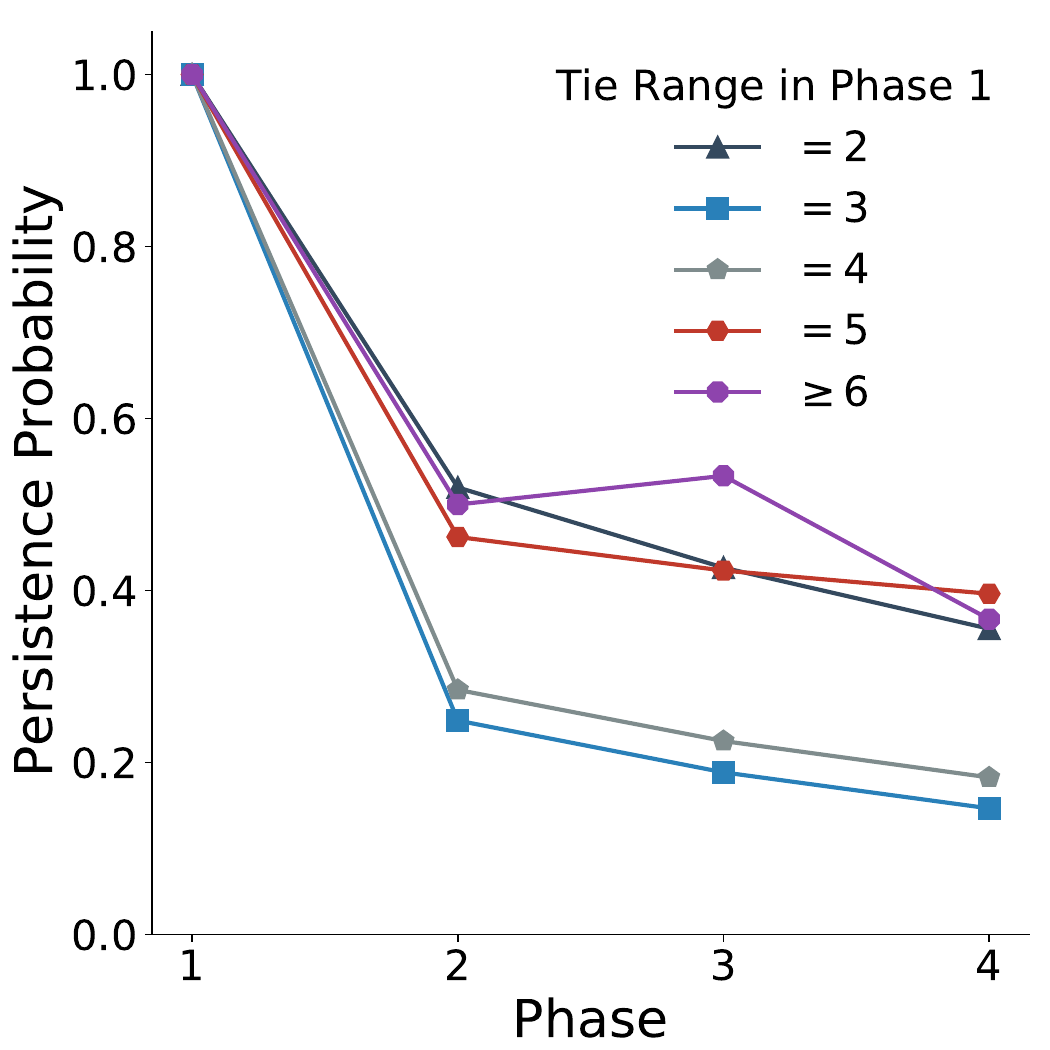}}
\quad
\subfloat[]{
	\includegraphics[width=0.48\linewidth]{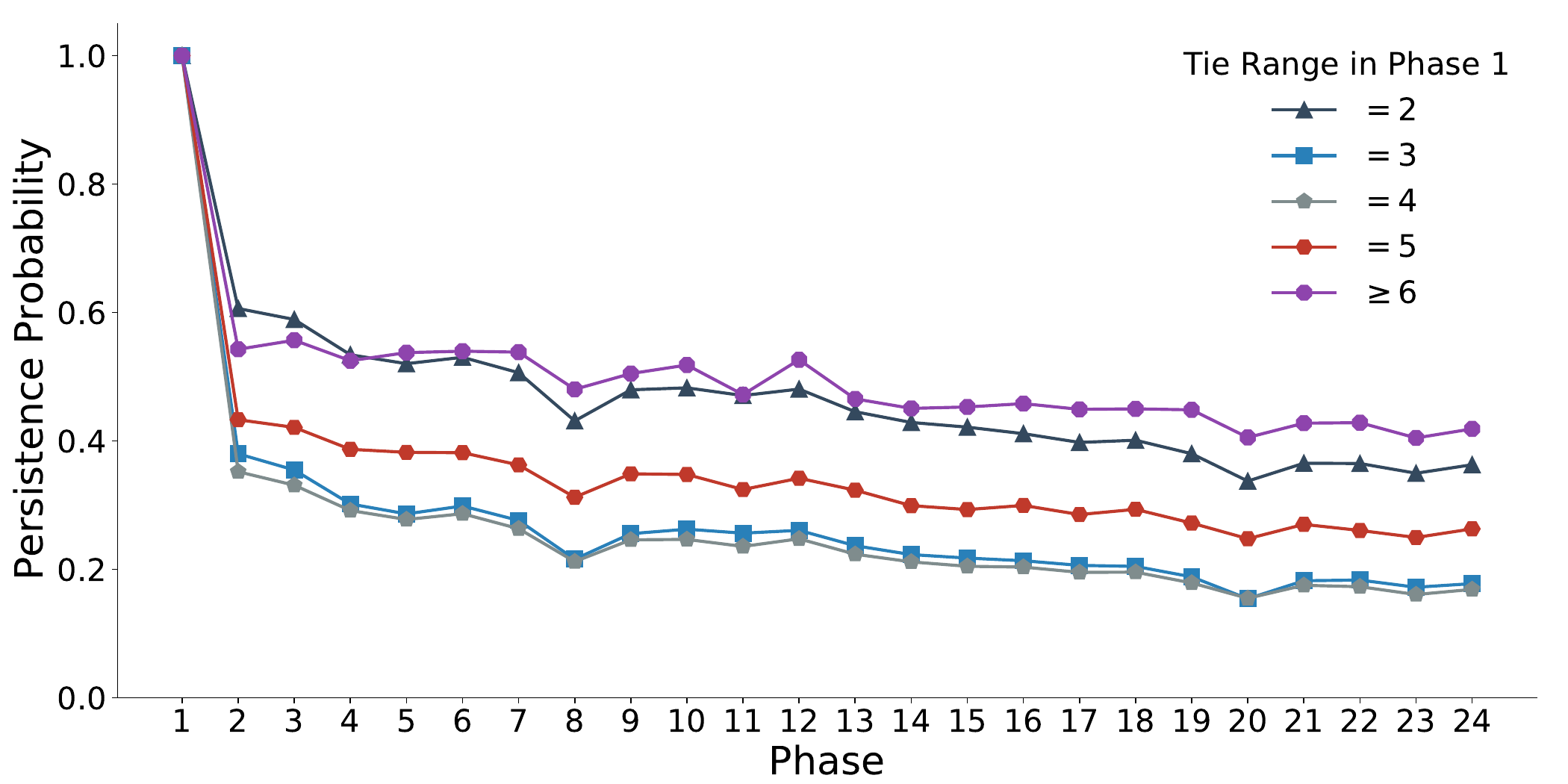}}
\vfill
\subfloat[]{
	\includegraphics[width=0.24\linewidth]{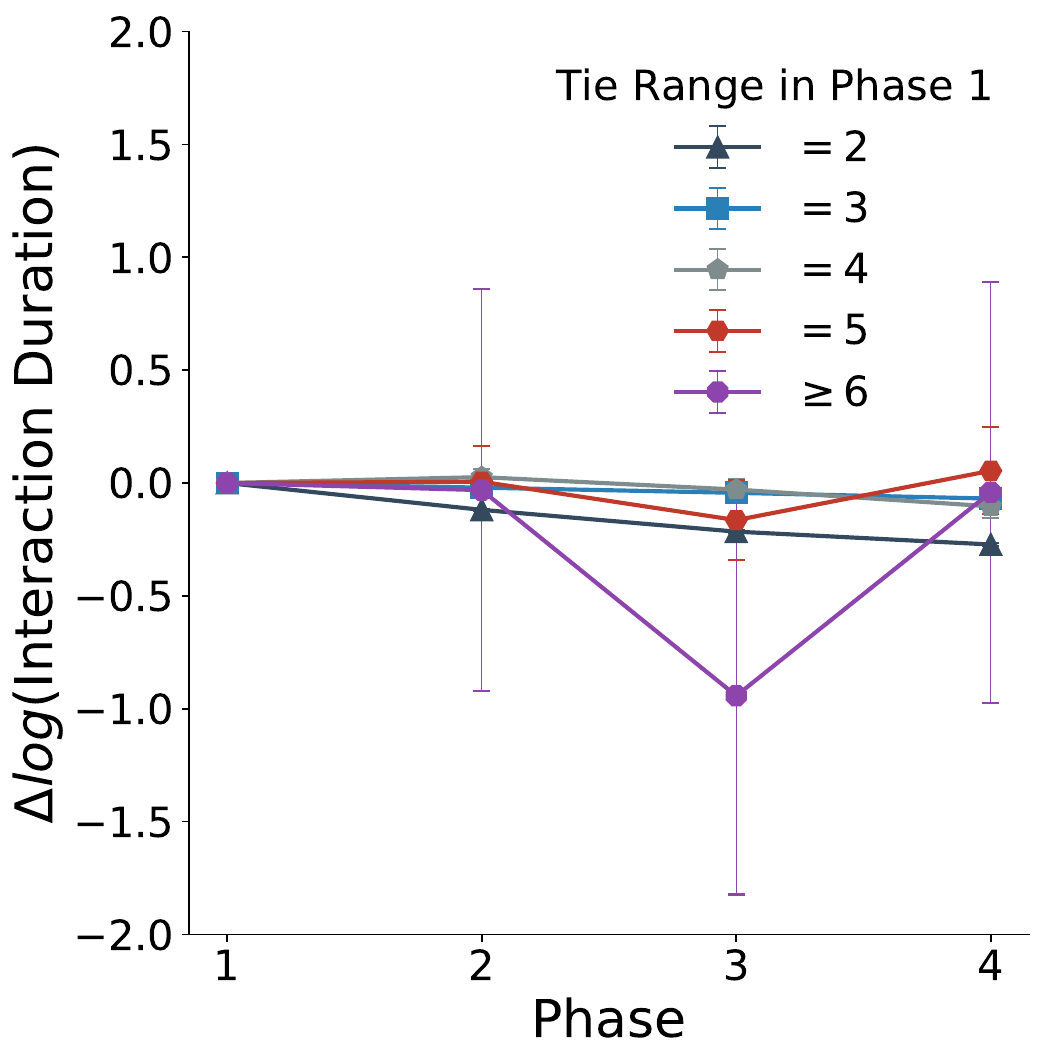}}
\quad
\subfloat[]{
	\includegraphics[width=0.48\linewidth]{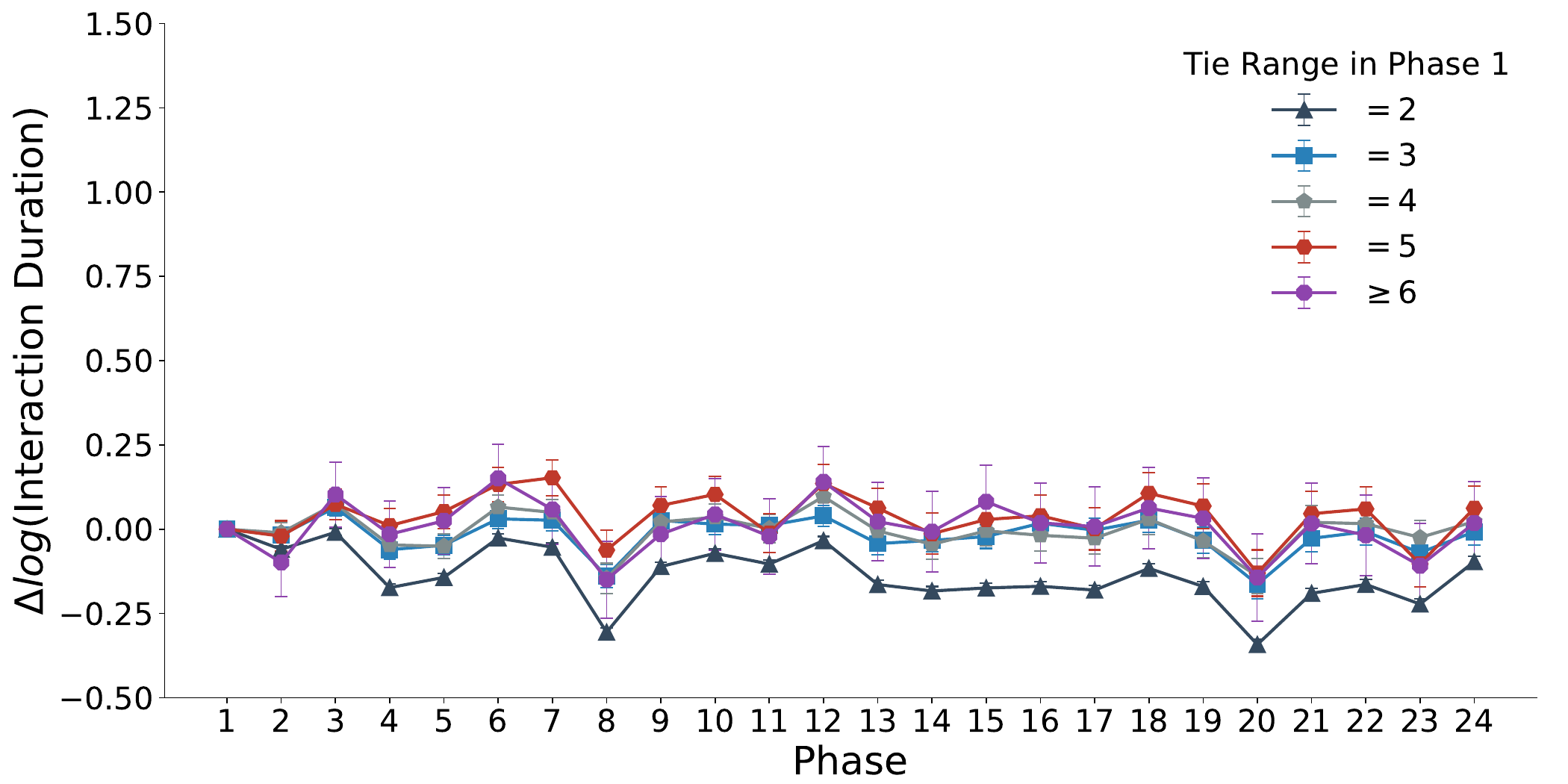}}
\vfill
\subfloat[]{
	\includegraphics[width=0.24\linewidth]{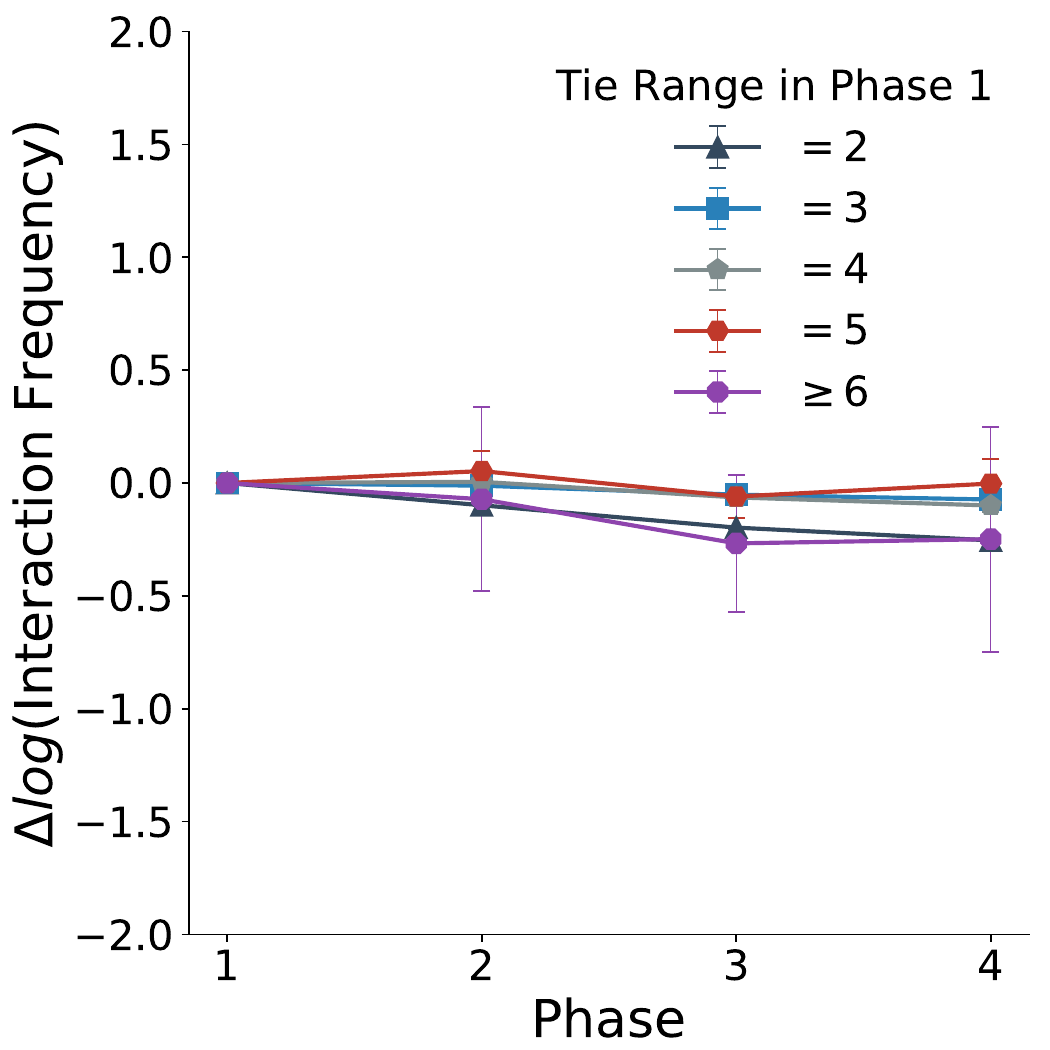}}
\quad
\subfloat[]{
	\includegraphics[width=0.48\linewidth]{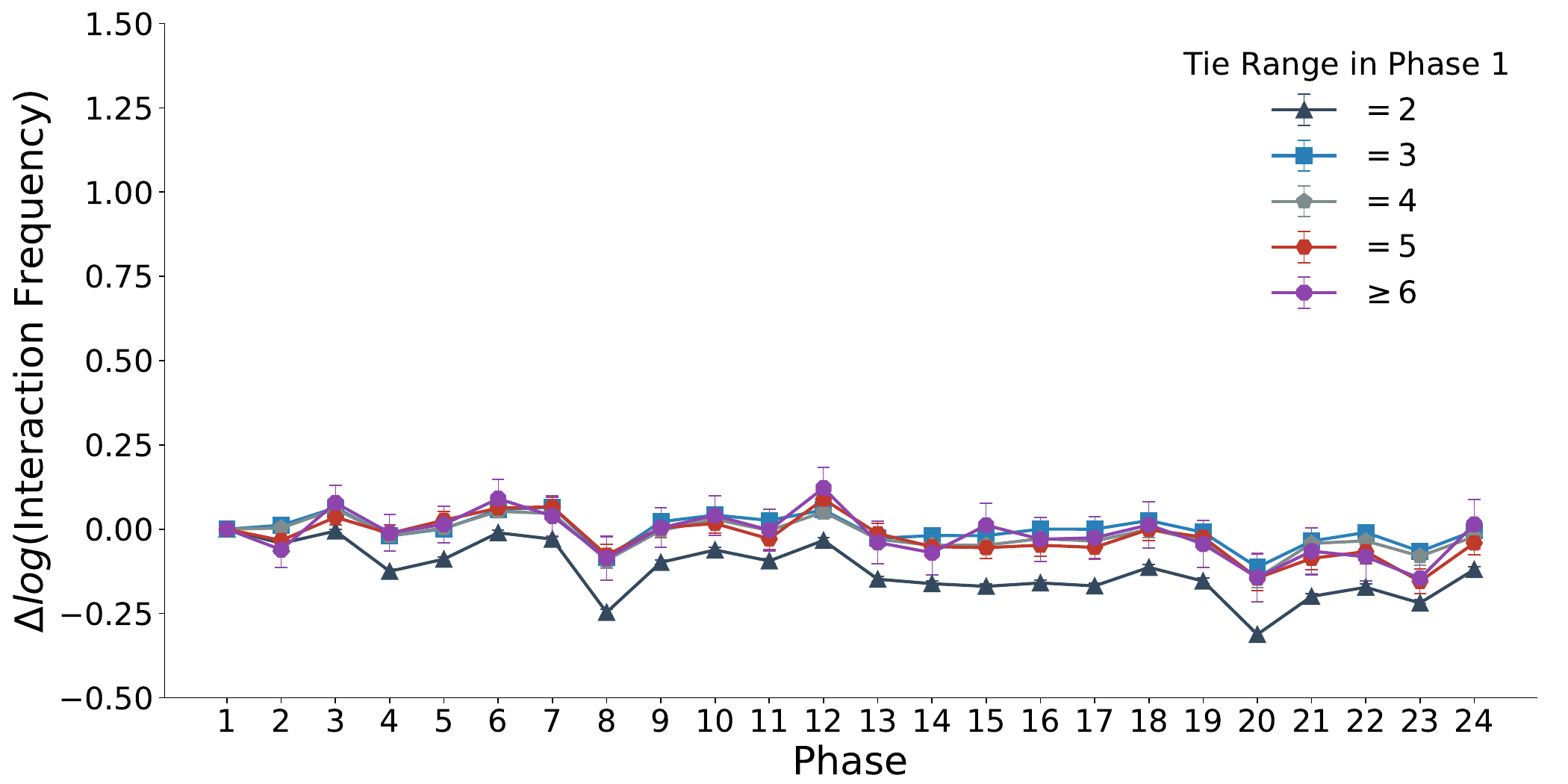}}
    \caption{\textbf{Dynamics of persistence probability and interaction increments conditional on that a tie exists in phase 1.} Either a semi-year (\textbf{a,c\&e}) or a month (\textbf{b,d\&f}) is set as the time window. Interaction duration is measured in seconds. All ties are classified according to their tie range in the first phase. Error bars are 95\% confidence intervals for the mean $\Delta \log$ interaction duration and frequency (assuming normal distribution). Note that error bars are sometimes smaller than the data point markers.}
\label{fig:Fig.S2}
\end{figure*}

\begin{figure*}
    \captionsetup[subfigure]{labelformat=simple, font={small}}
    \centering
    \subfloat[]{
	\includegraphics[width=0.49\linewidth]{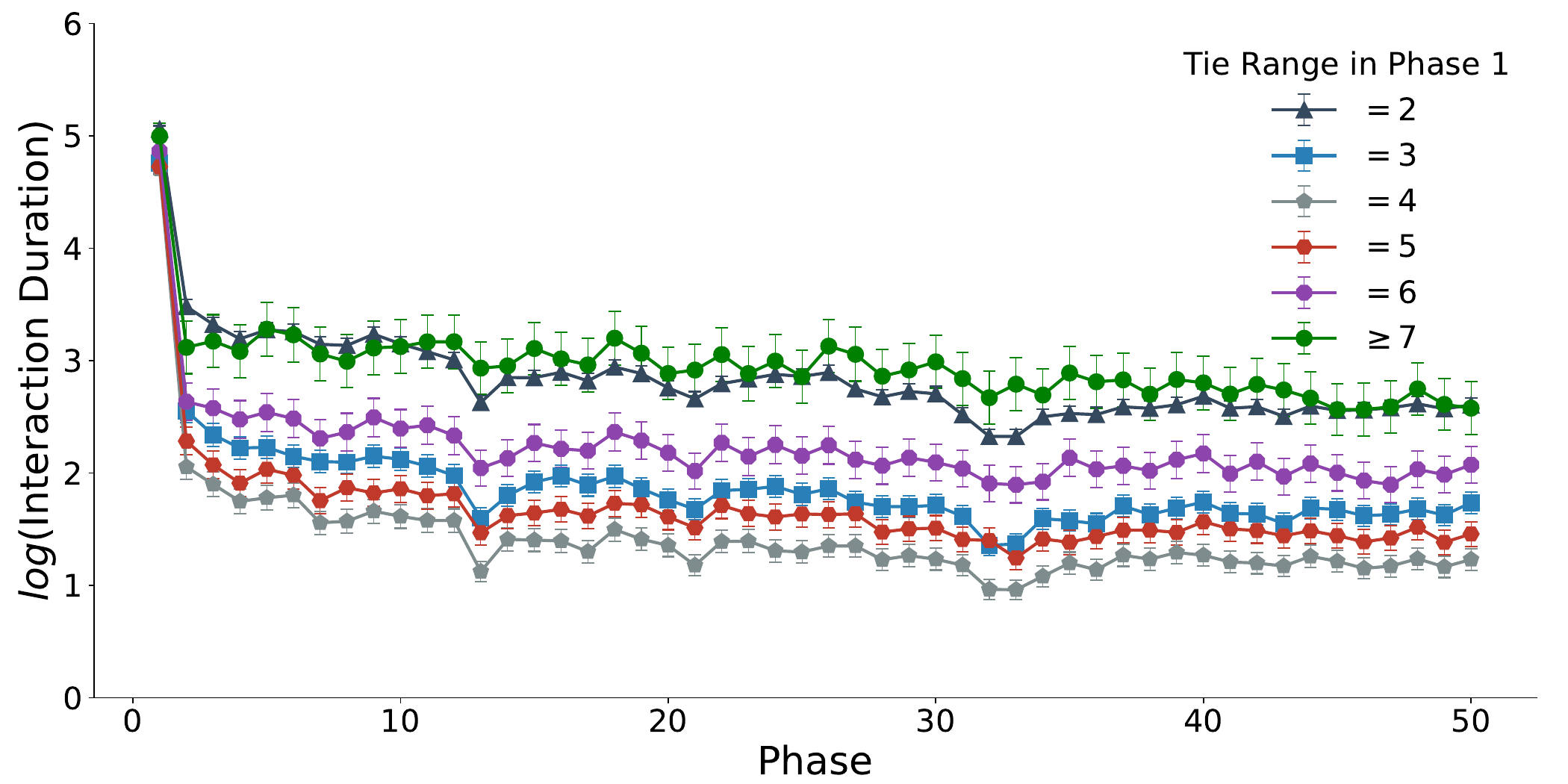}}
	\subfloat[]{
	\includegraphics[width=0.49\linewidth]{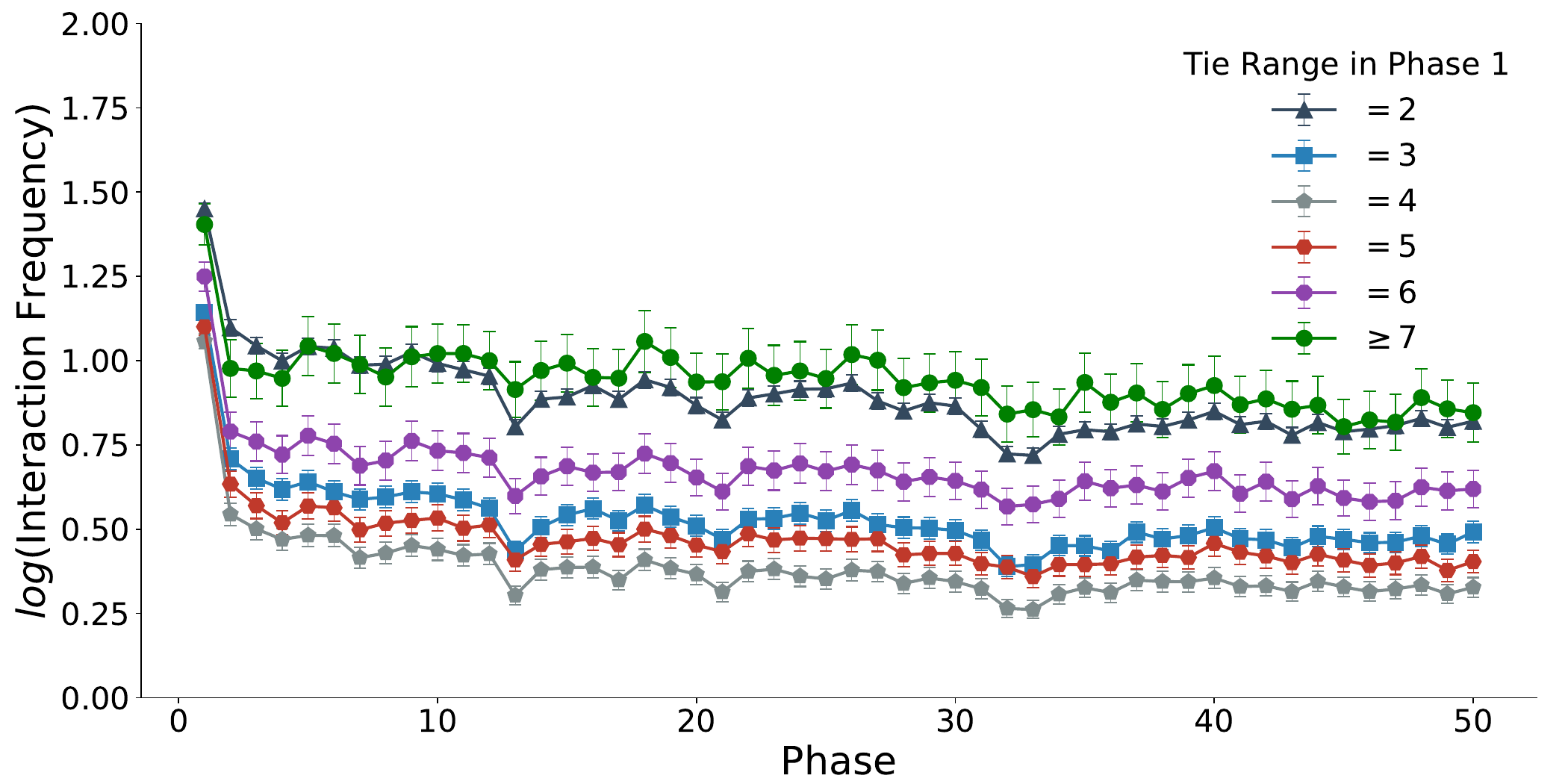}}
    \caption{\textbf{Dynamics of tie strength throughout weekly snapshots.} Tie strength is measured by interaction duration (\textbf{a}; the total duration of the calls in seconds) and interaction frequency (\textbf{b}; the number of calls or texts). Each phase represents a week. We take logarithms ($\log$) for both interaction duration and frequency. All ties are classified according to their tie range in the first phase. The curves represent the average ($\log$) interaction duration or frequency conditional on that a tie exists in phase 1 with the given tie range. Error bars are 95\% confidence intervals for the mean $\log$ interaction duration and frequency (assuming normal distribution).}
    \label{fig:S3}
\end{figure*}

\begin{figure*}
    \centering
	\includegraphics[width=0.3\linewidth]{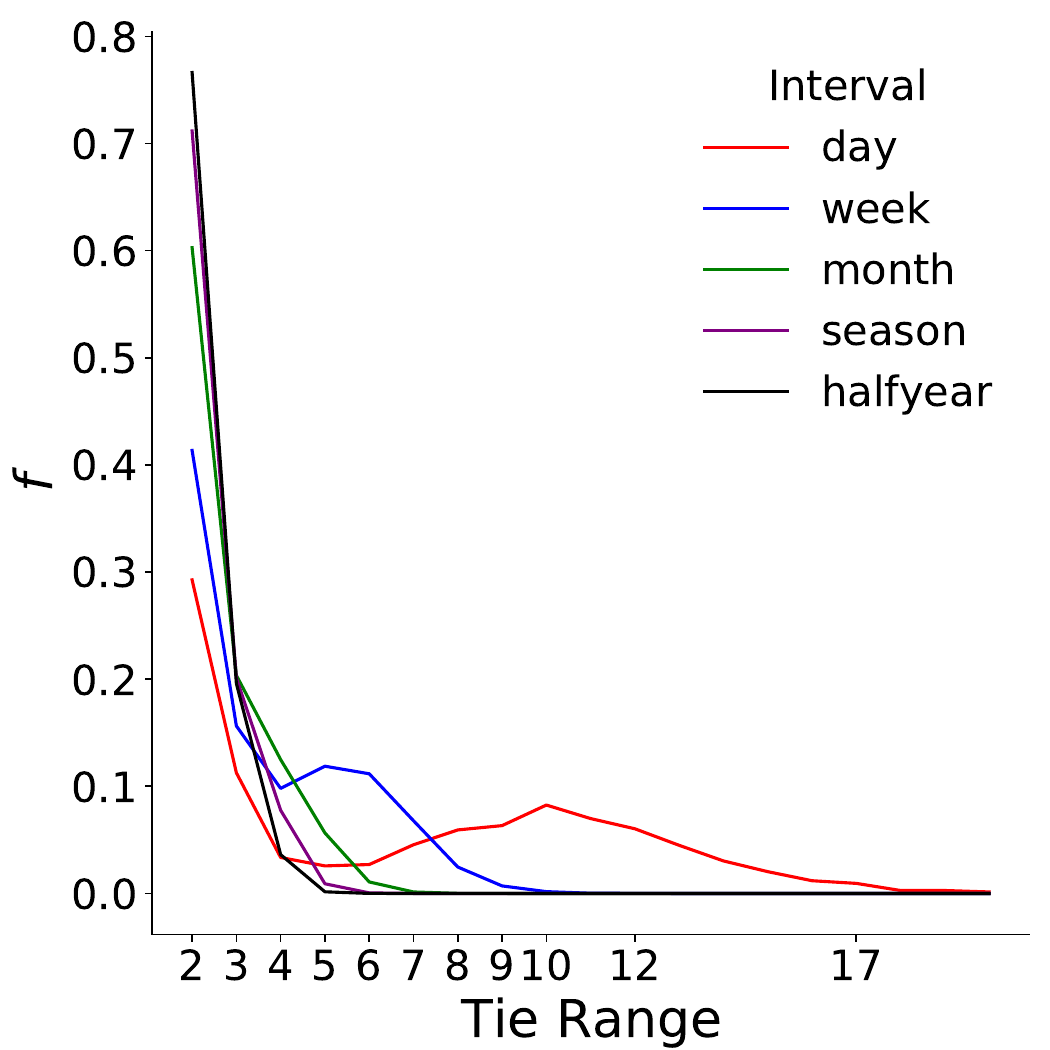}
    \caption{\textbf{Distributions of tie range with respect to different time windows.} Each curve corresponds to a time window. $x$-axis represents the tie range of a tie and $y$-axis is the probability mass of each tie range under that time window. $f$ is the probability mass function.}
    \label{fig:S4}
\end{figure*}

\begin{figure*}
    \captionsetup[subfigure]{labelformat=simple, font={small}}
    \centering
    \subfloat[]{
	\includegraphics[width=0.49\linewidth]{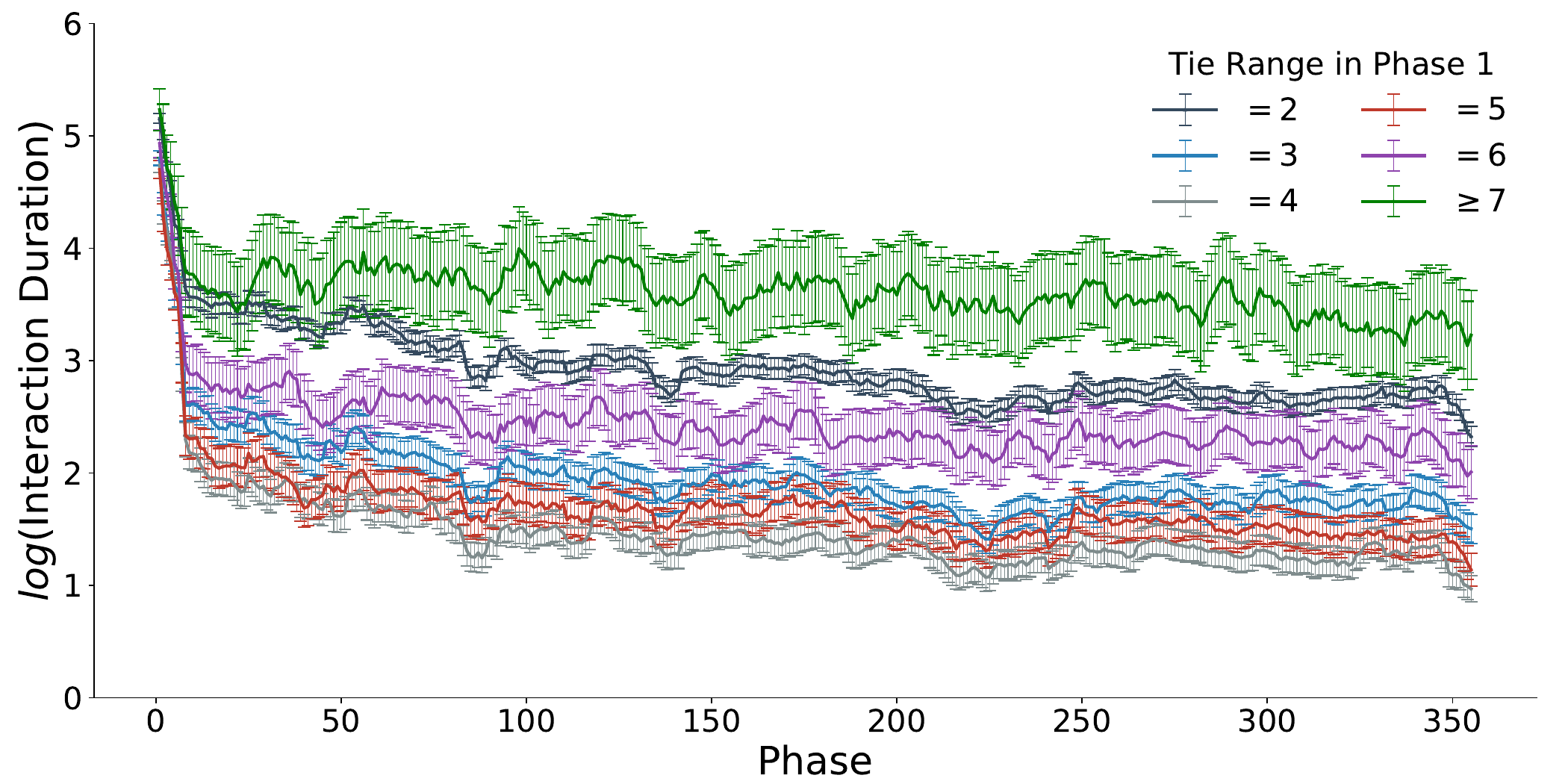}}
	\subfloat[]{
	\includegraphics[width=0.49\linewidth]{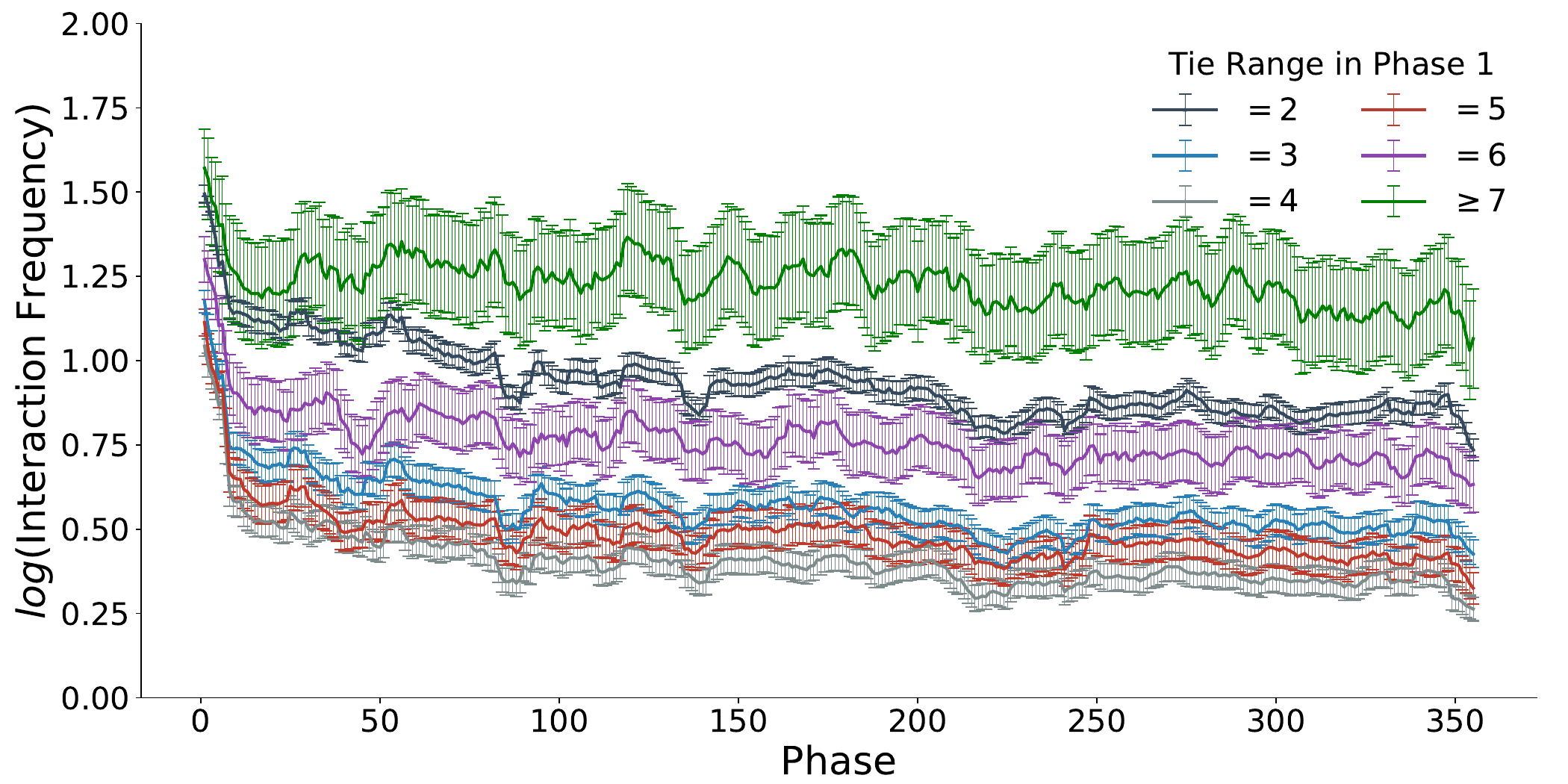}}
    \caption{\textbf{Dynamics of tie strength throughout a year.} Each phase represents a sliding window of seven days, but we move the window day by day. Tie strength is measured by interaction duration (\textbf{a}; the total duration of the calls in seconds) and interaction frequency (\textbf{b}; the number of calls or texts). We take logarithms ($\log$) for both interaction duration and frequency. All ties are classified according to their tie range in the first phase. The curves represent the average ($\log$) interaction duration or frequency conditional on a tie existing in phase 1 with the given tie range. Error bars are 95\% confidence intervals for the $\log$ means of interaction duration and frequency (assuming normal distribution).}
    \label{fig:S5}
\end{figure*}

\begin{figure*}
\captionsetup[subfigure]{labelformat=simple, font={small}}
\centering
\subfloat[]{
	\includegraphics[width=0.3\linewidth]{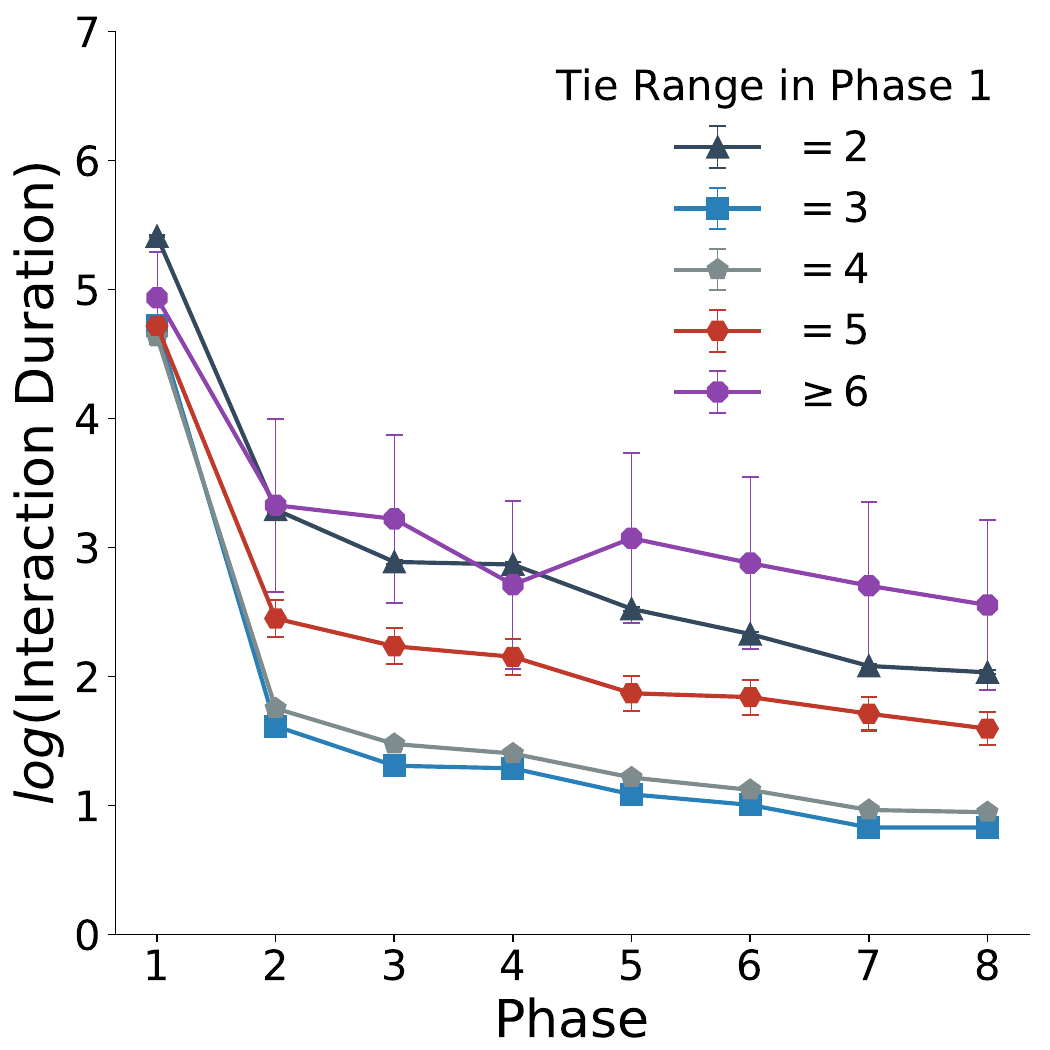}}
\quad
\subfloat[]{
	\includegraphics[width=0.3\linewidth]{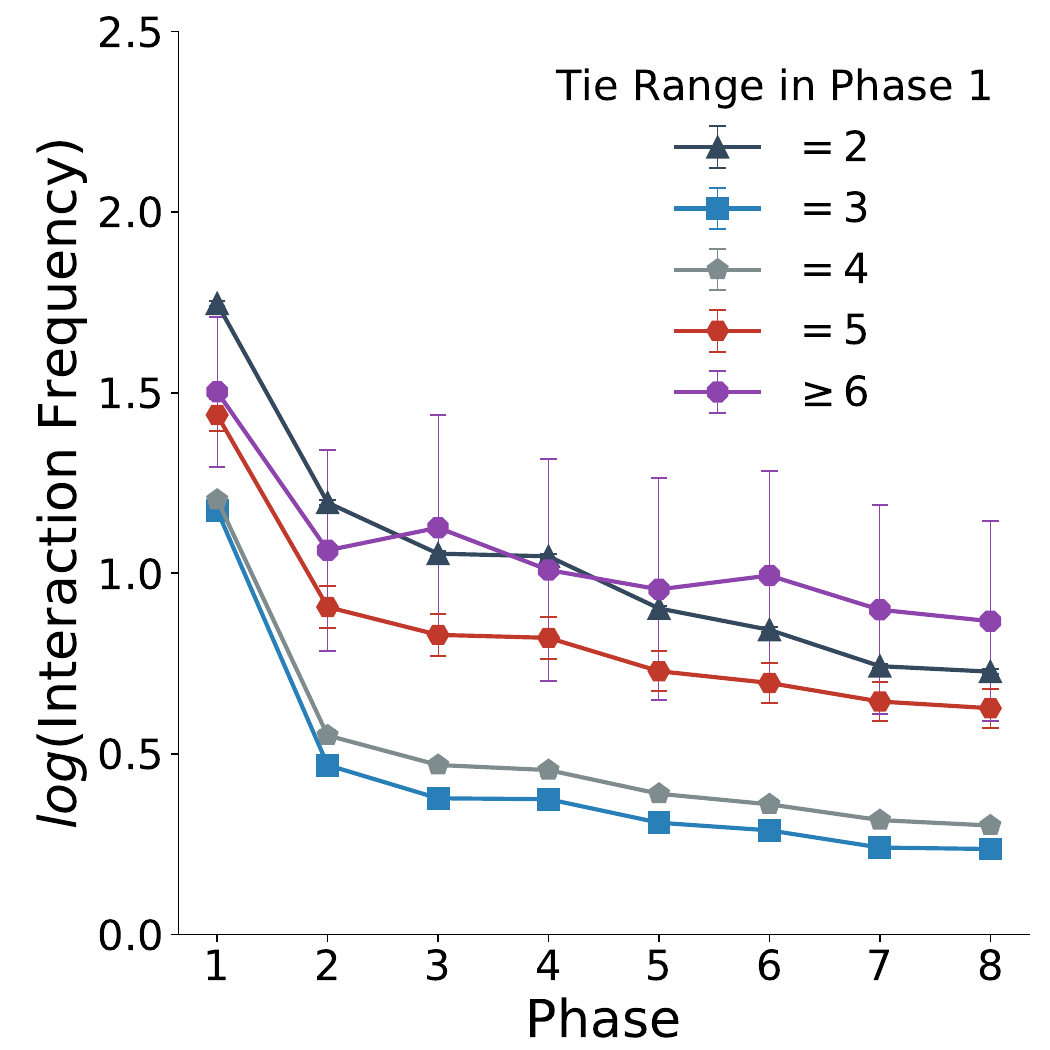}}
\vfill
\subfloat[]{
	\includegraphics[width=0.3\linewidth]{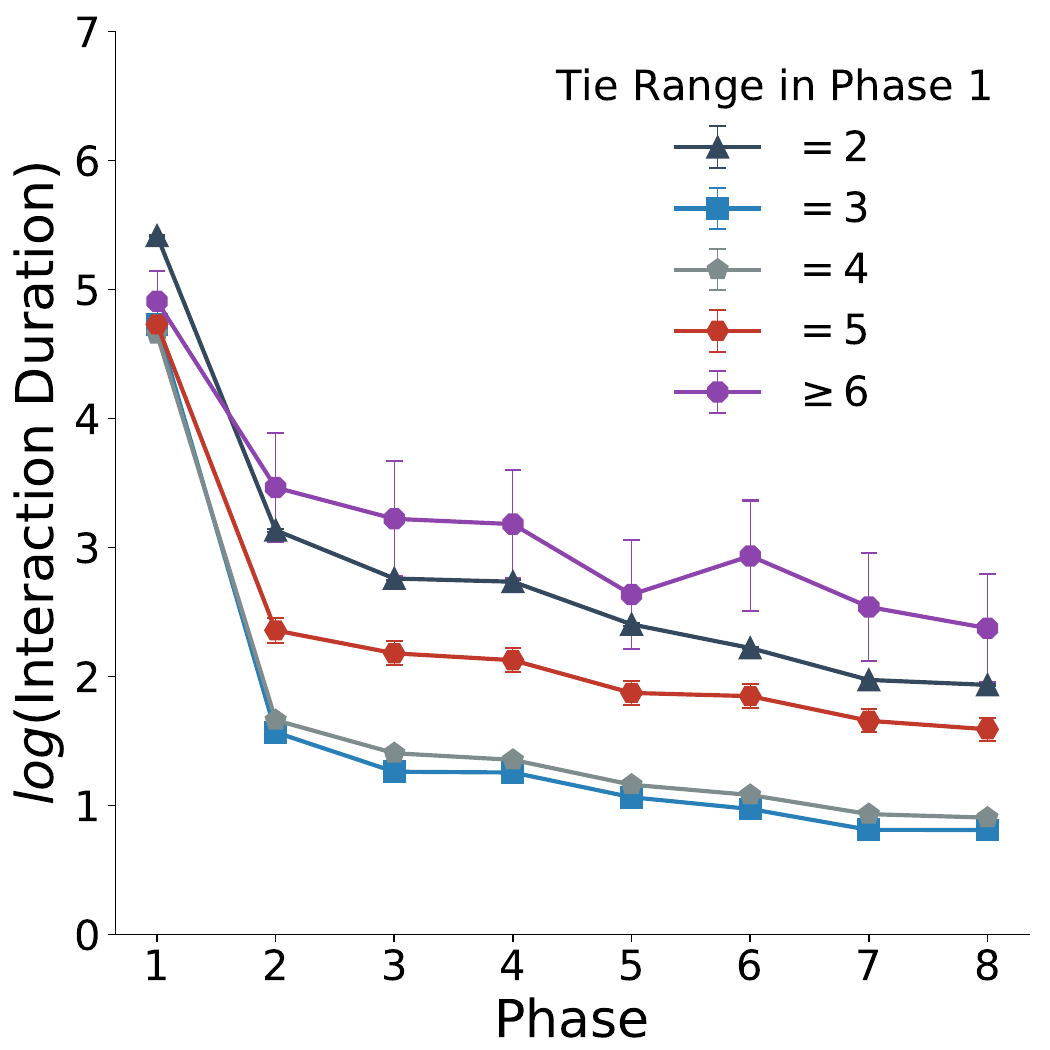}}
\quad
\subfloat[]{
	\includegraphics[width=0.3\linewidth]{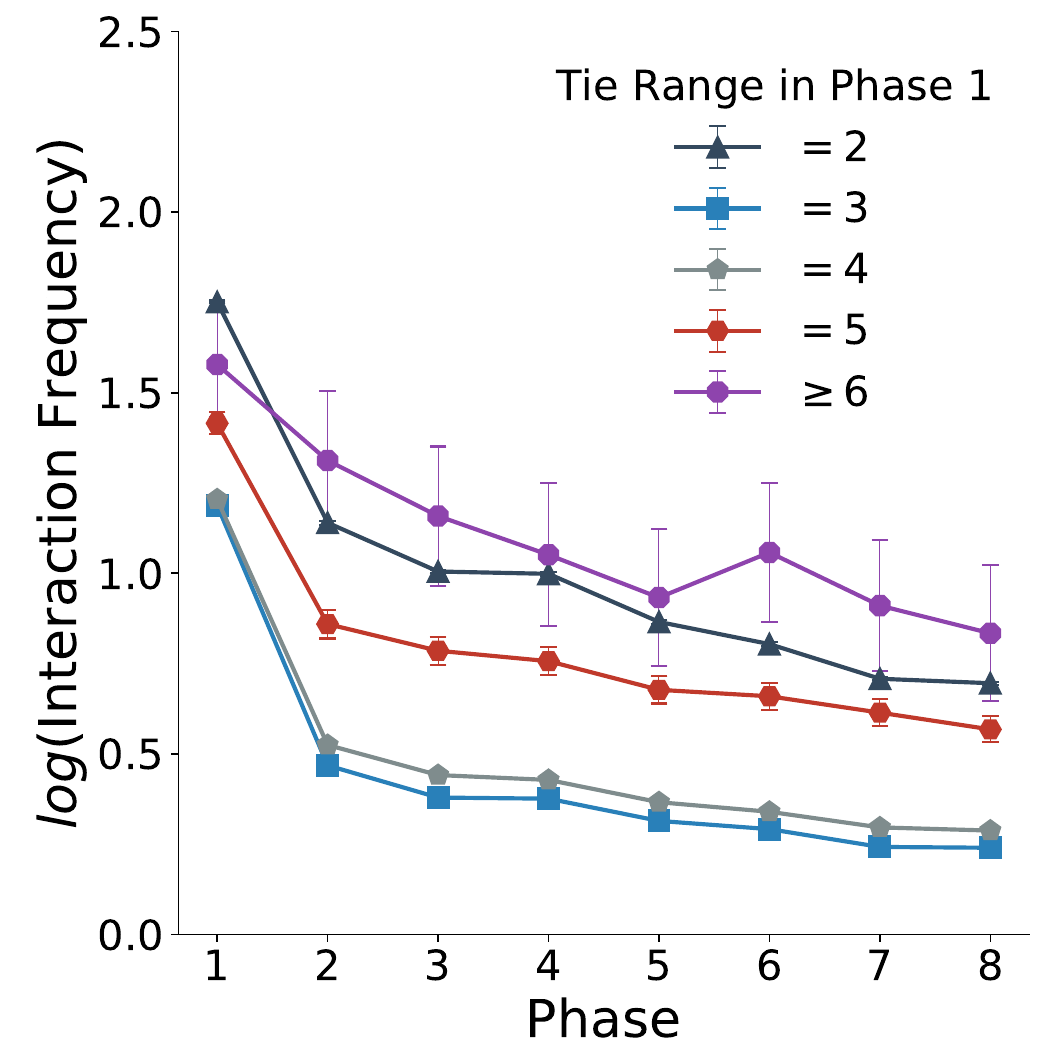}}
    \caption{\textbf{Sensitivity check by randomly dropping a proportion (5\%) of nodes or edges.} Dynamics of tie strength (\textbf{a\&c}: interaction duration; \textbf{b\&d}: interaction frequency) throughout the eight snapshots with a proportion (5\%) of nodes (\textbf{a\&b}) or edges (\textbf{c\&d}) removed randomly. Each phase represents a season (three months). Interaction duration is measured in seconds. We take logarithms ($\log$) for both interaction duration and frequency. All ties are classified according to their tie range in the first phase. The curves represent the average ($\log$) interaction duration or frequency conditional on that a tie exists in phase 1 with the given tie range. Error bars are 95\% confidence intervals for the mean $\log$ interaction duration and frequency (assuming normal distribution). Note that error bars are sometimes smaller than the data point markers.}
\label{fig:Fig.S6}
\end{figure*}

\begin{figure*}
    \captionsetup[subfigure]{labelformat=simple, font={small}}
    \centering
    \subfloat[]{
    \includegraphics[width=0.49\linewidth]{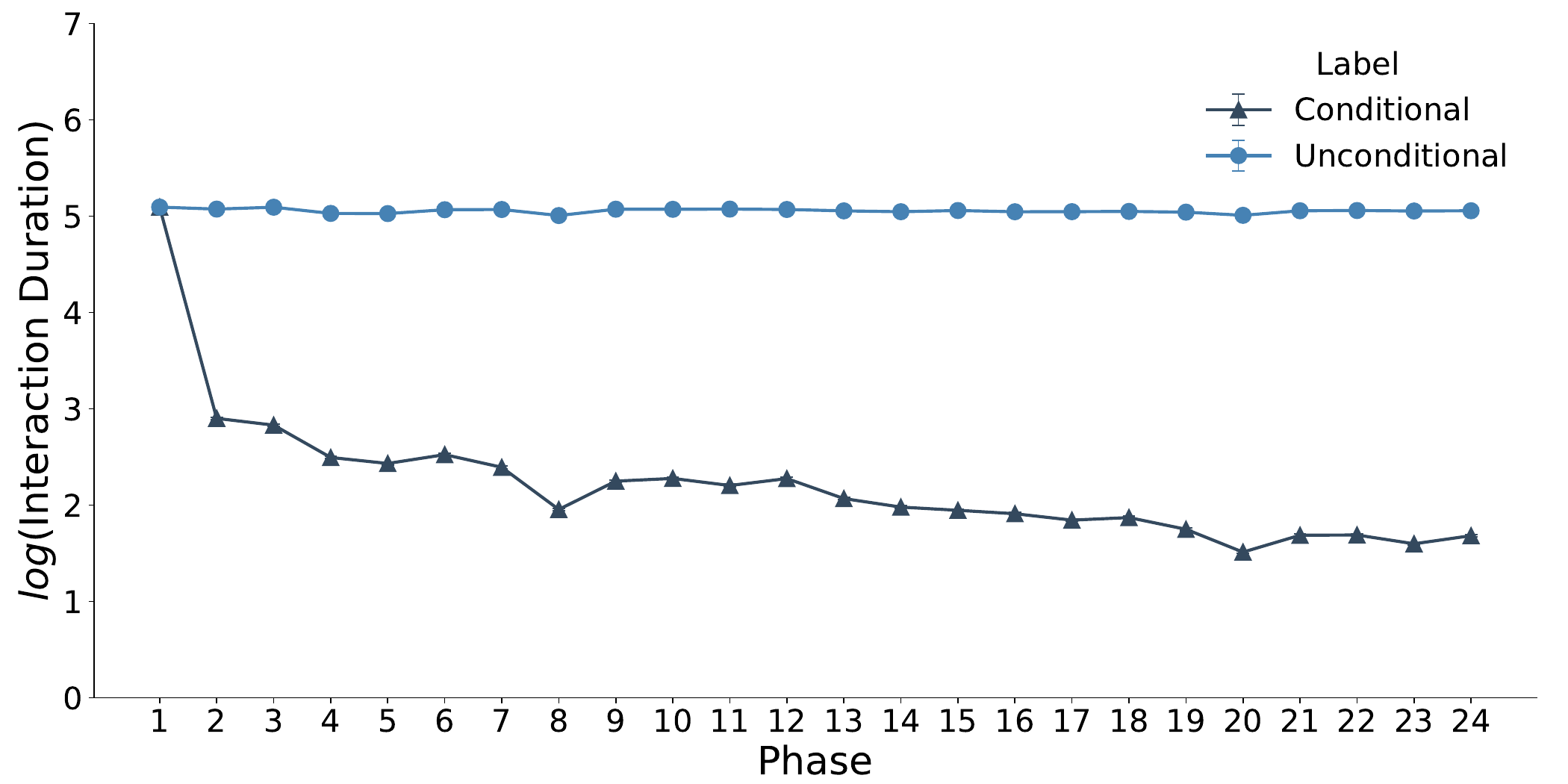}}~
    \subfloat[]{
    \includegraphics[width=0.49\linewidth]{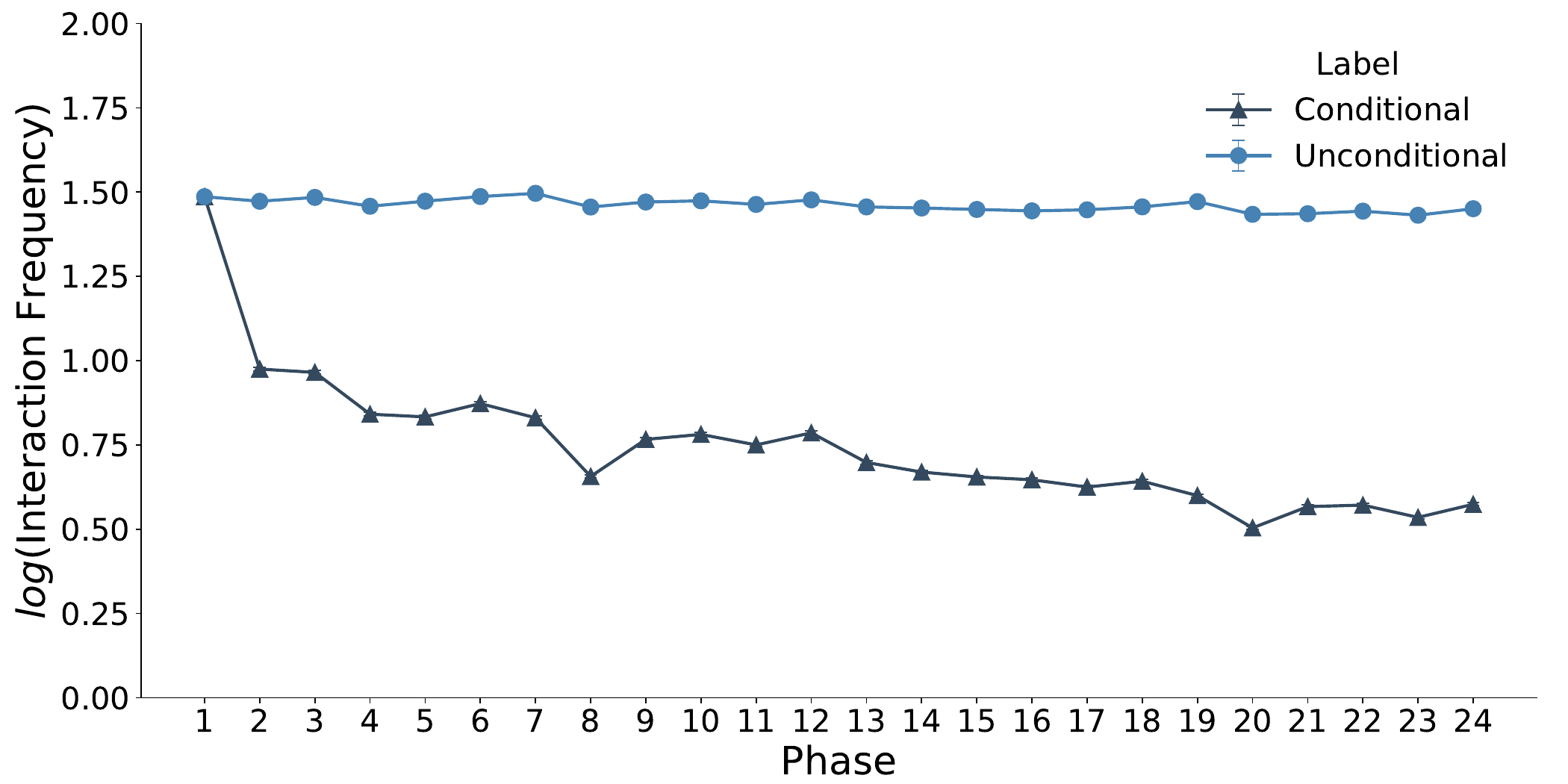}}
    \caption{\textbf{Dynamics of tie strength throughout two years.} Tie strength is measured by interaction duration (\textbf{a}; the total duration of the calls in seconds) and interaction frequency (\textbf{b}; the number of calls or texts). Each phase represents a month. We take logarithms ($\log$) for both interaction duration and frequency. All ties are classified according to their tie range in the first phase. We respectively plot the conditional $E[y_t|y_1>0]$ and unconditional ( $E[y_t]$). $y_t$ denotes the interactions at phase $t$. Error bars are 95\% confidence intervals for the mean $\log$ interaction duration and frequency (assuming normal distribution). Note that error bars are sometimes smaller than the data point markers.}
    \label{fig:Fig.S7}
\end{figure*}

\begin{figure*}
\captionsetup[subfigure]{labelformat=simple, font={small}}
\centering
\subfloat[]{
	\includegraphics[width=0.3\linewidth]{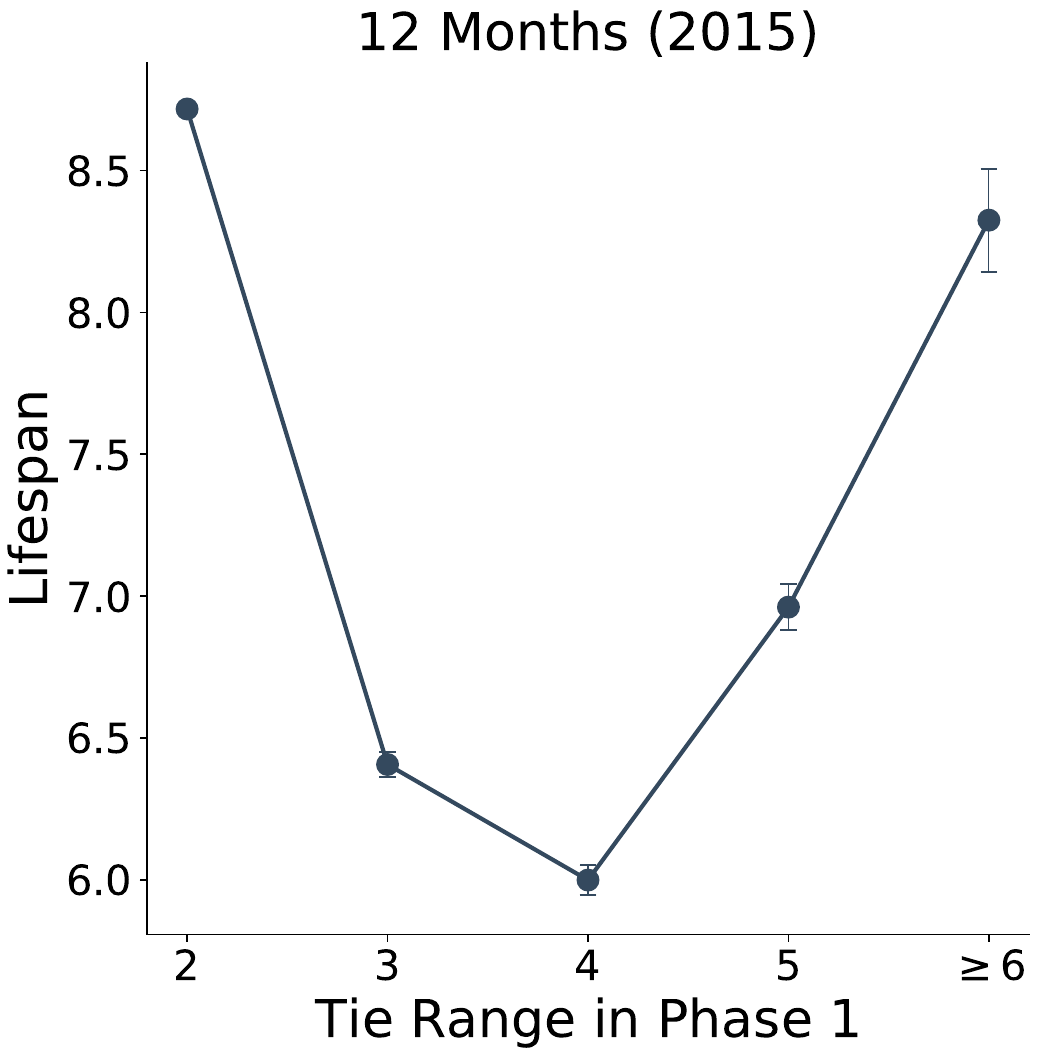}}
\quad
\subfloat[]{
	\includegraphics[width=0.3\linewidth]{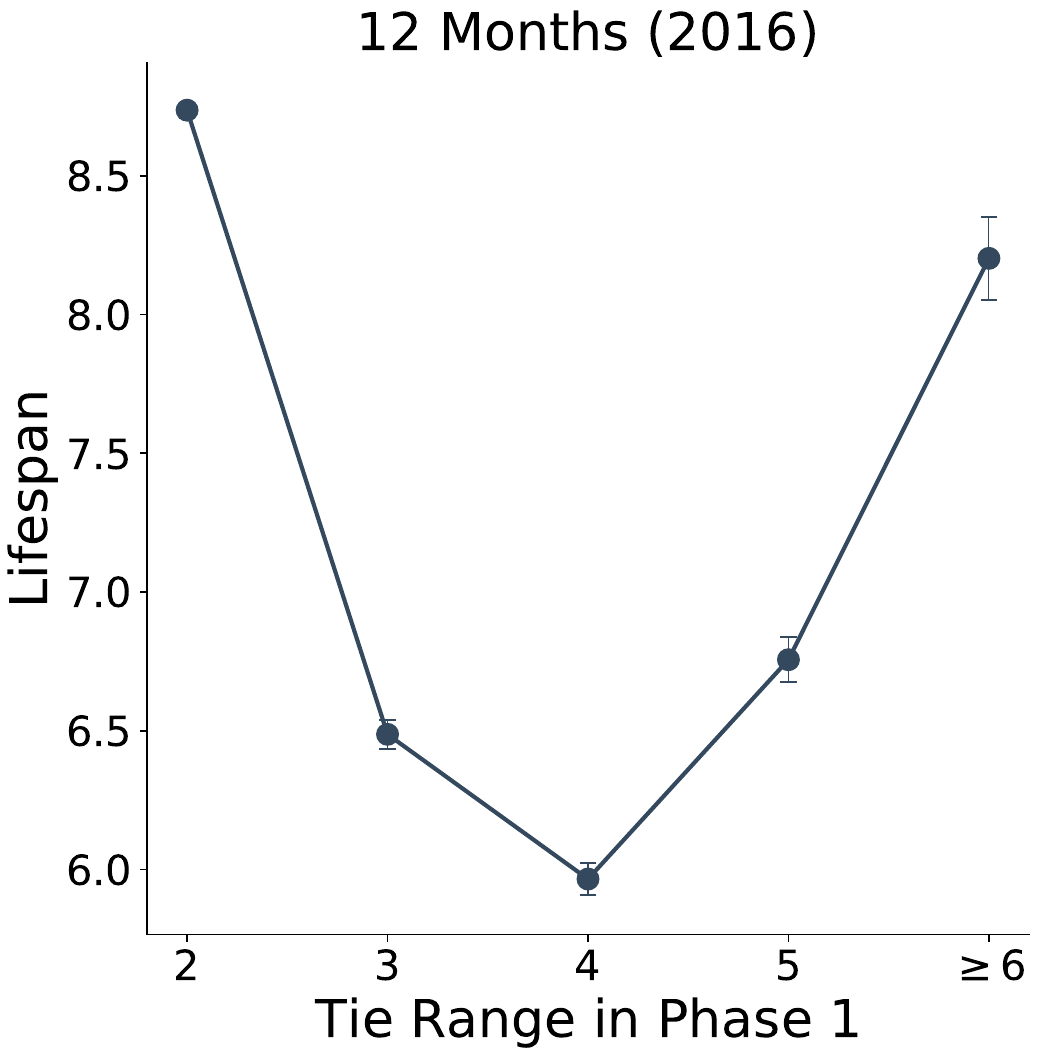}}
\vfill
\subfloat[]{
	\includegraphics[width=0.3\linewidth]{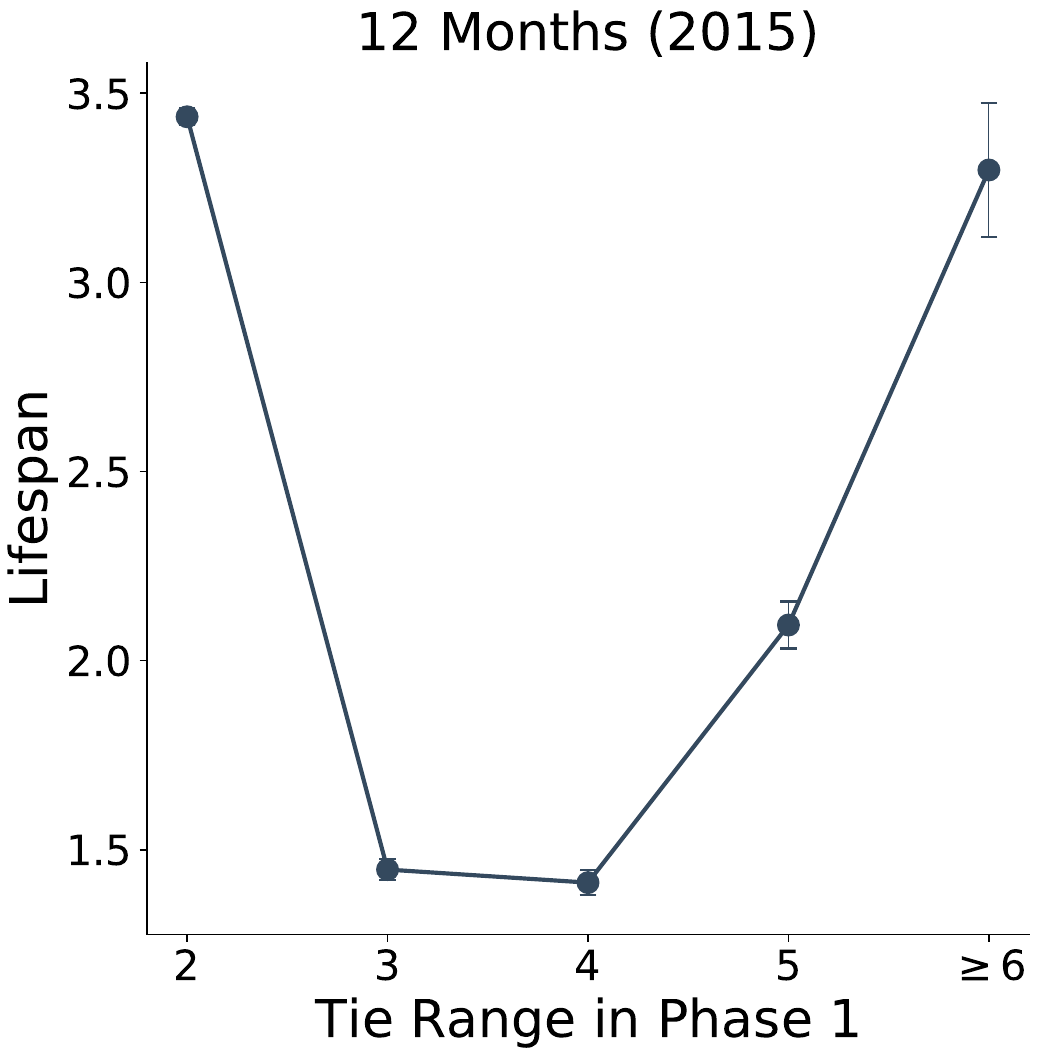}}
\quad
\subfloat[]{
	\includegraphics[width=0.3\linewidth]{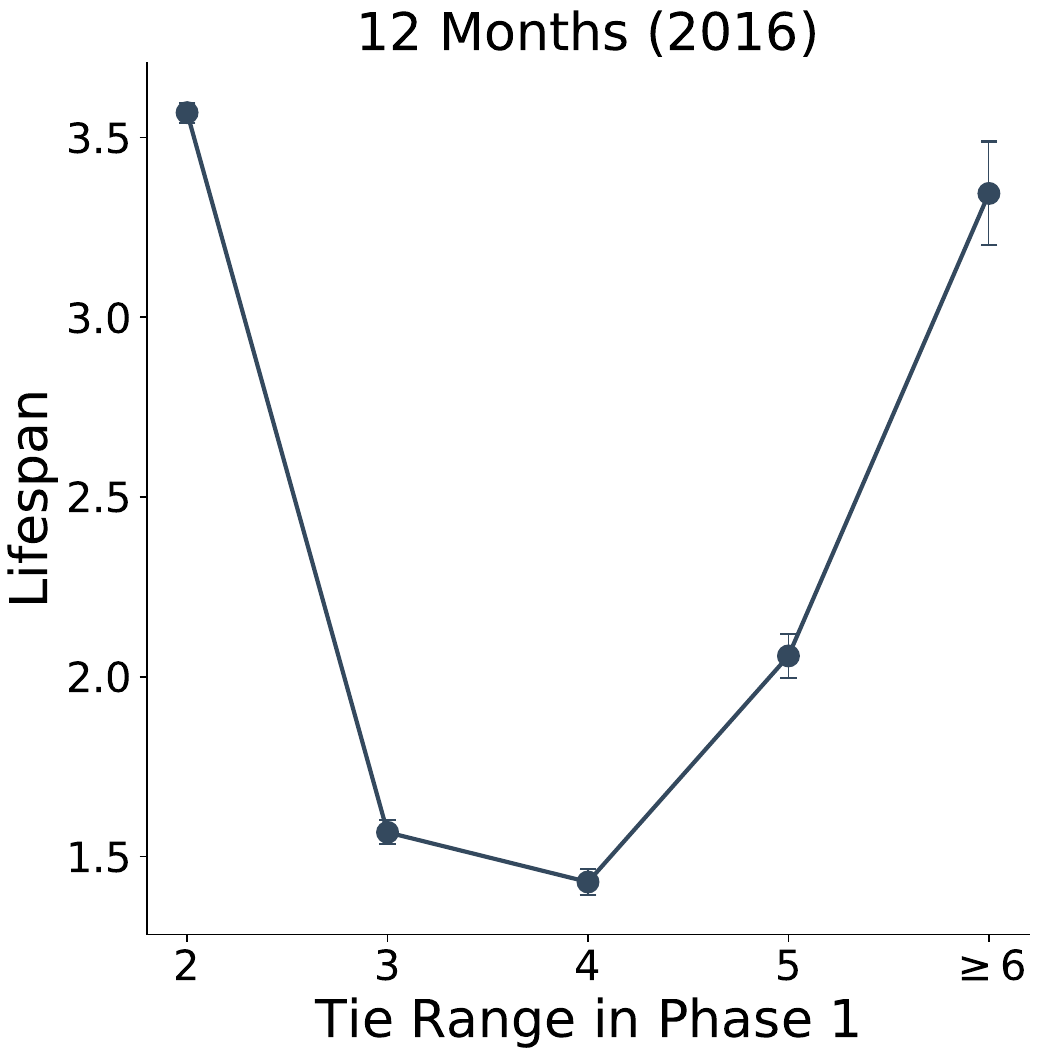}}
    \caption{\textbf{Lifespans without termination and with termination of ties with different tie ranges.} Note that when defining the lifespan, we explore two choices: \textbf{a\&b} a social tie has interactions in the first and the last phases no matter whether they have interactions in the phases in between; \textbf{c\&d} the social tie has to have interactions for every phase within the lifespan. The former choice considers the ties being re-established after termination, while the latter one does not. Lifespan is measured in months. We examine the two years separately. Error bars are 95\% confidence intervals for lifespans of social ties that exist in the first month (assuming normal distribution). Note that error bars are sometimes smaller than the data point markers.}
\label{fig:Fig.S8}
\end{figure*}

\begin{figure*}
    \centering
    \includegraphics[width=0.15\linewidth]{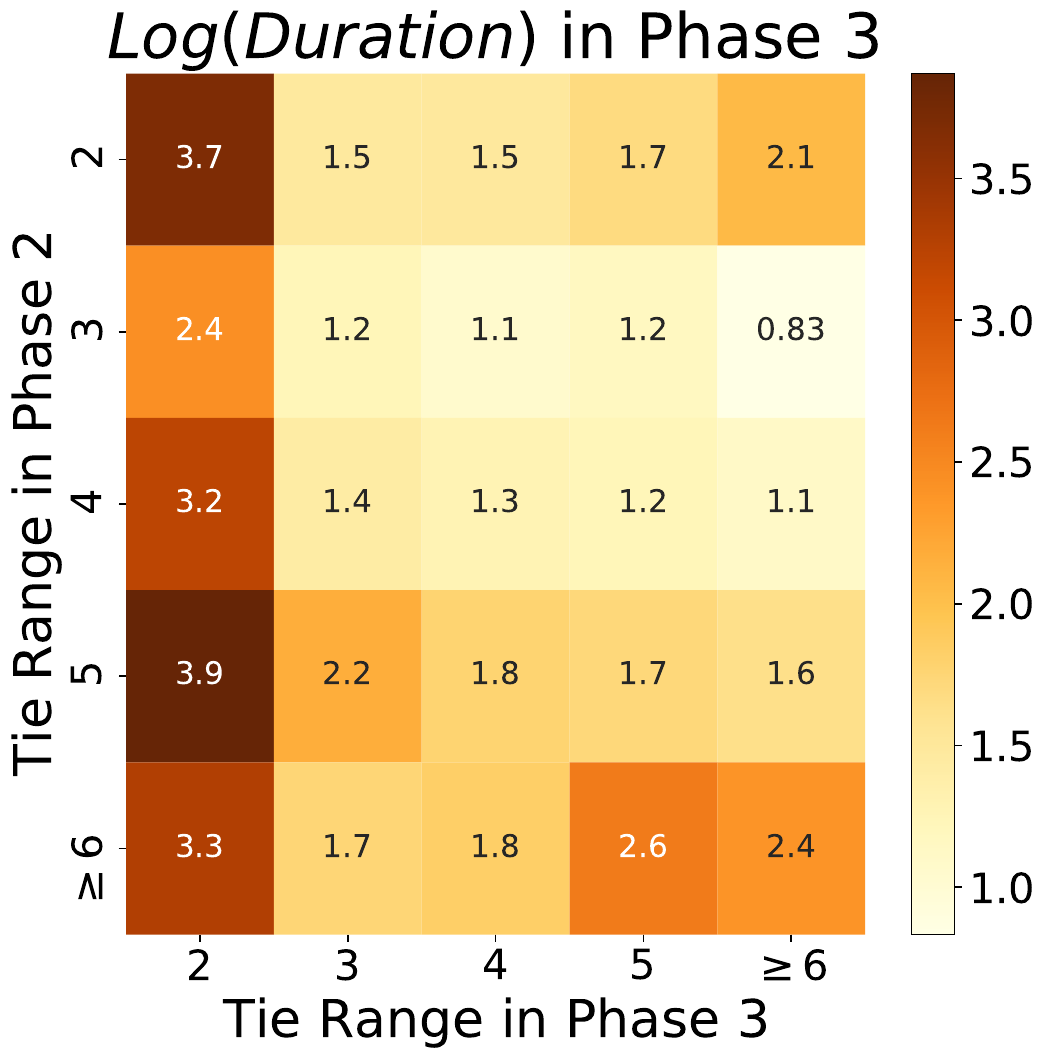}~
    \includegraphics[width=0.15\linewidth]{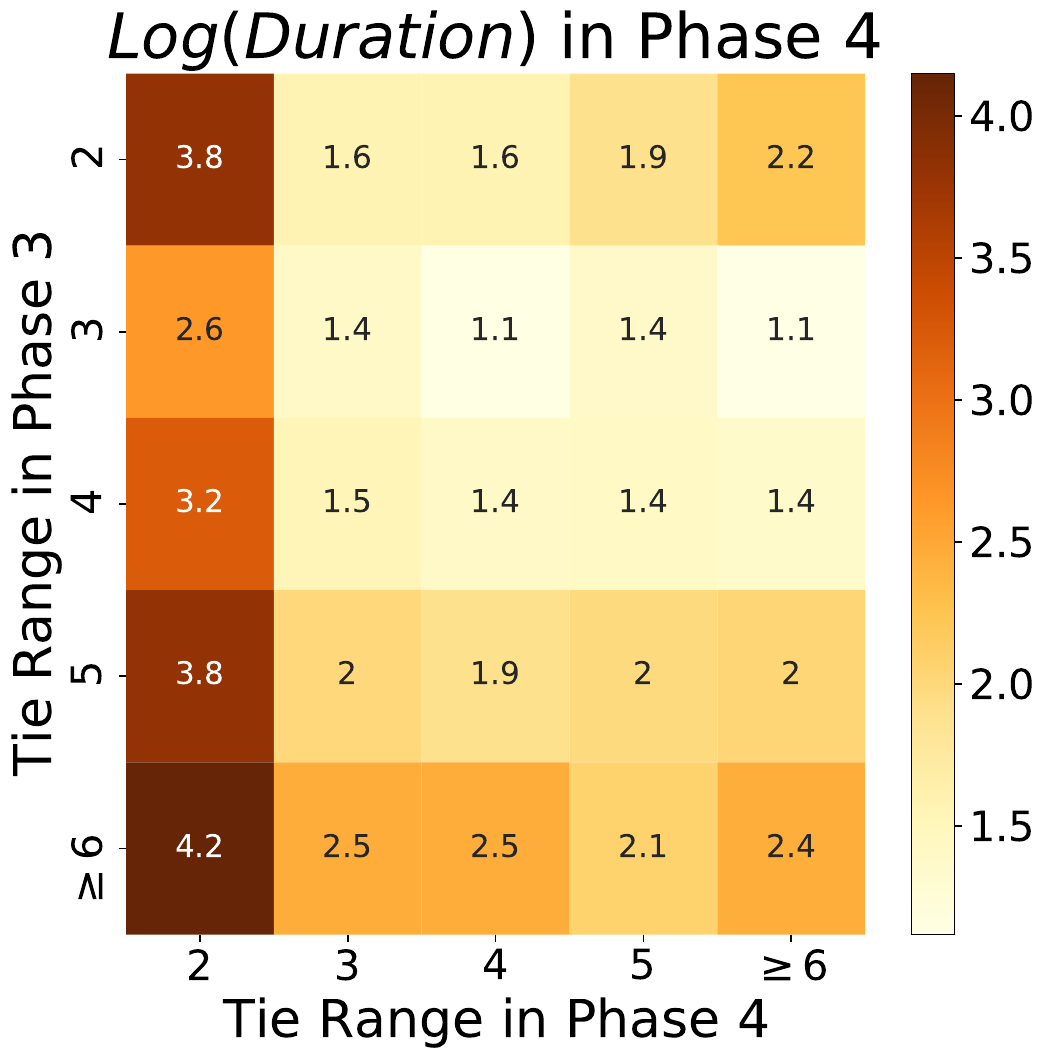}~
    \includegraphics[width=0.15\linewidth]{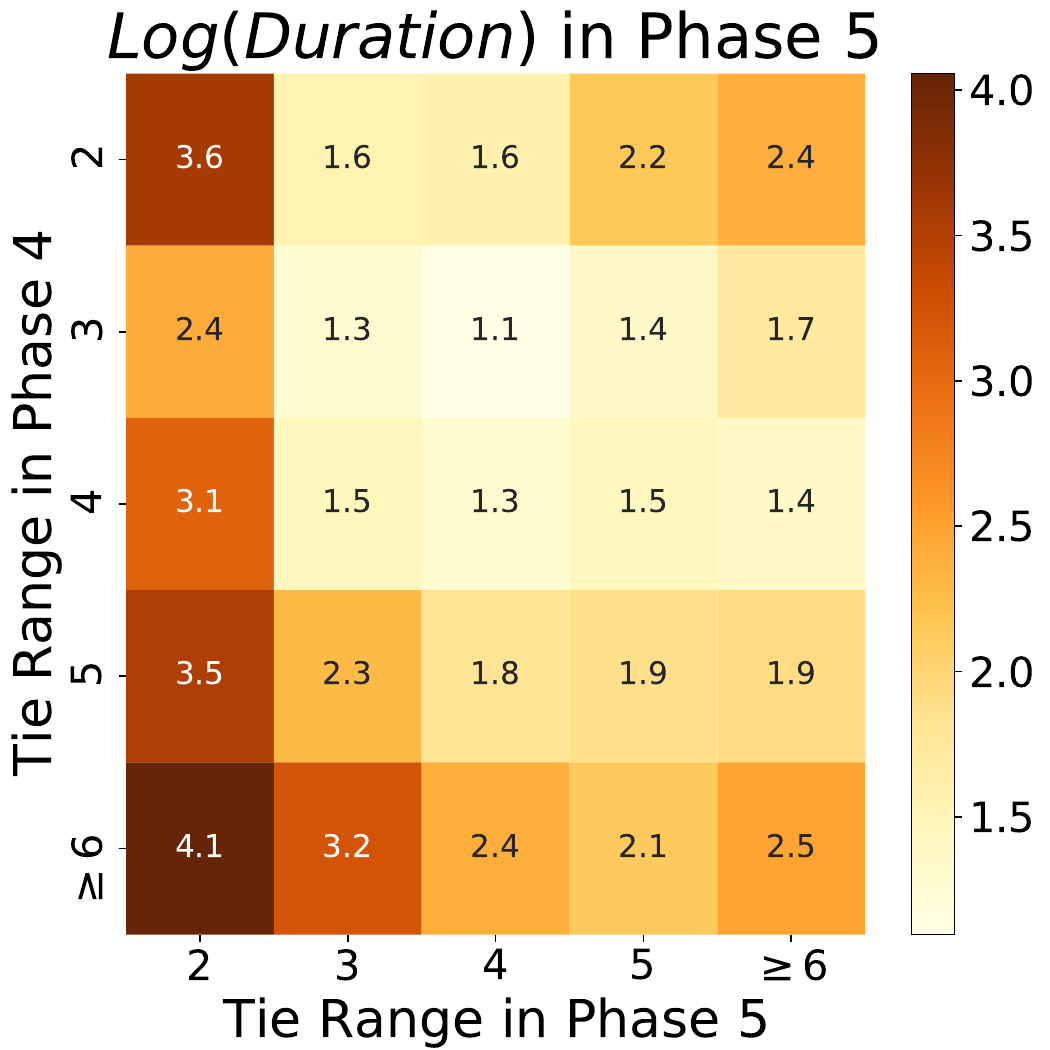}~
    \includegraphics[width=0.15\linewidth]{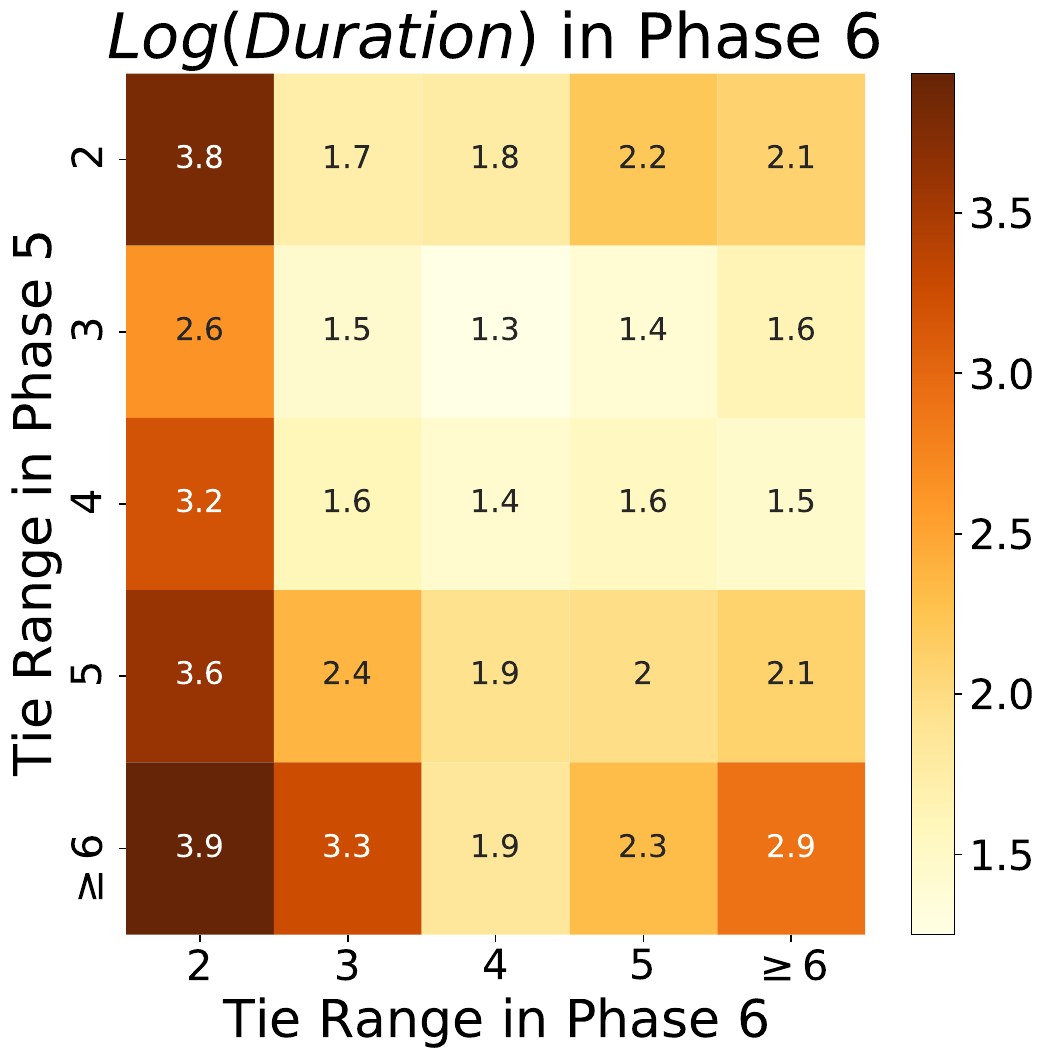}~
    \includegraphics[width=0.15\linewidth]{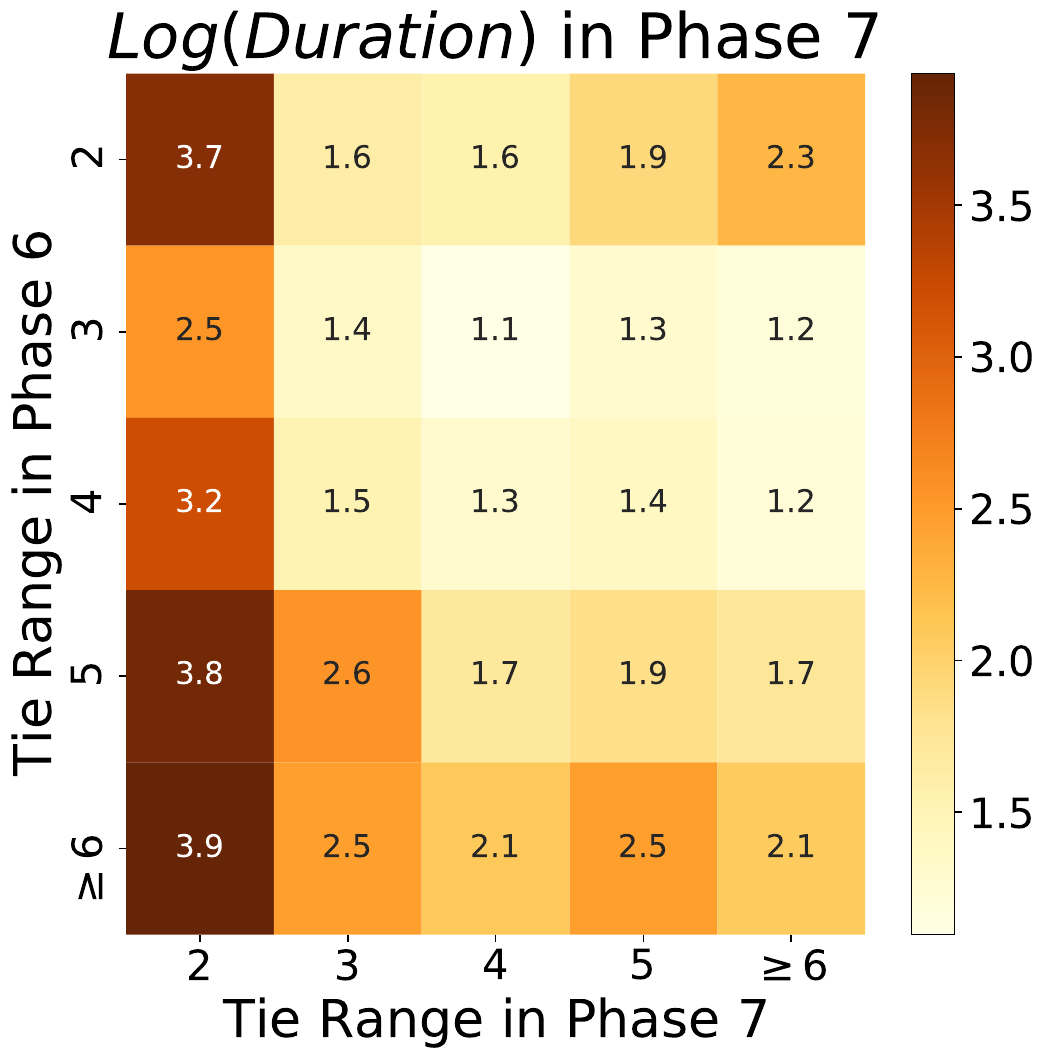}~
    \includegraphics[width=0.15\linewidth]{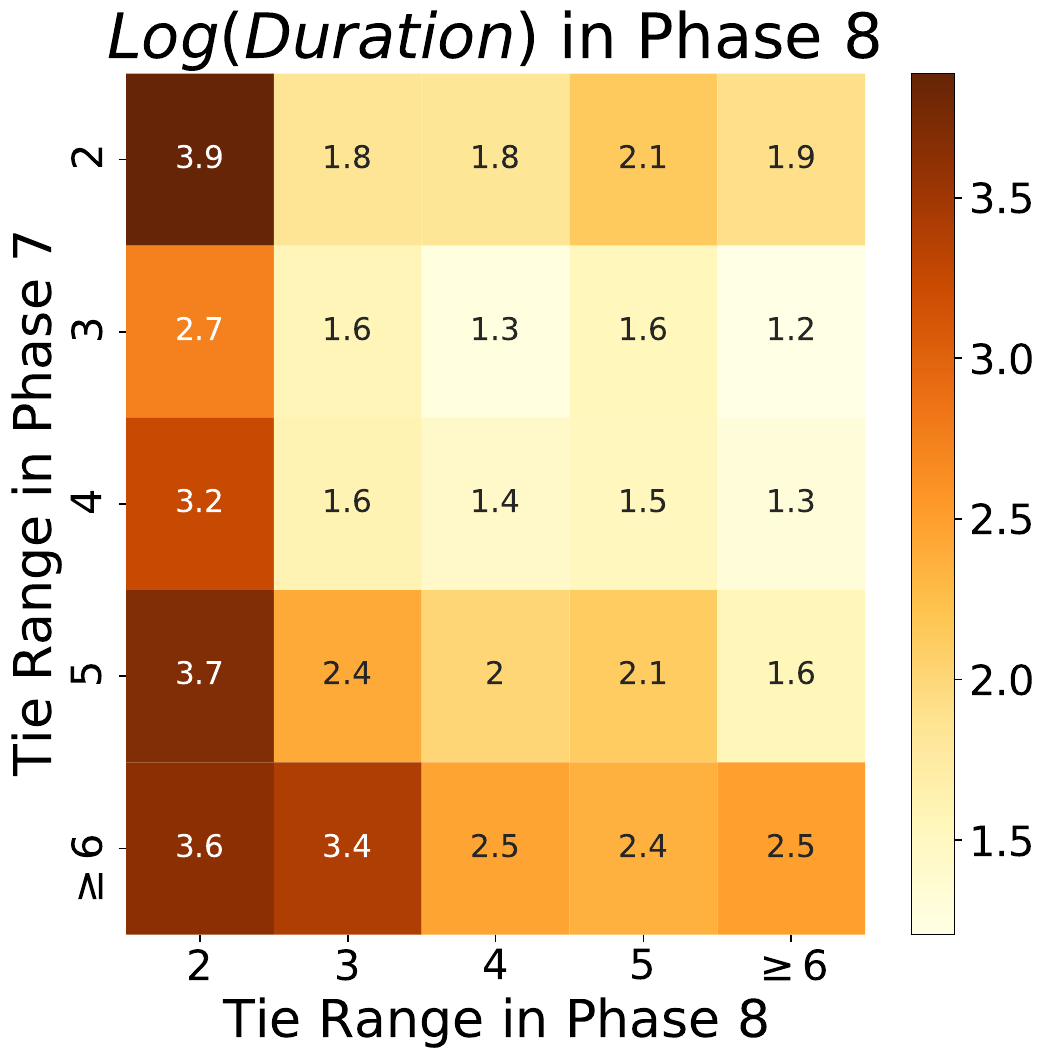}
    \vfill
    \includegraphics[width=0.15\linewidth]{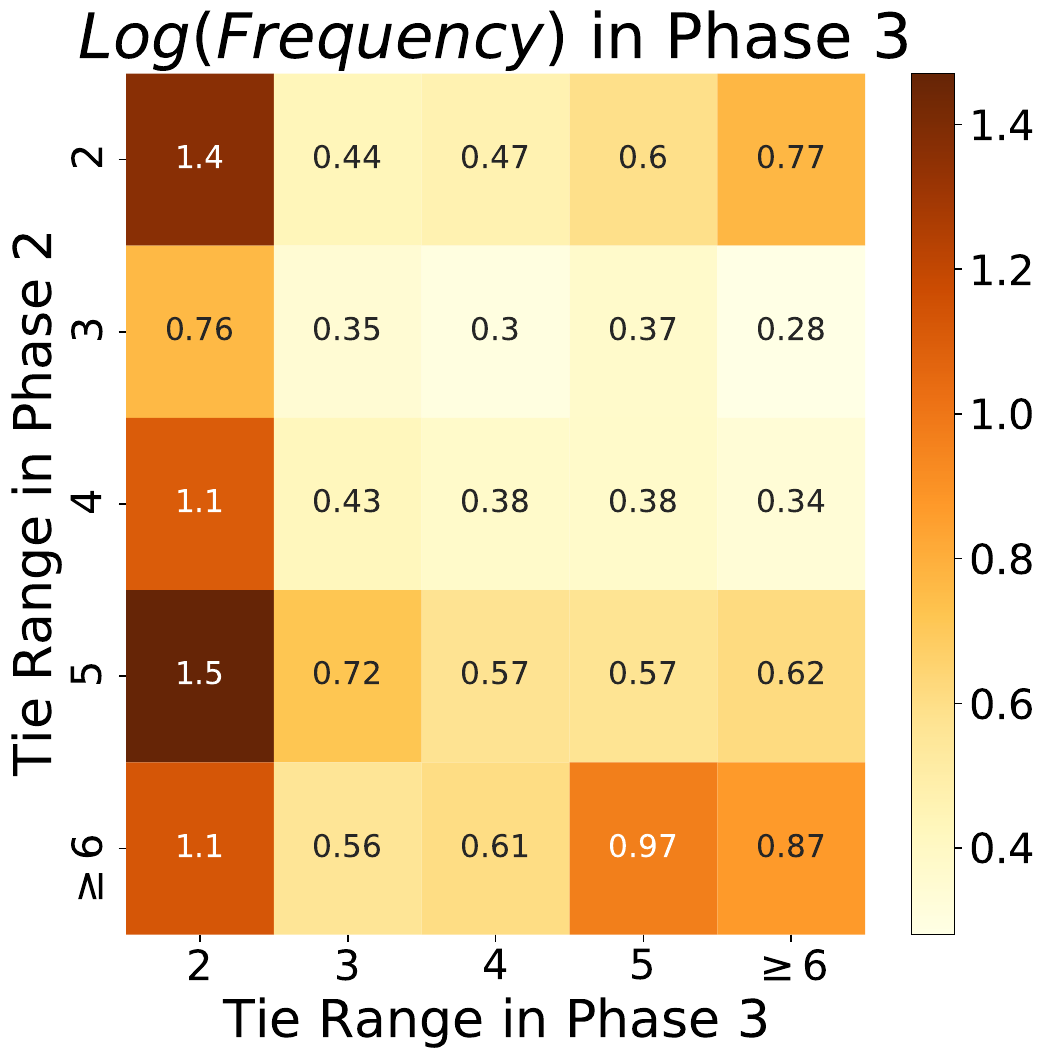}~
    \includegraphics[width=0.15\linewidth]{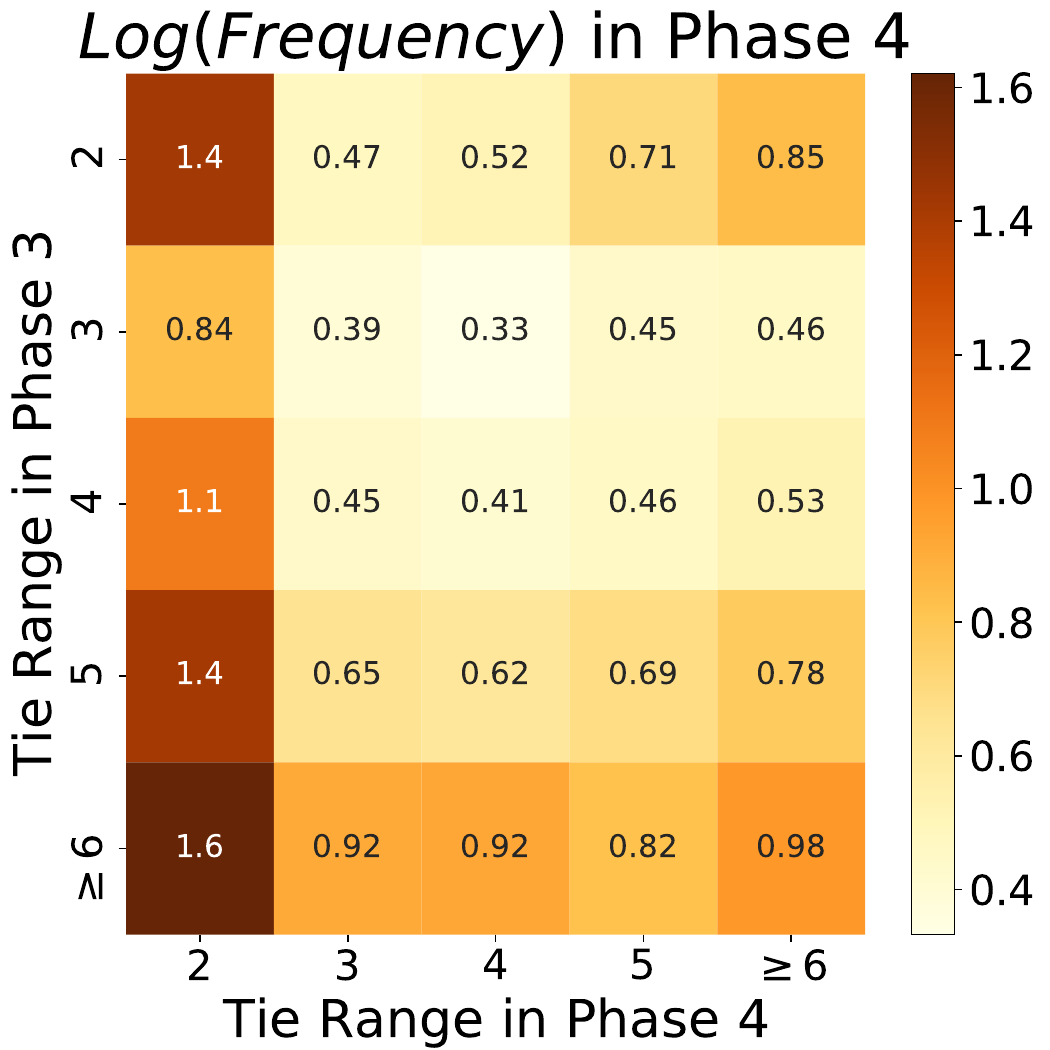}~
    \includegraphics[width=0.15\linewidth]{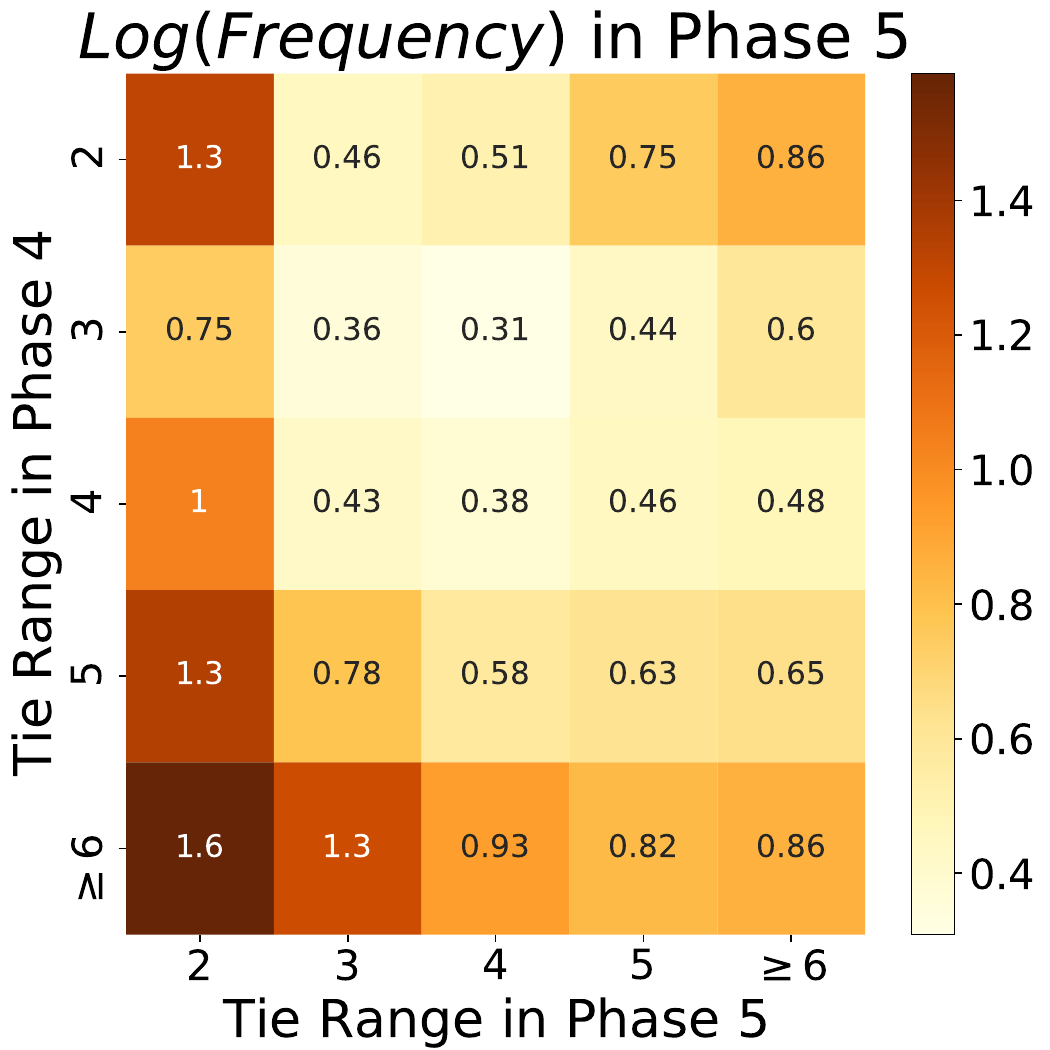}~
    \includegraphics[width=0.15\linewidth]{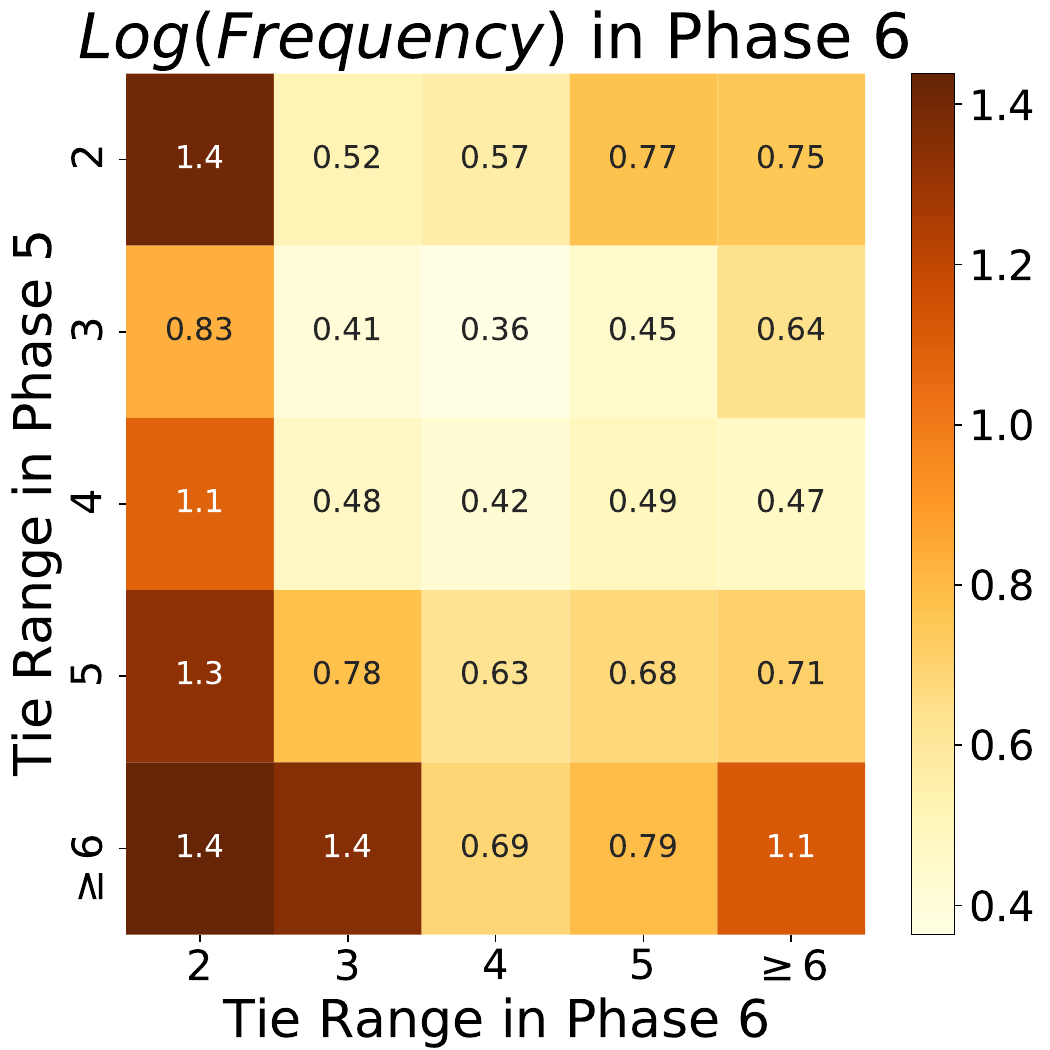}~
    \includegraphics[width=0.15\linewidth]{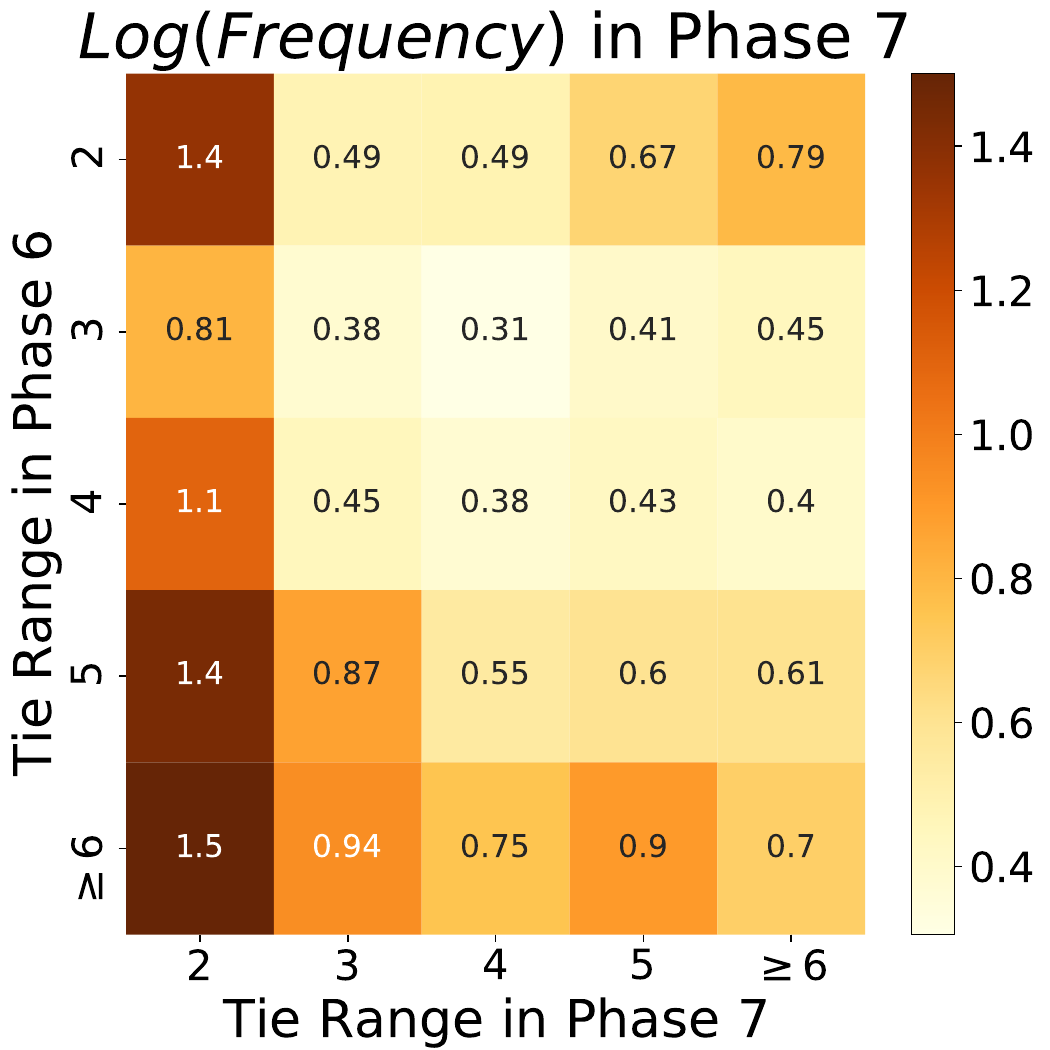}~
    \includegraphics[width=0.15\linewidth]{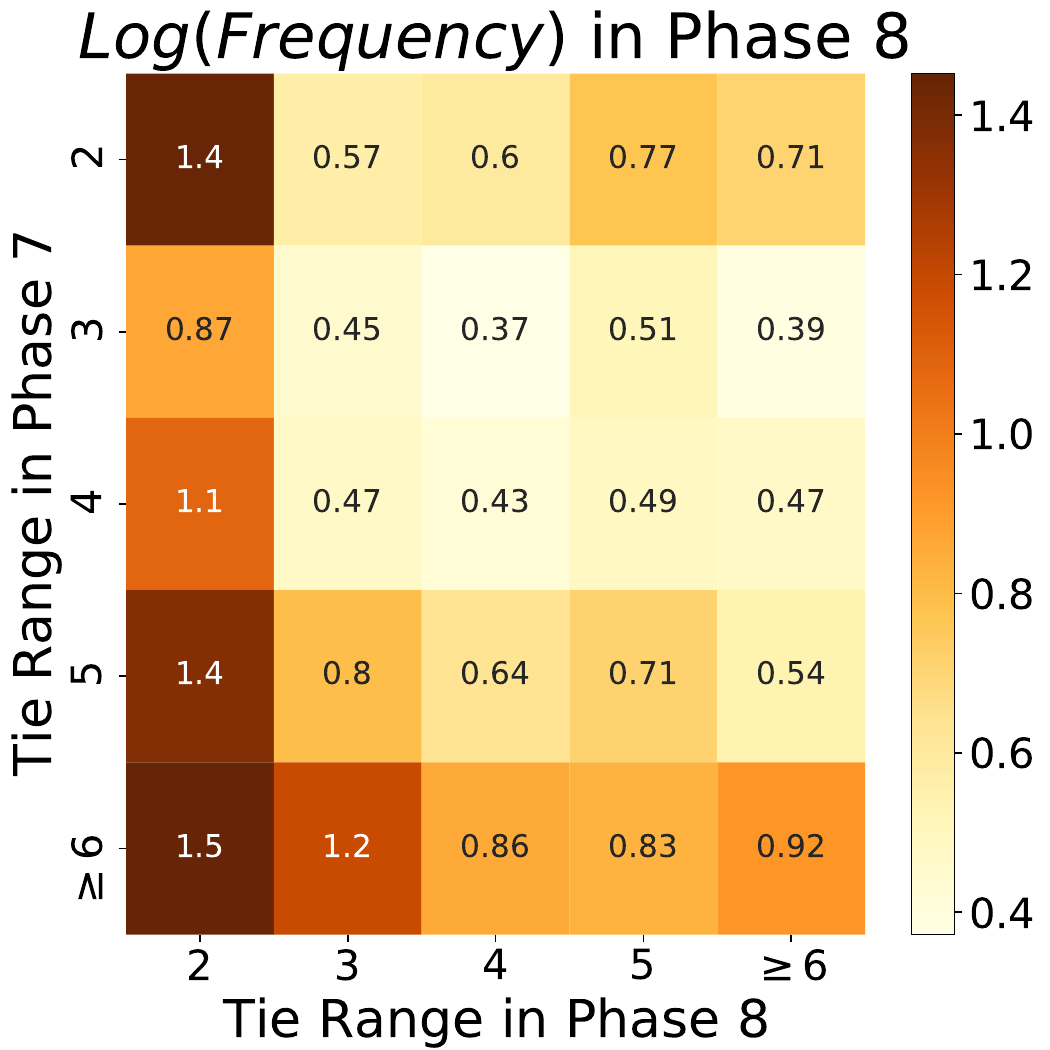}
    \vfill
    \includegraphics[width=0.15\linewidth]{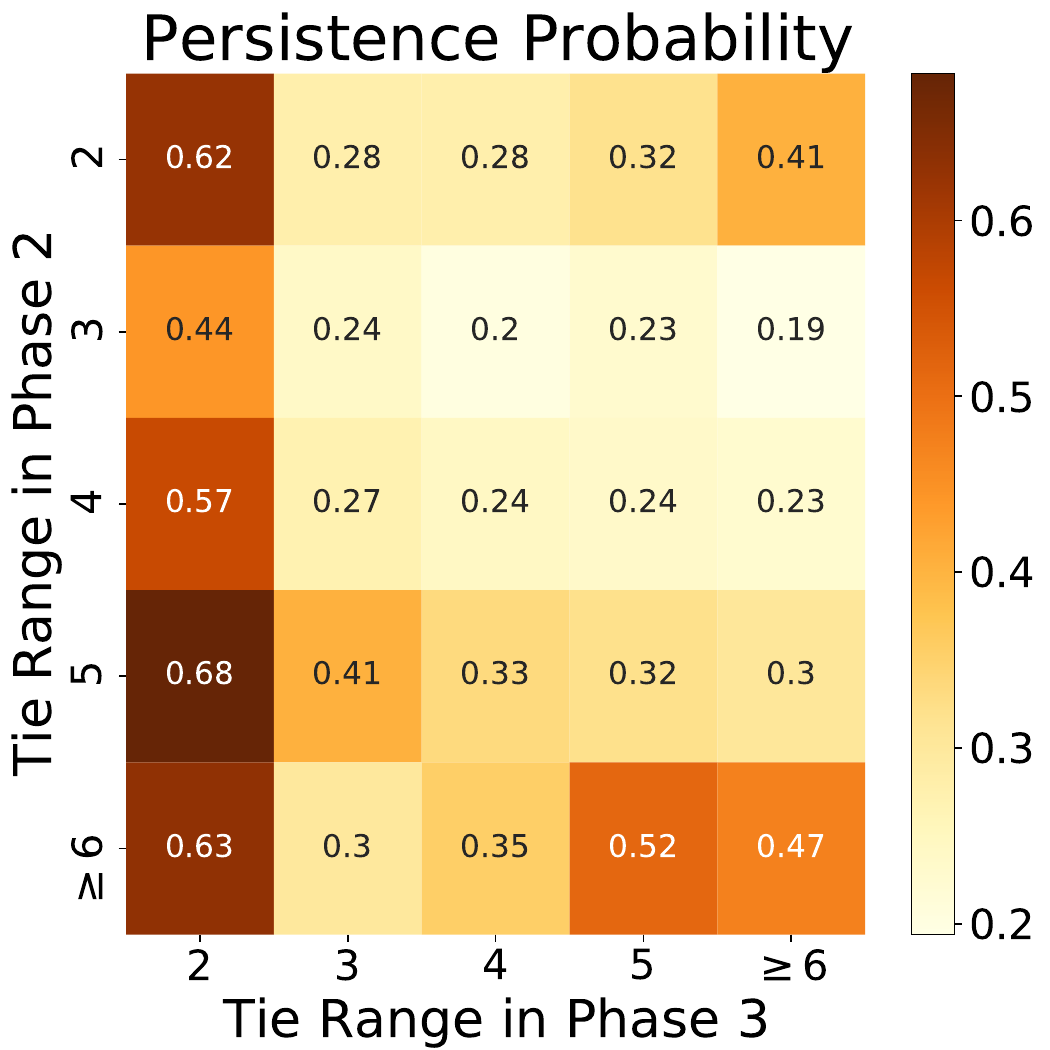}~
    \includegraphics[width=0.15\linewidth]{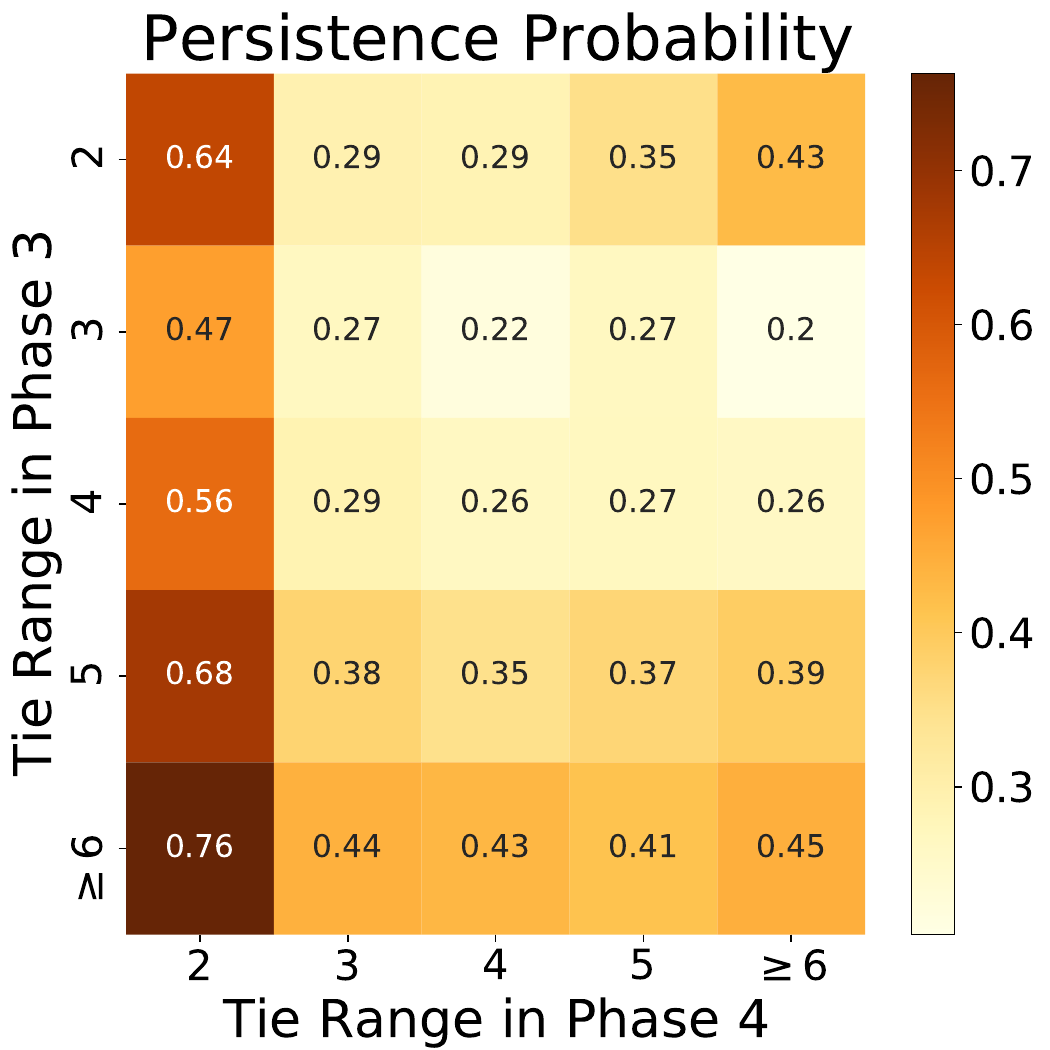}~
    \includegraphics[width=0.15\linewidth]{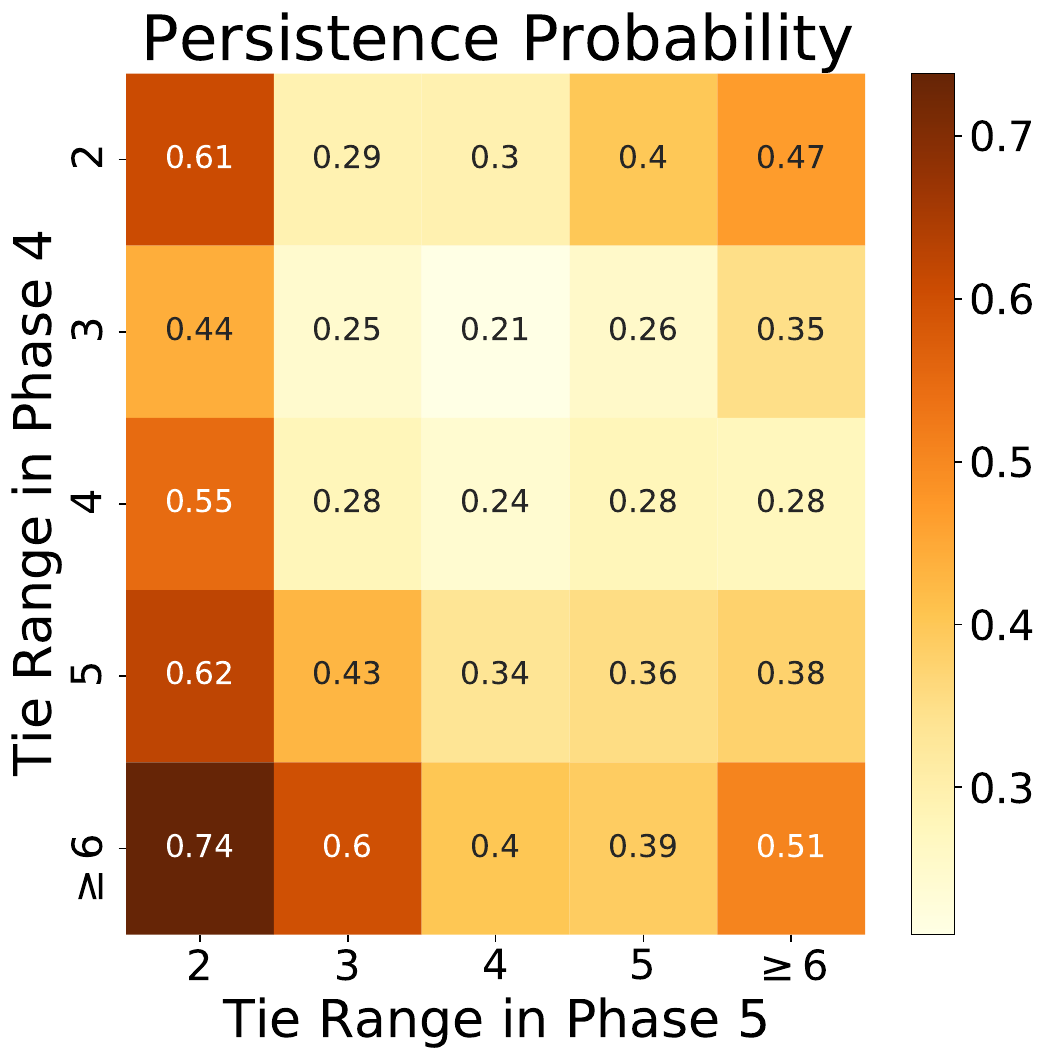}~
    \includegraphics[width=0.15\linewidth]{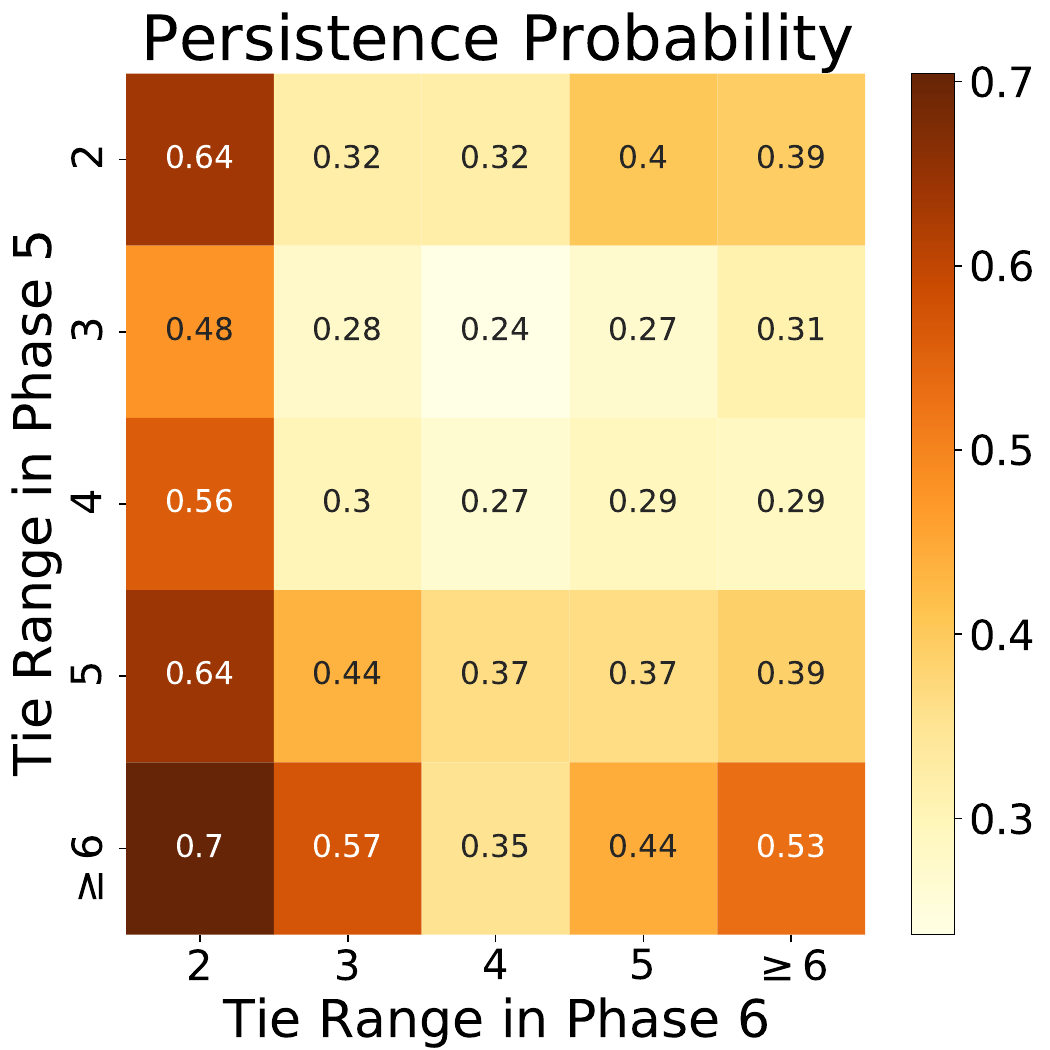}~
    \includegraphics[width=0.15\linewidth]{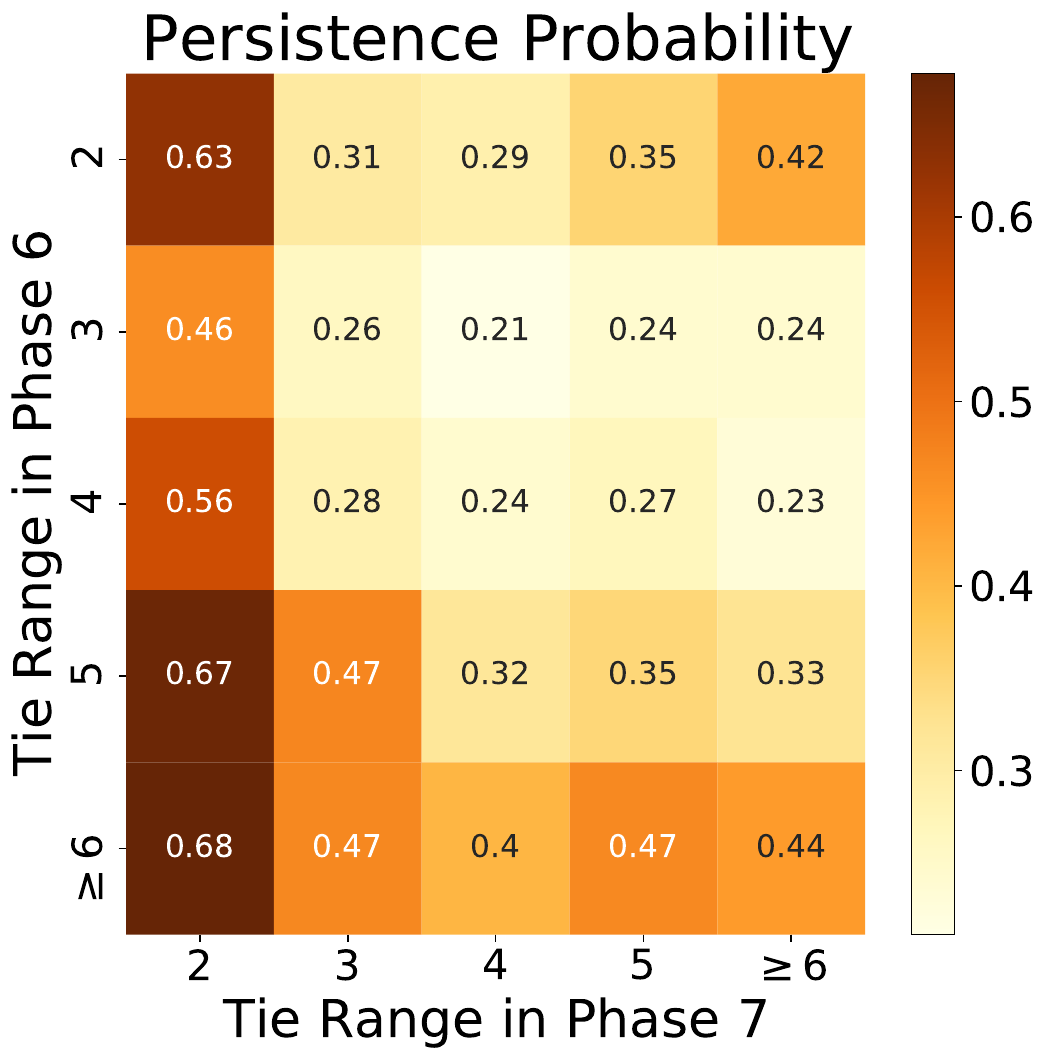}~
    \includegraphics[width=0.15\linewidth]{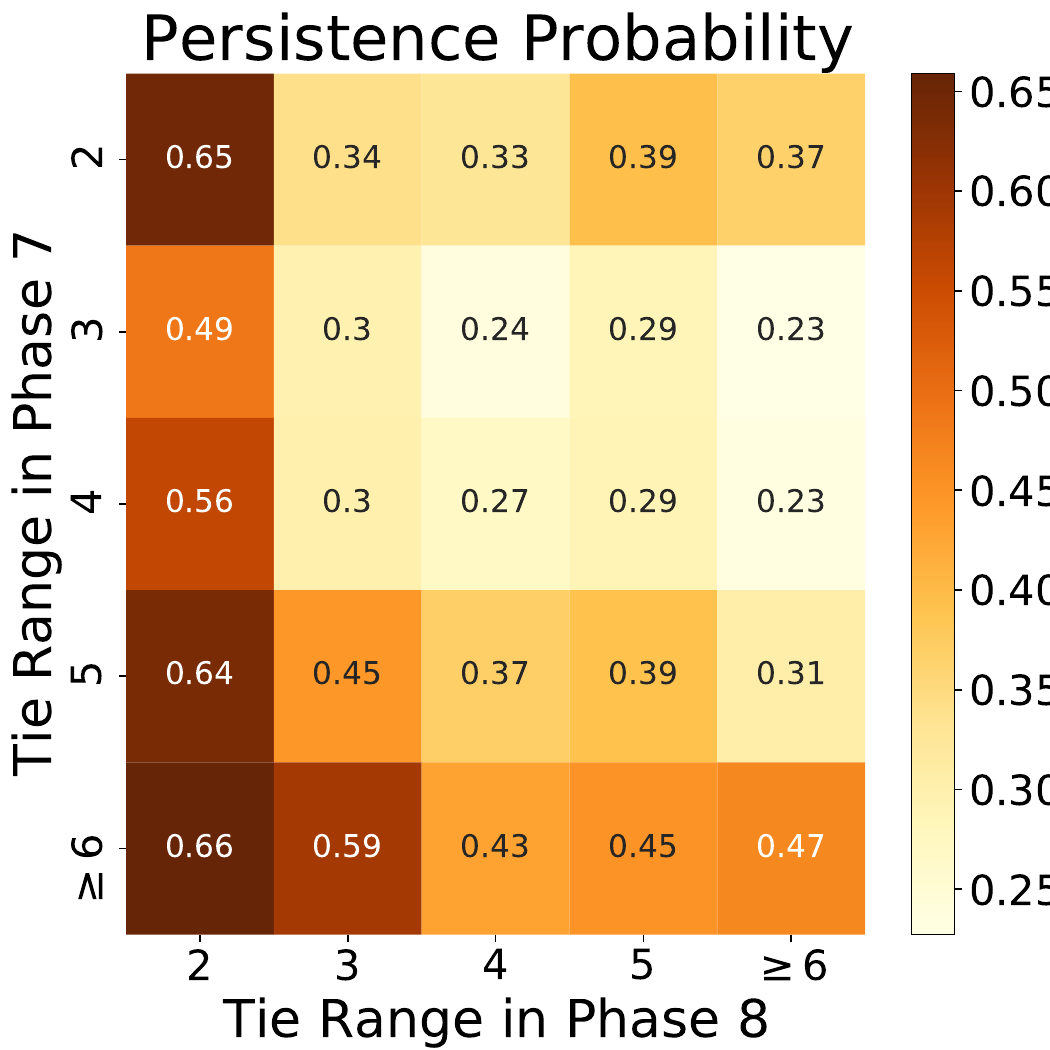}
    \caption{\textbf{Interaction duration, frequency and persistent probability (bottom row) in the next phase when tie range evolves.} The y-axis and x-axis represent tie range of social ties in phase $t$ and in phase $t+1$, respectively. Interaction duration is measured by call volume in seconds. Interaction frequency is the number of calls or texts. Persistence probability is defined as the probability of social ties persisting from phase $t$ to phase $t+1$. The numbers on the cells indicate the mean ($\log$) interaction duration (top row), the mean ($\log$) interaction frequency (middle row), and persistence probability (bottom row).}
    \label{fig:Fig.S9}
\end{figure*}

\begin{figure*}
    \centering
    \includegraphics[width=0.3\linewidth]{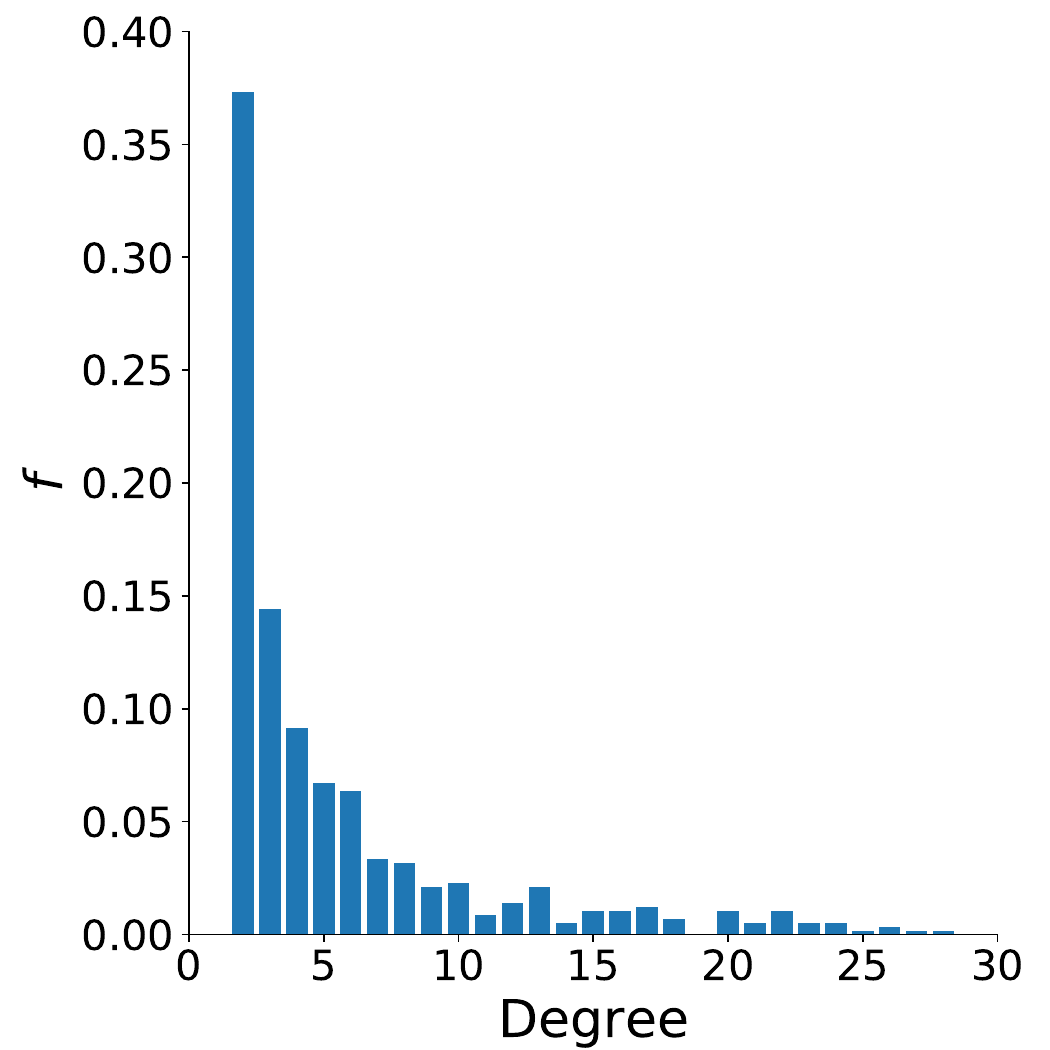}
    \caption{\textbf{Distribution of long range ties (tie range $\geq 6$) with respect to node degree.} $f$ is the probability mass function.}
    \label{fig:Fig.S10}
\end{figure*}

\begin{figure*}
\captionsetup[subfigure]{labelformat=simple, font={small}}
\centering
\subfloat[]{
	\includegraphics[width=0.3\linewidth]{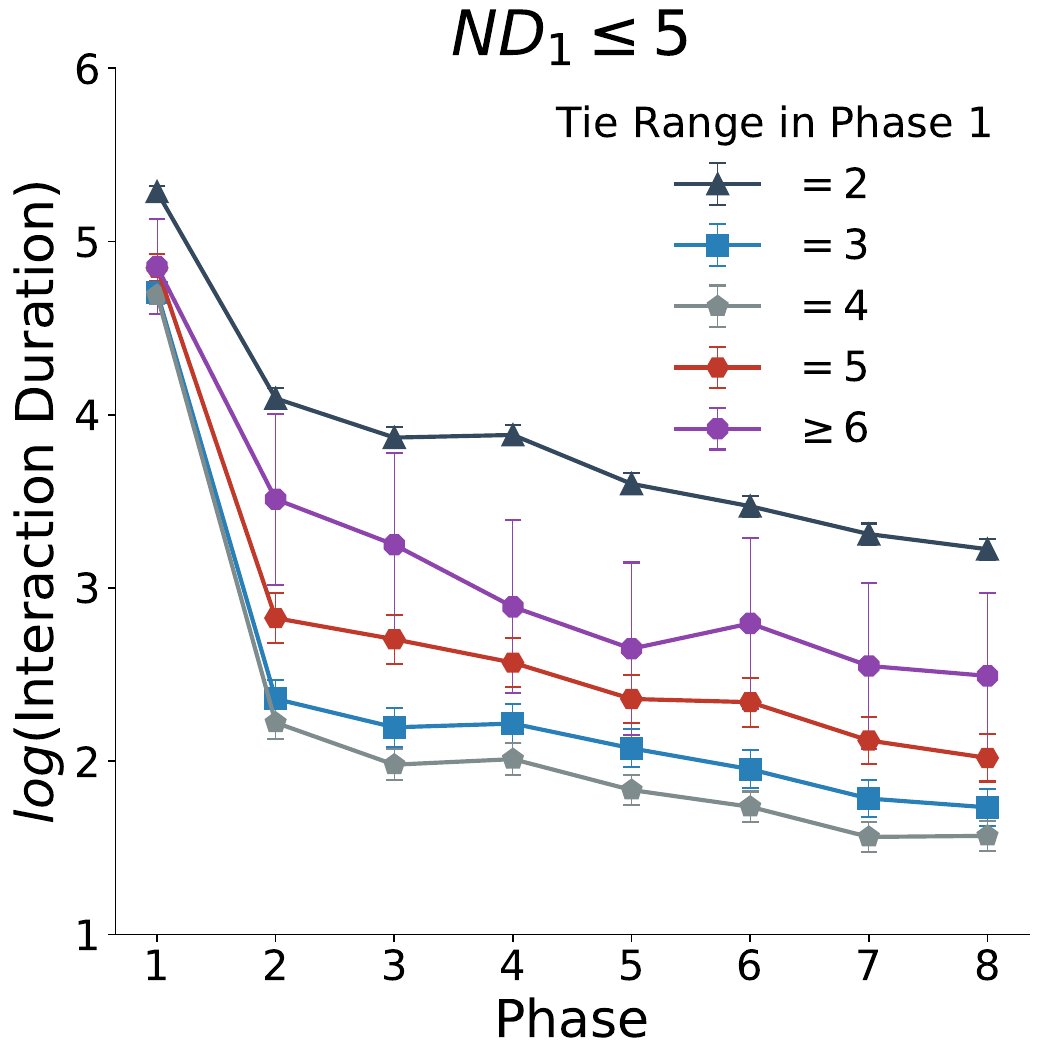}}
\quad
\subfloat[]{
	\includegraphics[width=0.3\linewidth]{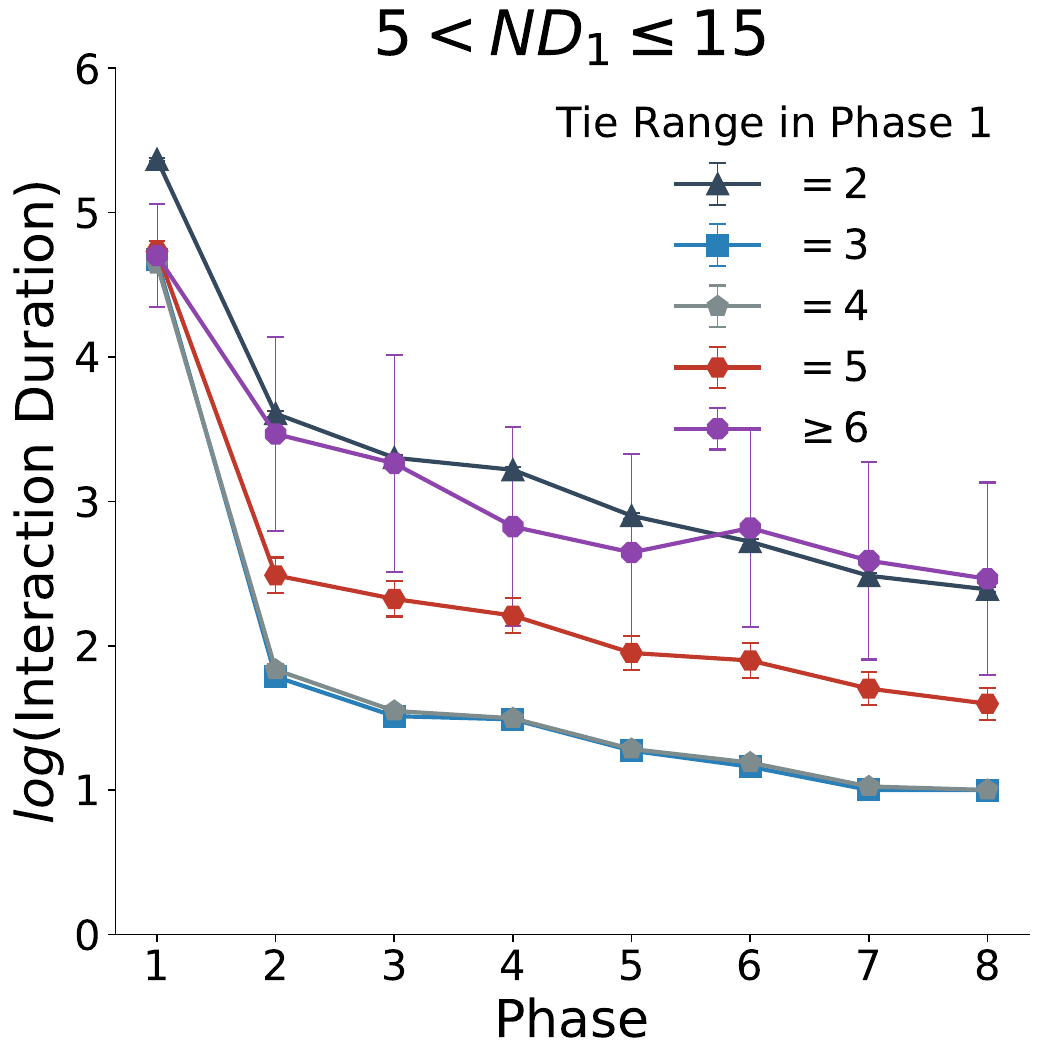}}
\quad
\subfloat[]{
	\includegraphics[width=0.3\linewidth]{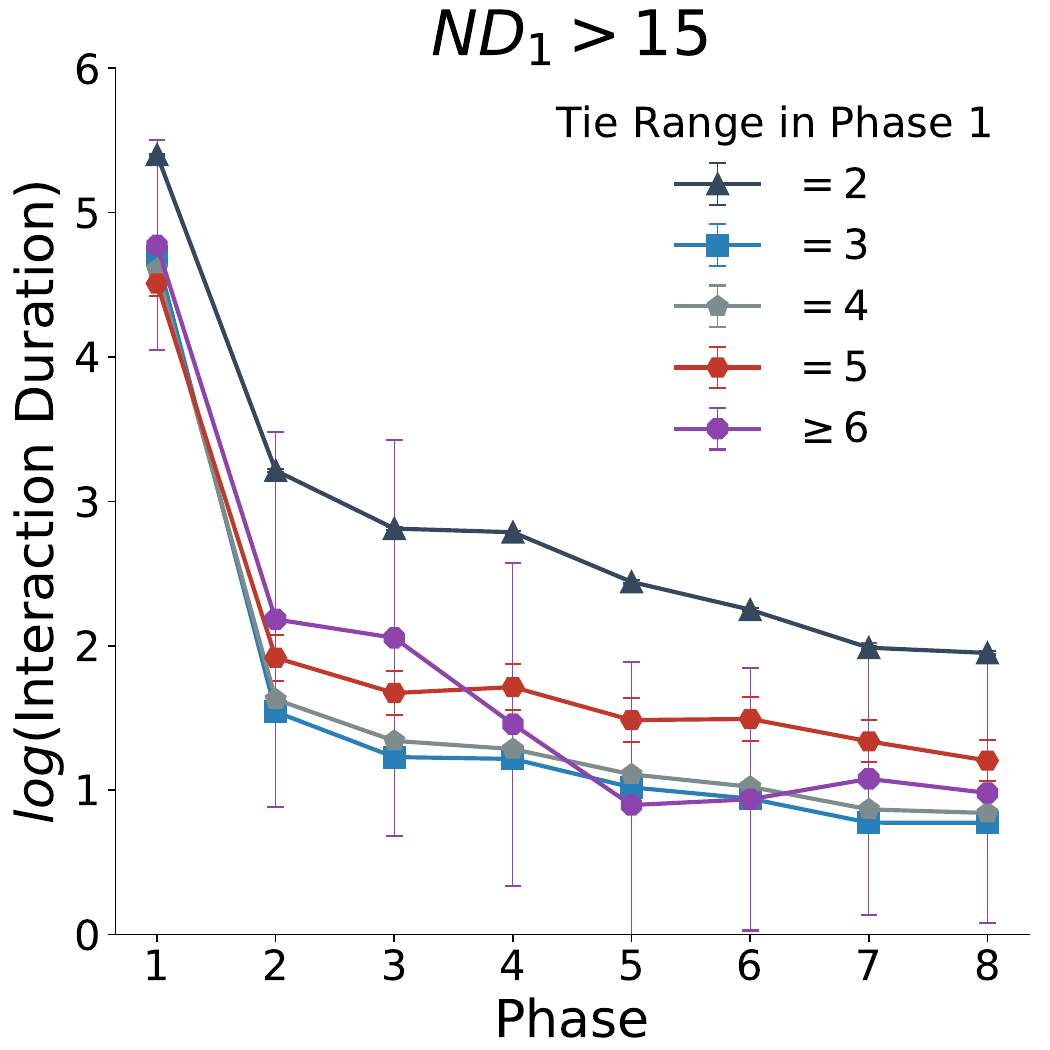}}
\vfill
\subfloat[]{
	\includegraphics[width=0.3\linewidth]{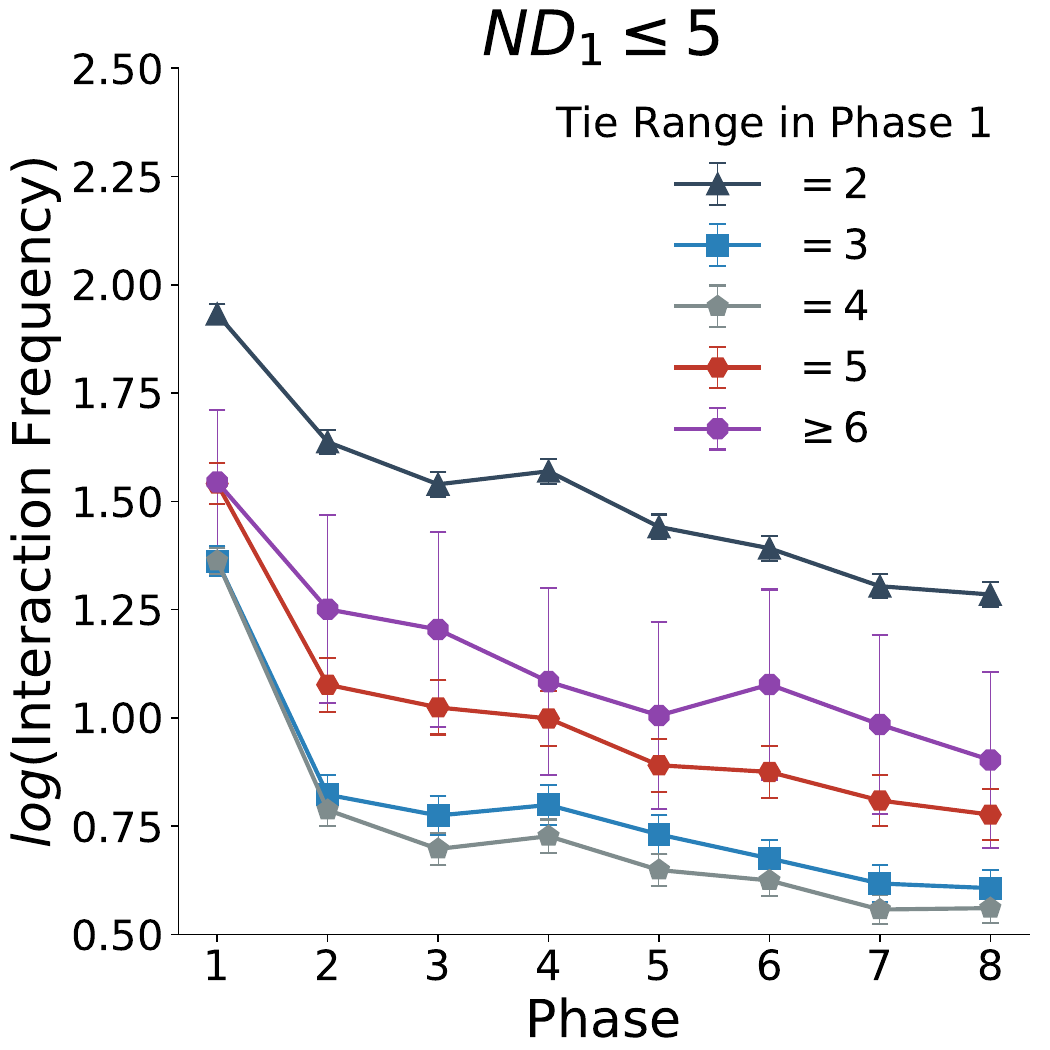}}
\quad
\subfloat[]{
	\includegraphics[width=0.3\linewidth]{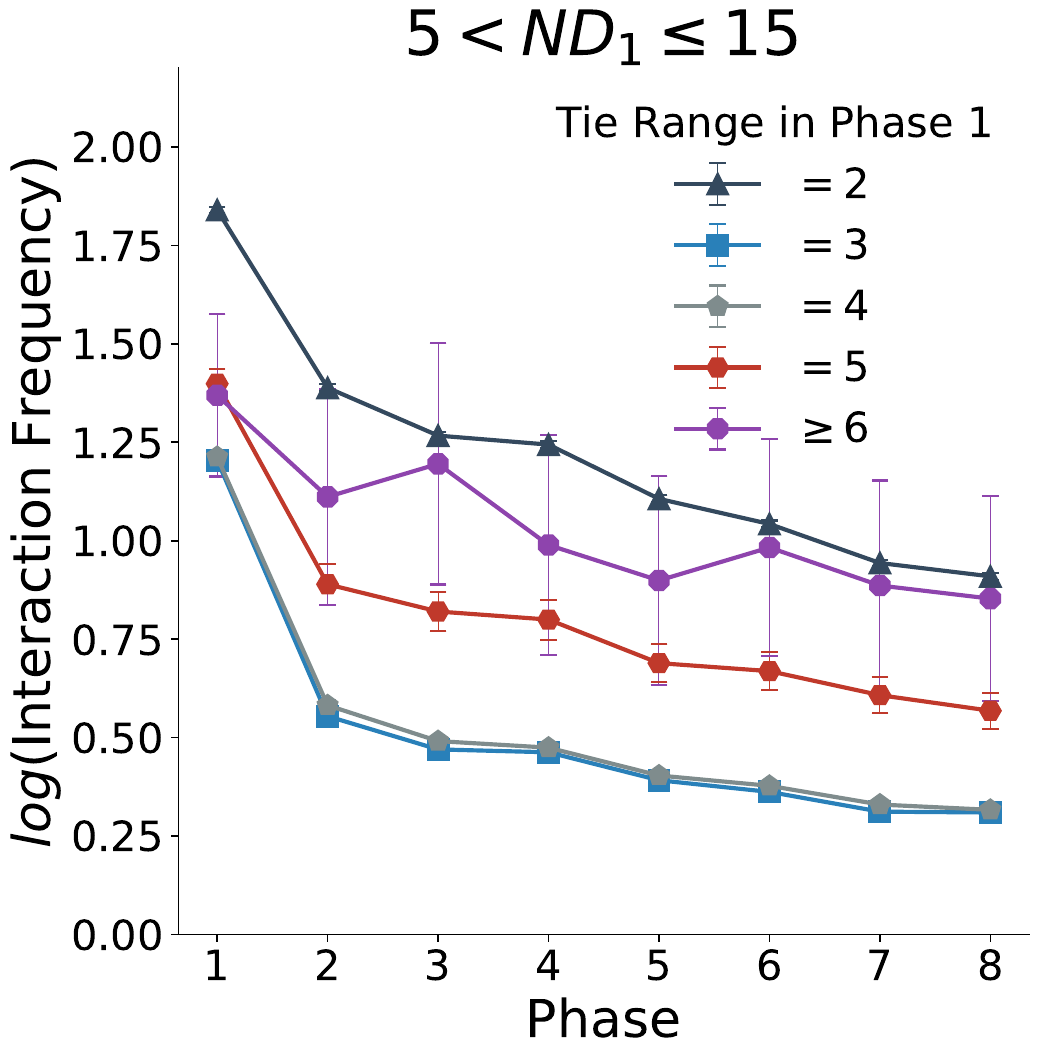}}
\quad
\subfloat[]{
	\includegraphics[width=0.3\linewidth]{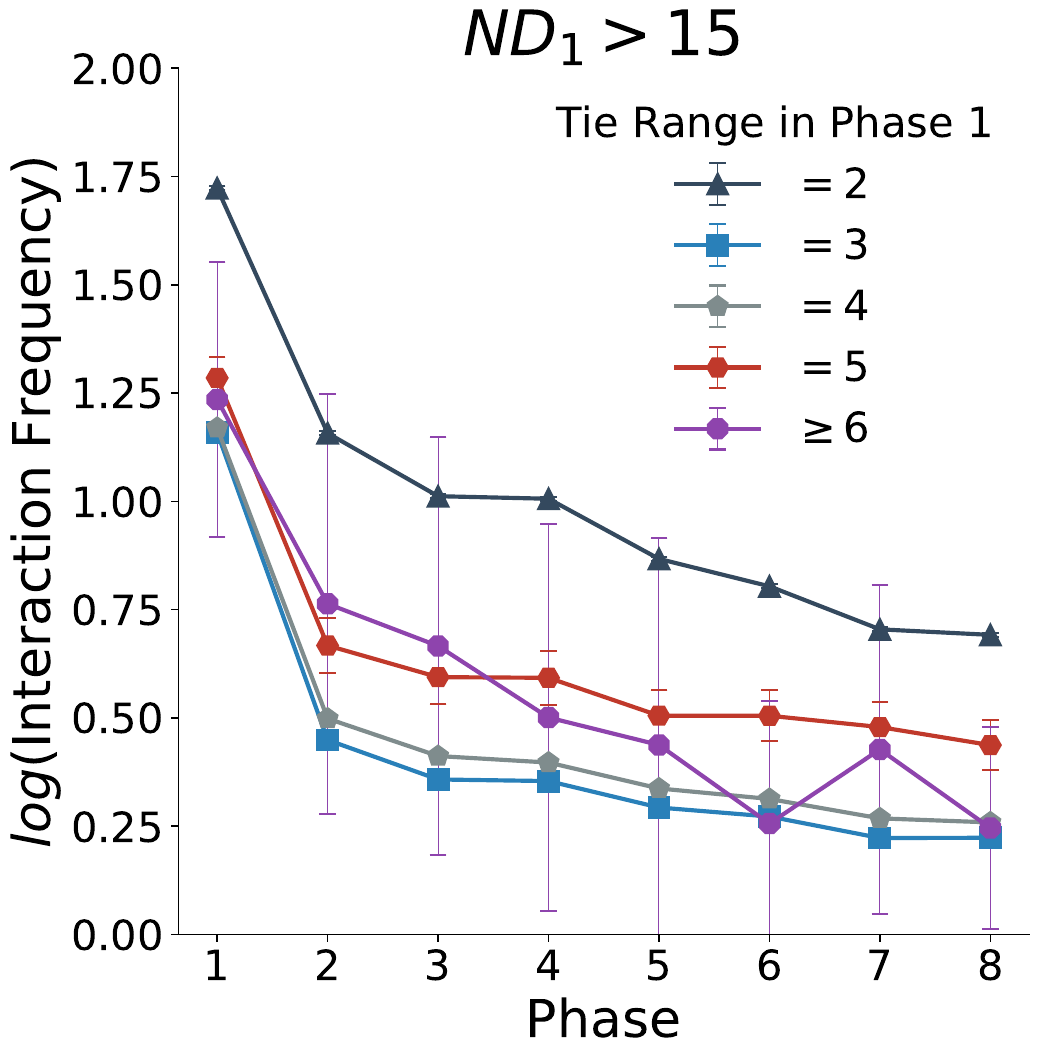}}
    \caption{\textbf{Dynamics of tie strength with different ranges when we examine degree subgroups.} Tie strength is measured by interaction duration (\textbf{a-c}; the total duration of the calls in seconds) and interaction frequency (\textbf{d-f}; the number of calls or texts). $ND_{1}$ indicates node degree in phase 1. The medium node degree of the snapshot in phase 1 is 12. Each phase represents a season (three months). We take logarithms ($\log$) for both interaction duration and frequency. All ties are classified according to their tie range in the first phase. The curves represent the average ($\log$) interaction duration or frequency conditional on that a tie exists in phase 1 with the given tie range. Error bars are 95\% confidence intervals for the mean $\log$ interaction duration and frequency (assuming normal distribution). Note that error bars are sometimes smaller than the data point markers.}
\label{fig:Fig.S11}
\end{figure*}

\begin{figure*}
\captionsetup[subfigure]{labelformat=simple, font={small}}
\centering
\subfloat[]{
	\includegraphics[width=0.3\linewidth]{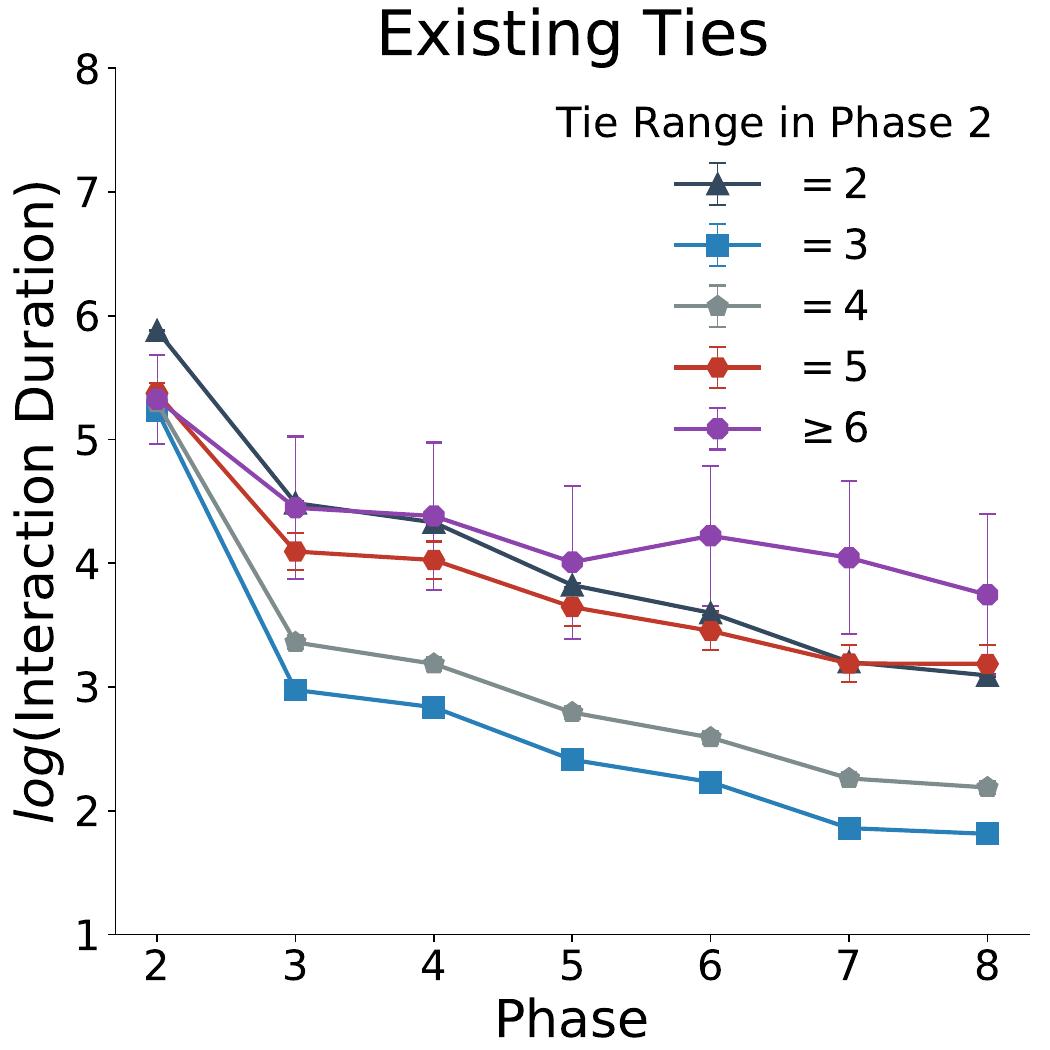}}
\quad
\subfloat[]{
	\includegraphics[width=0.3\linewidth]{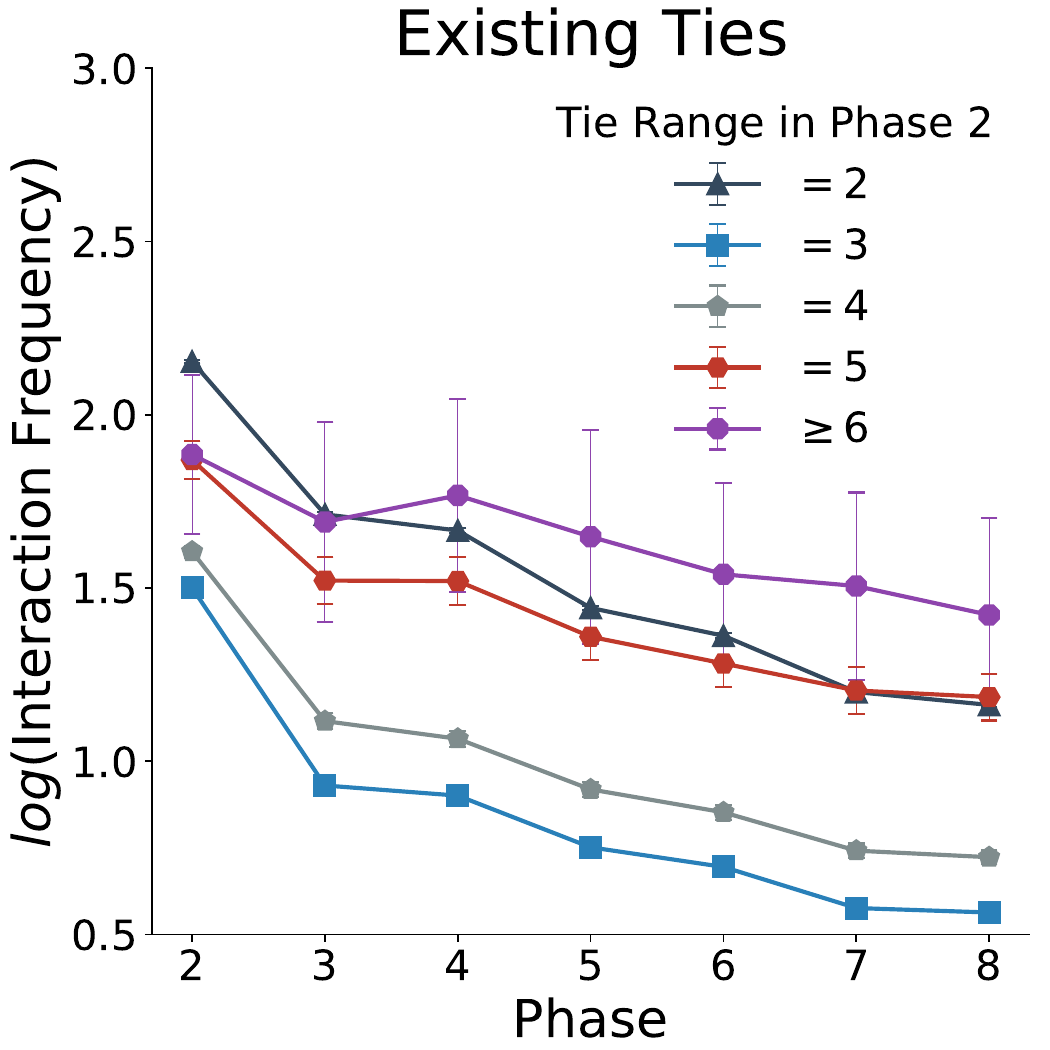}}
\quad
\subfloat[]{
	\includegraphics[width=0.3\linewidth]{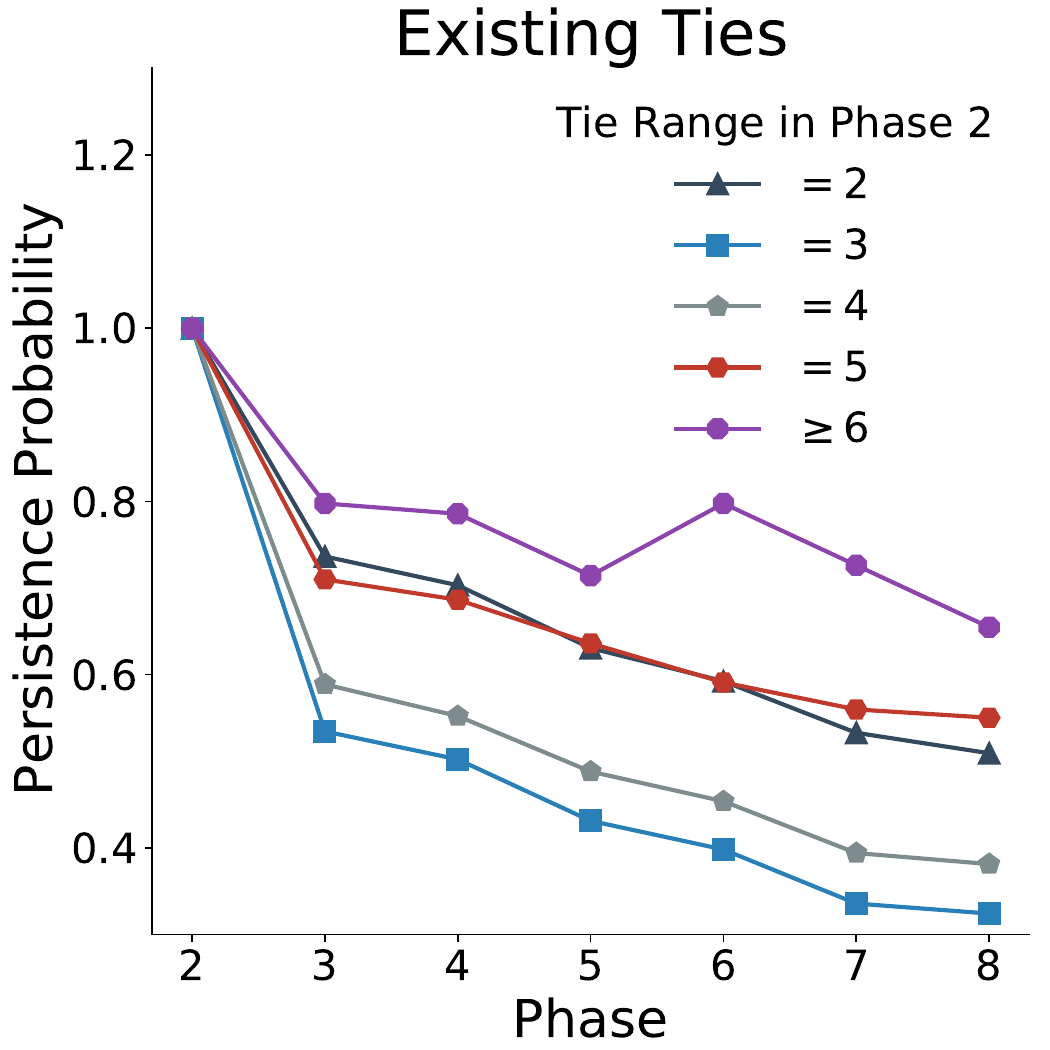}}	
\vfill
\subfloat[]{
	\includegraphics[width=0.3\linewidth]{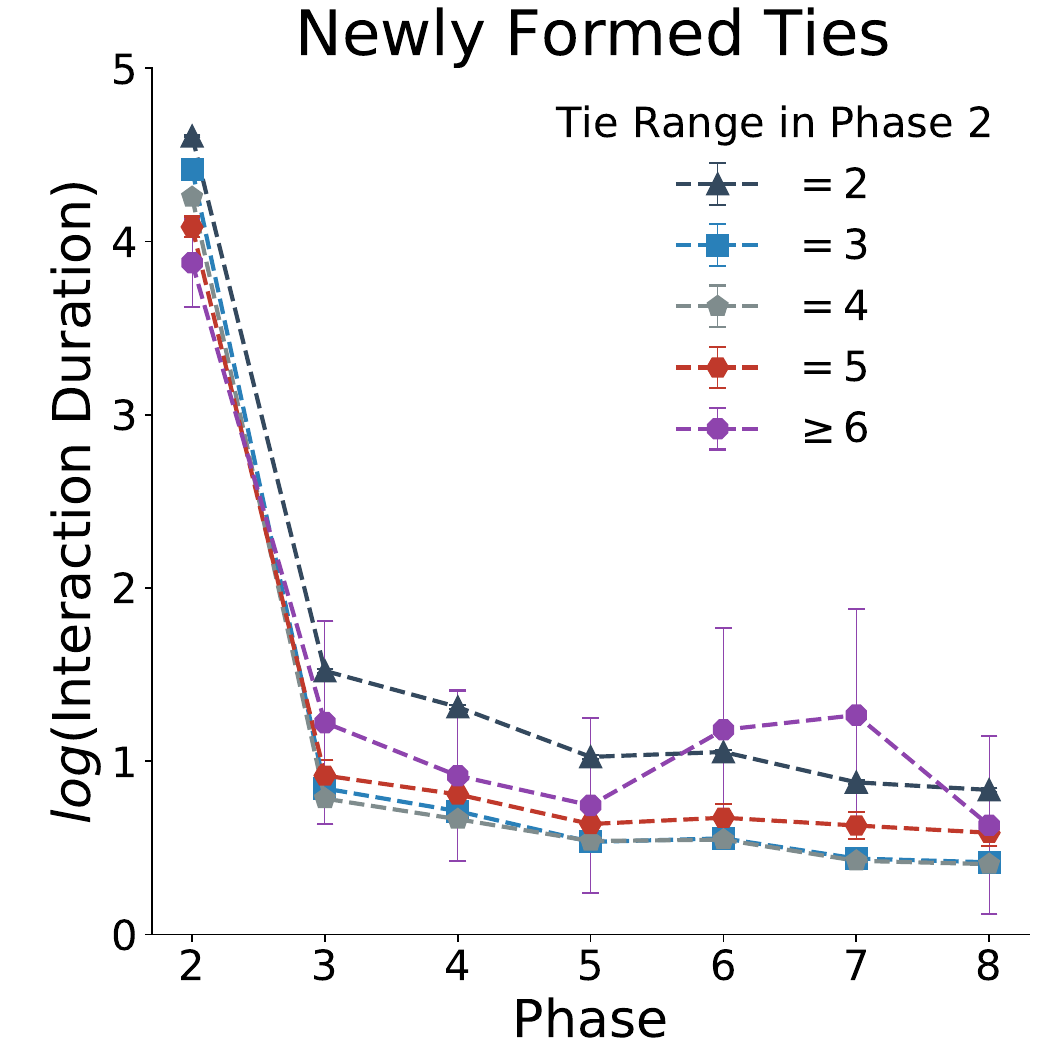}}
\quad
\subfloat[]{
	\includegraphics[width=0.3\linewidth]{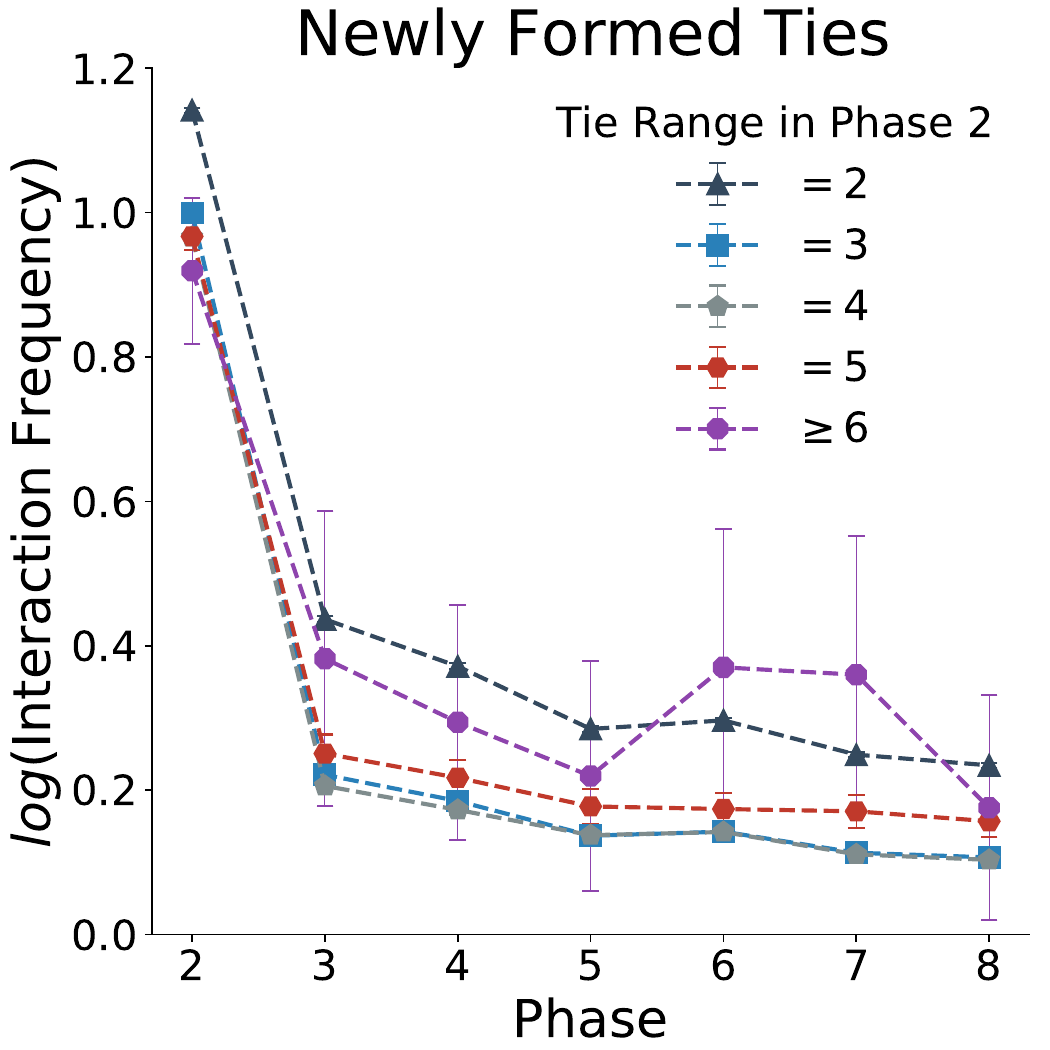}}
\quad
\subfloat[]{
	\includegraphics[width=0.3\linewidth]{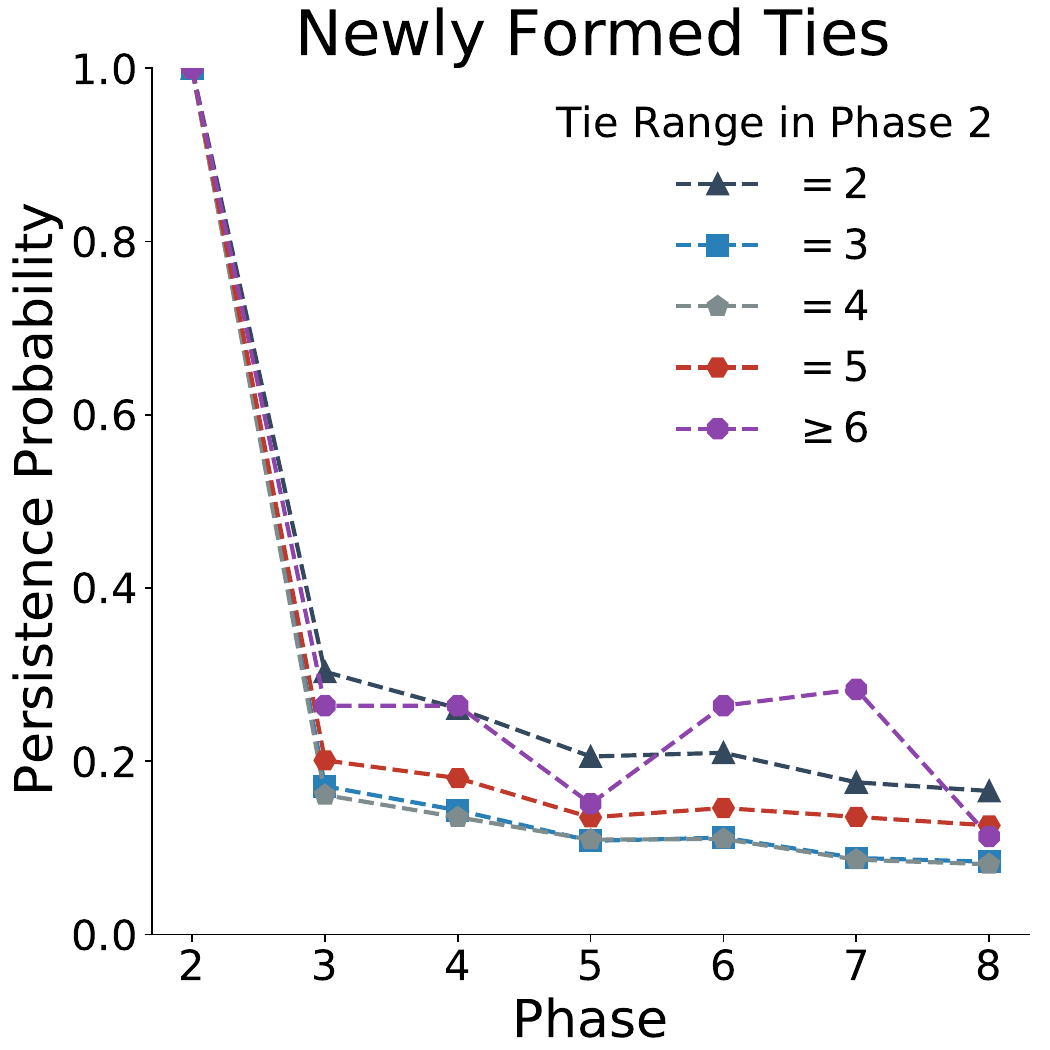}}
    \caption{\textbf{Dynamics of interaction frequency, interaction duration, and persistent probability of survival or newly-formed ties throughout the next seven phases conditional on that a tie exists in phase 2.} Each phase represents a season (three months). Interaction duration is measured in seconds. We take logarithms ($\log$) for both interaction duration and frequency. All ties are classified according to their tie range in the first phase. The curves represent the average ($\log$) interaction duration or frequency conditional on that a tie exists in phase 1 with the given tie range. Error bars are 95\% confidence intervals for the mean $\log$ interaction duration and frequency (assuming normal distribution). Note that error bars are sometimes smaller than the data point markers.}
\label{fig:Fig.S12}
\end{figure*}

\begin{figure*}
    \captionsetup[subfigure]{labelformat=simple, font={small}}
    \centering
    \subfloat[]{
    \includegraphics[width=0.3\linewidth]{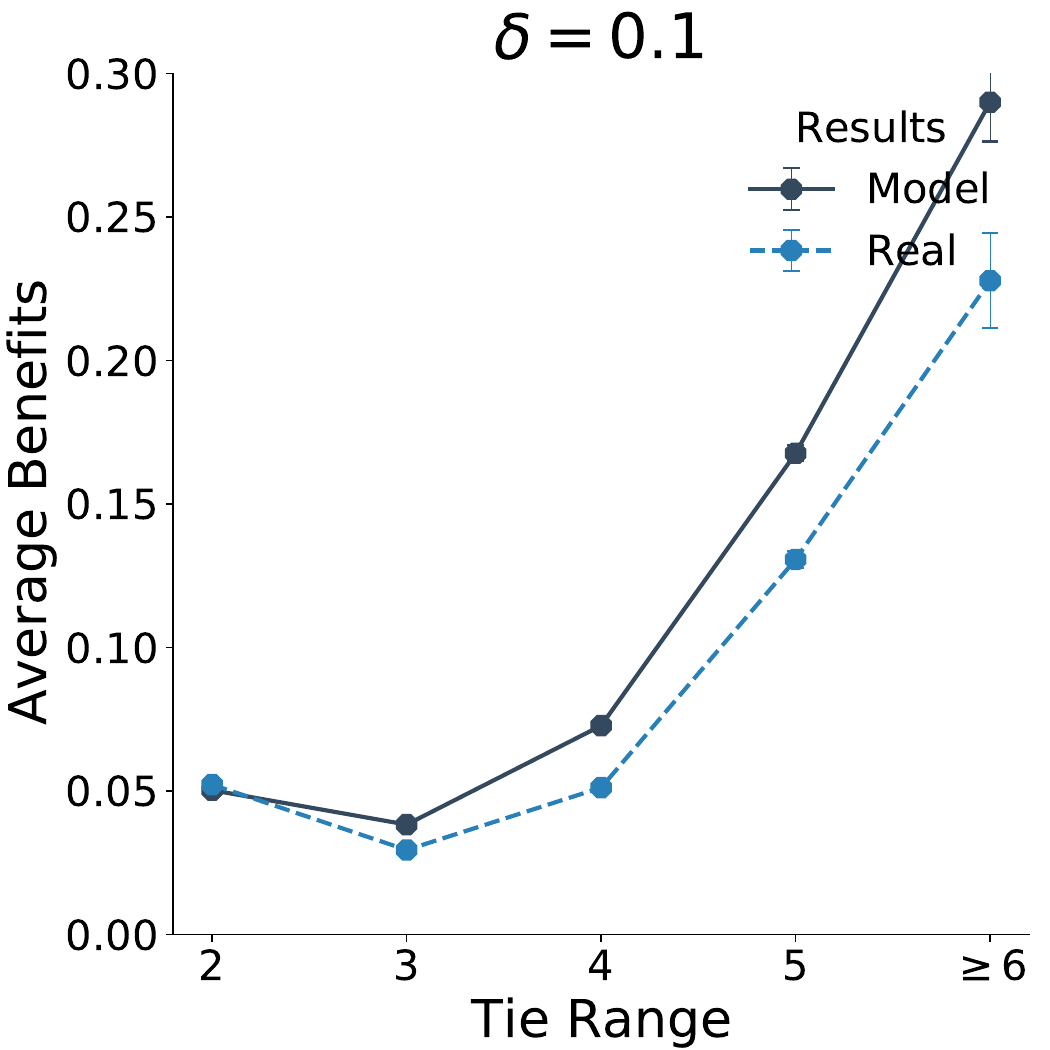}}~
    \subfloat[]{
    \includegraphics[width=0.3\linewidth]{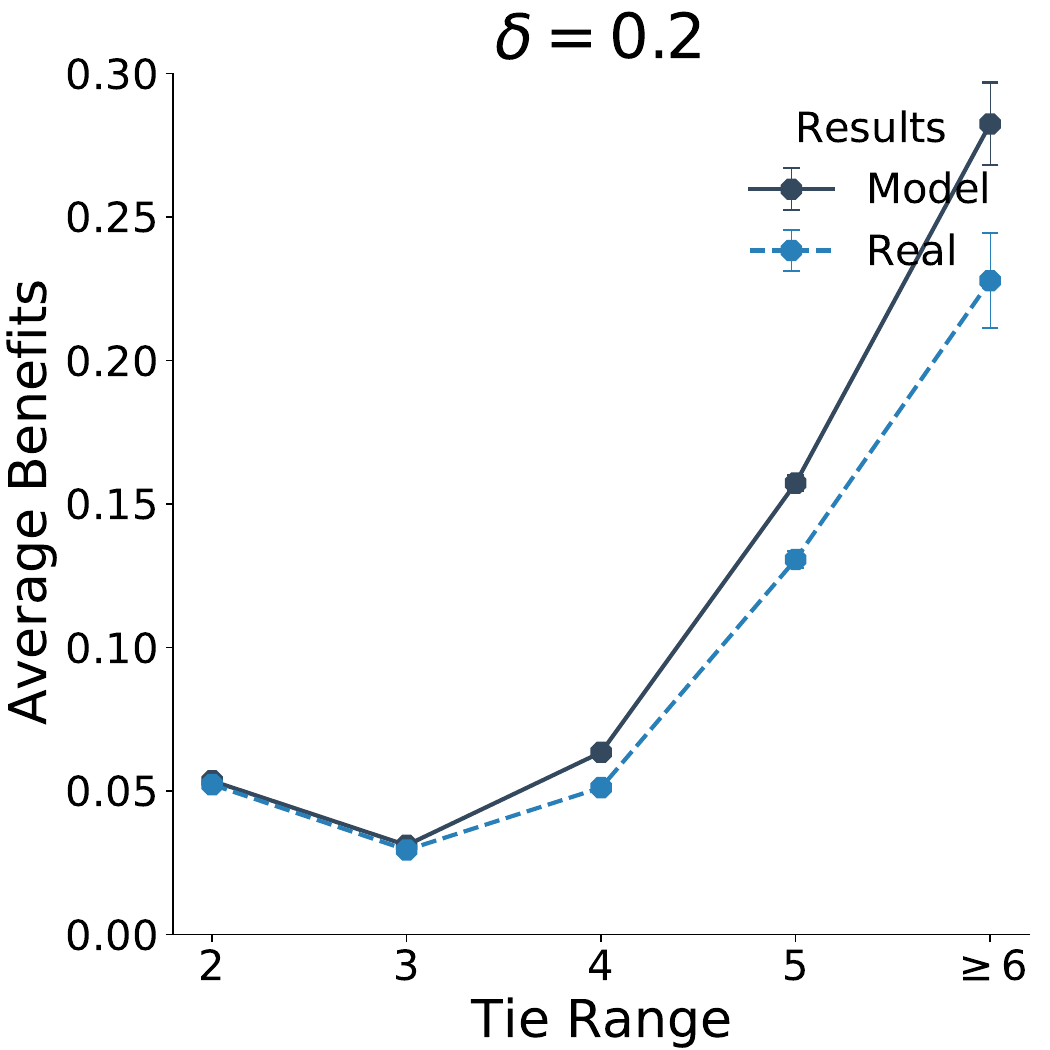}}~
    \subfloat[]{
    \includegraphics[width=0.3\linewidth]{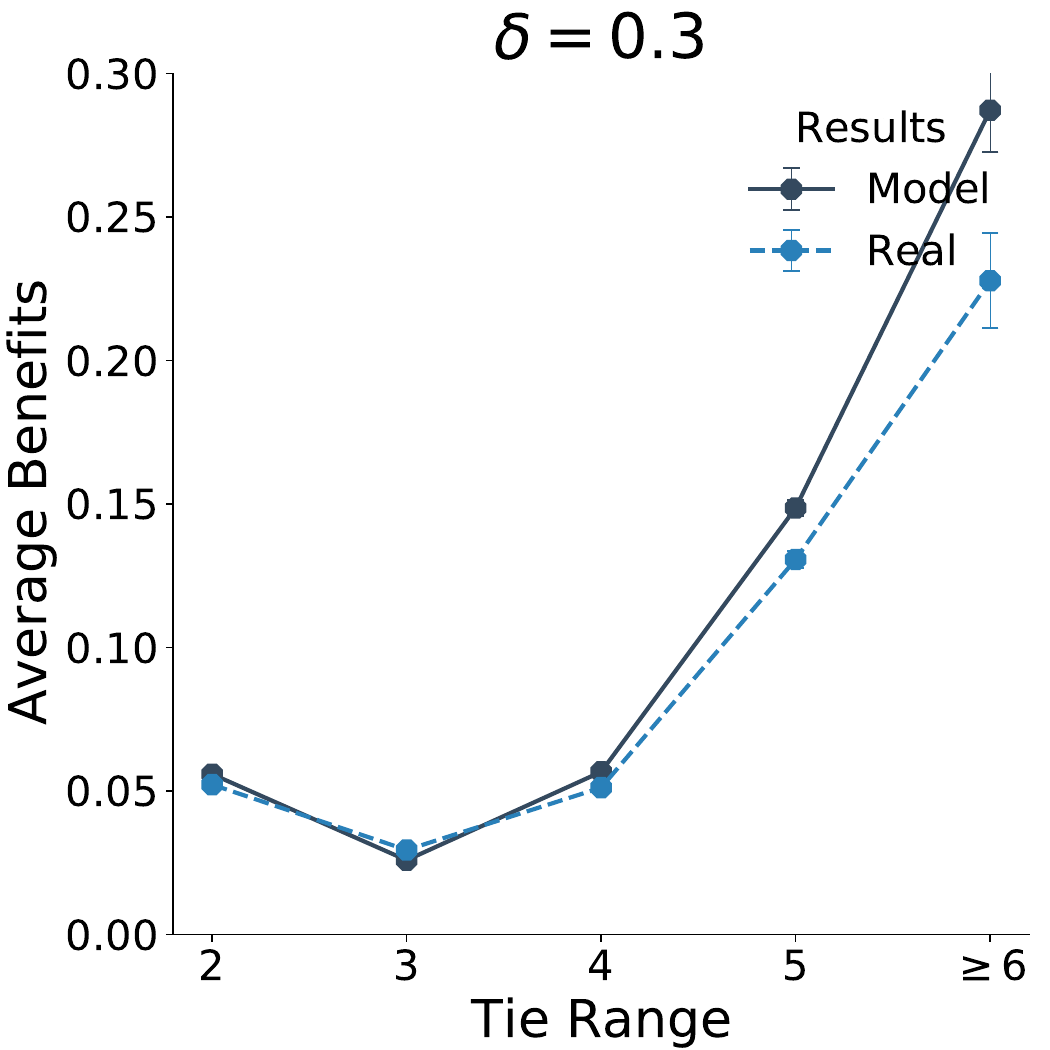}}
    \caption{\textbf{Choice of $\delta$ in the proposed model.} The navy curve is the average benefits calculated from the model while the blue curve is the average benefits calculated from empirical data. Error bars are 95\% confidence intervals for the benefits (assuming normal distribution). Note that error bars are sometimes smaller than the data point markers.}
    \label{fig:Fig.S13}
\end{figure*}

\begin{figure*}
    \centering
    \includegraphics[width=0.3\linewidth]{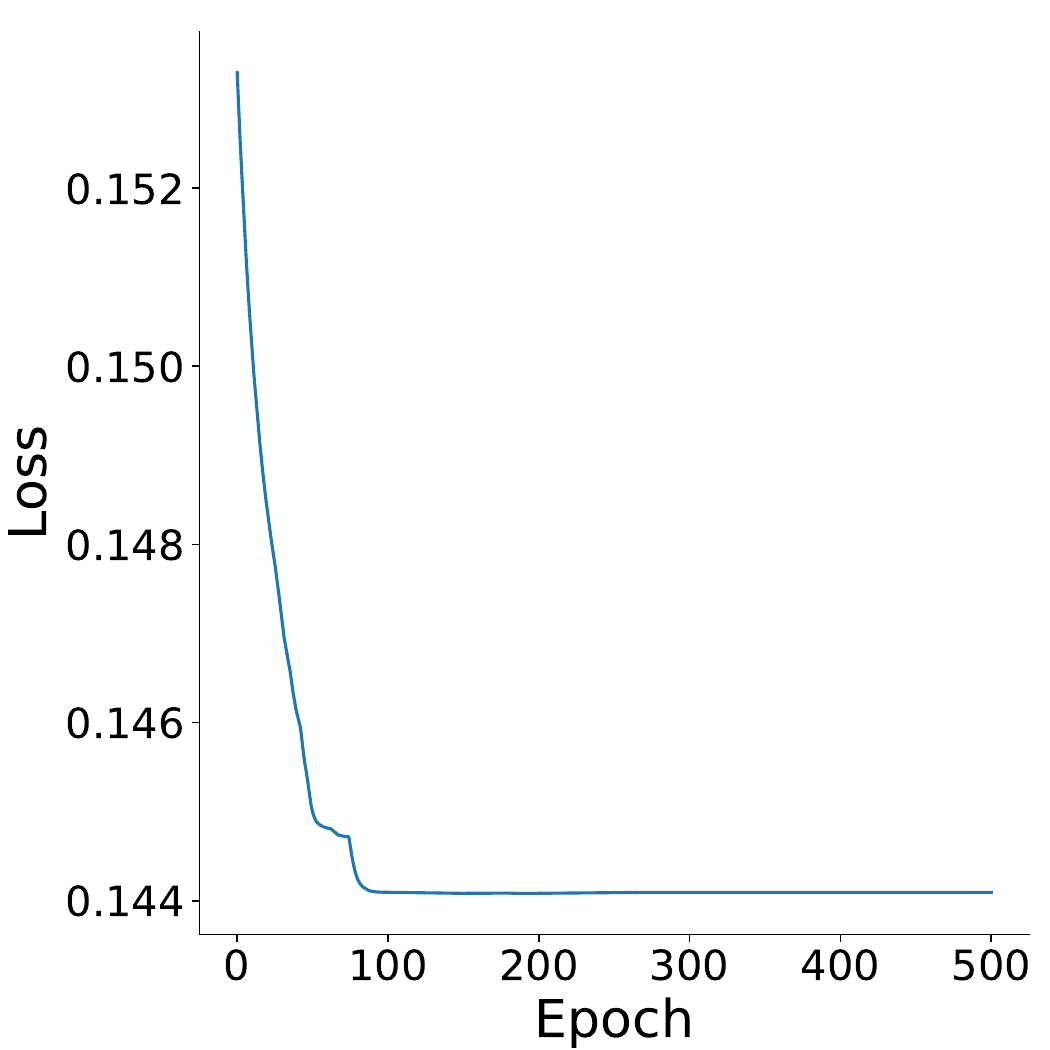}
    \caption{\textbf{The learning curve of our model training.} The navy curve is the average benefits calculated from the model while the blue curve is the average benefits calculated from empirical data. We see that after 100 epochs the learning curve converges.}
    \label{fig:Fig.S14}
\end{figure*}

\begin{figure*}
    \captionsetup[subfigure]{labelformat=simple, font={small}}
    \centering
    \subfloat[]{
    \includegraphics[width=0.24\linewidth]{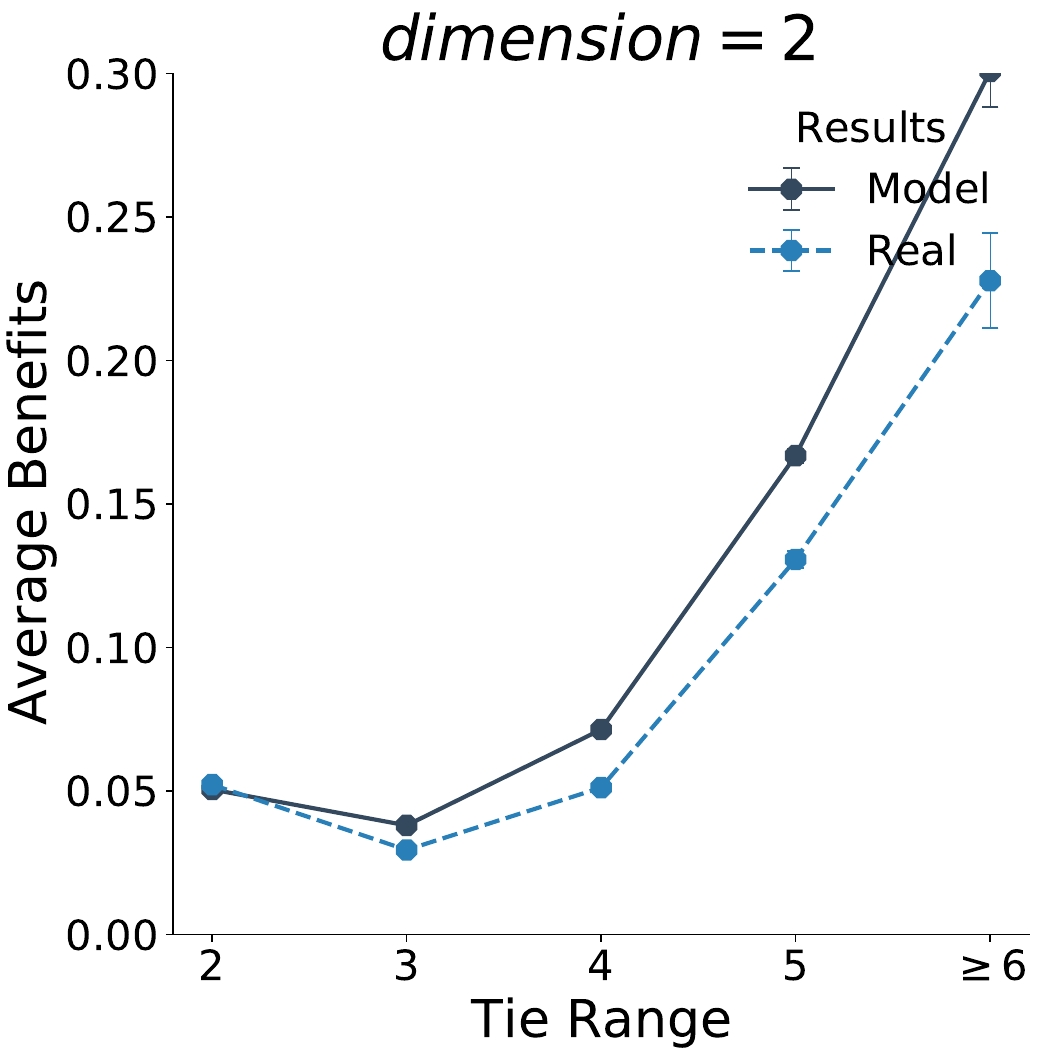}}~
    \subfloat[]{
    \includegraphics[width=0.24\linewidth]{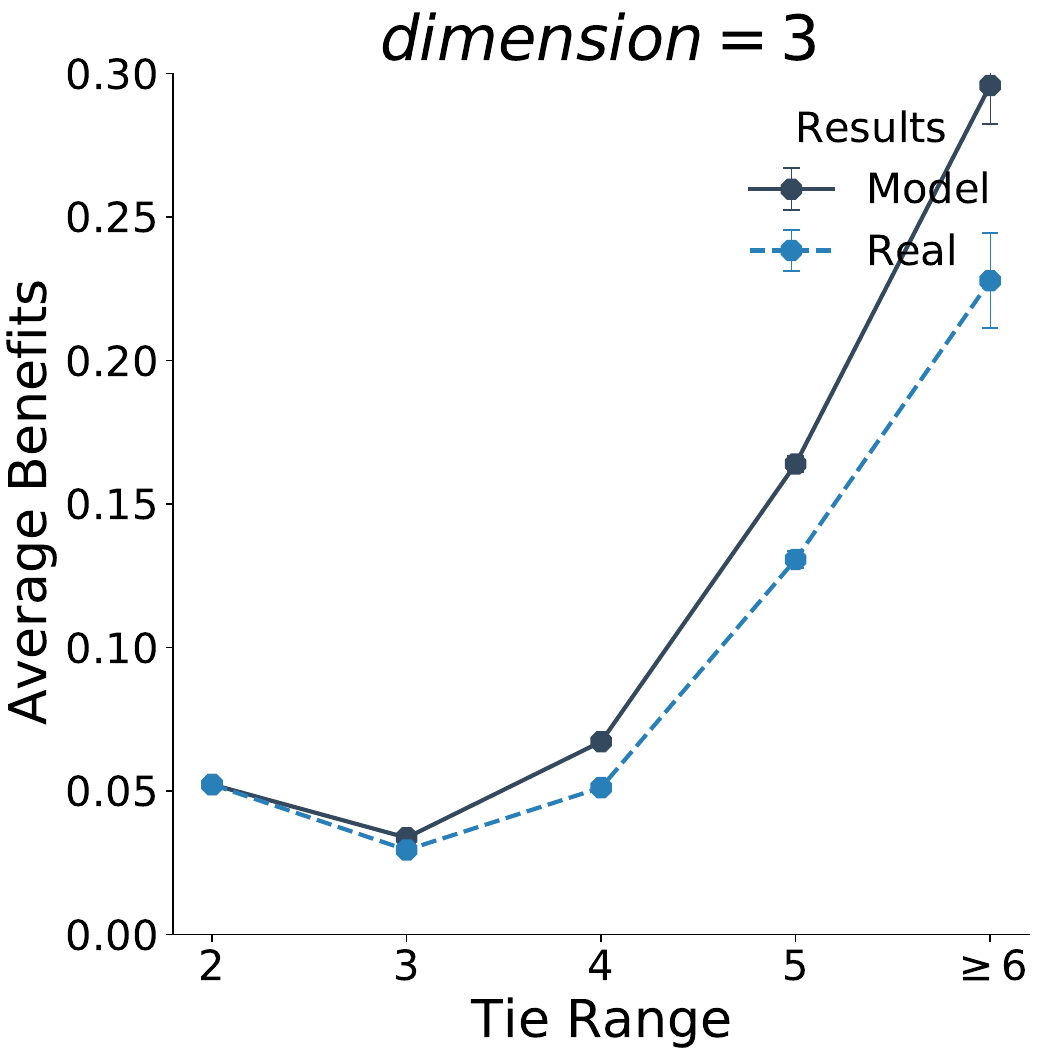}}~
    \subfloat[]{
    \includegraphics[width=0.24\linewidth]{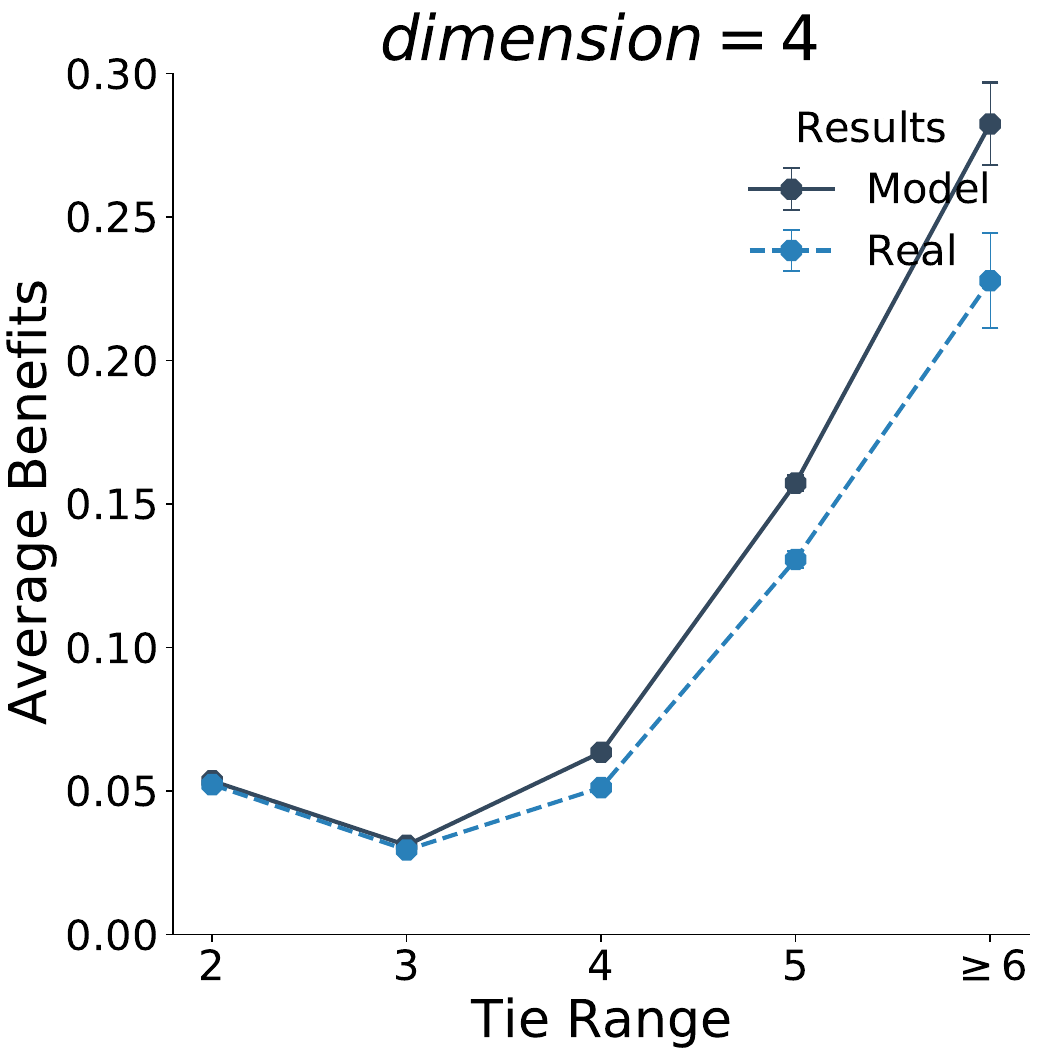}}~
    \subfloat[]{
    \includegraphics[width=0.24\linewidth]{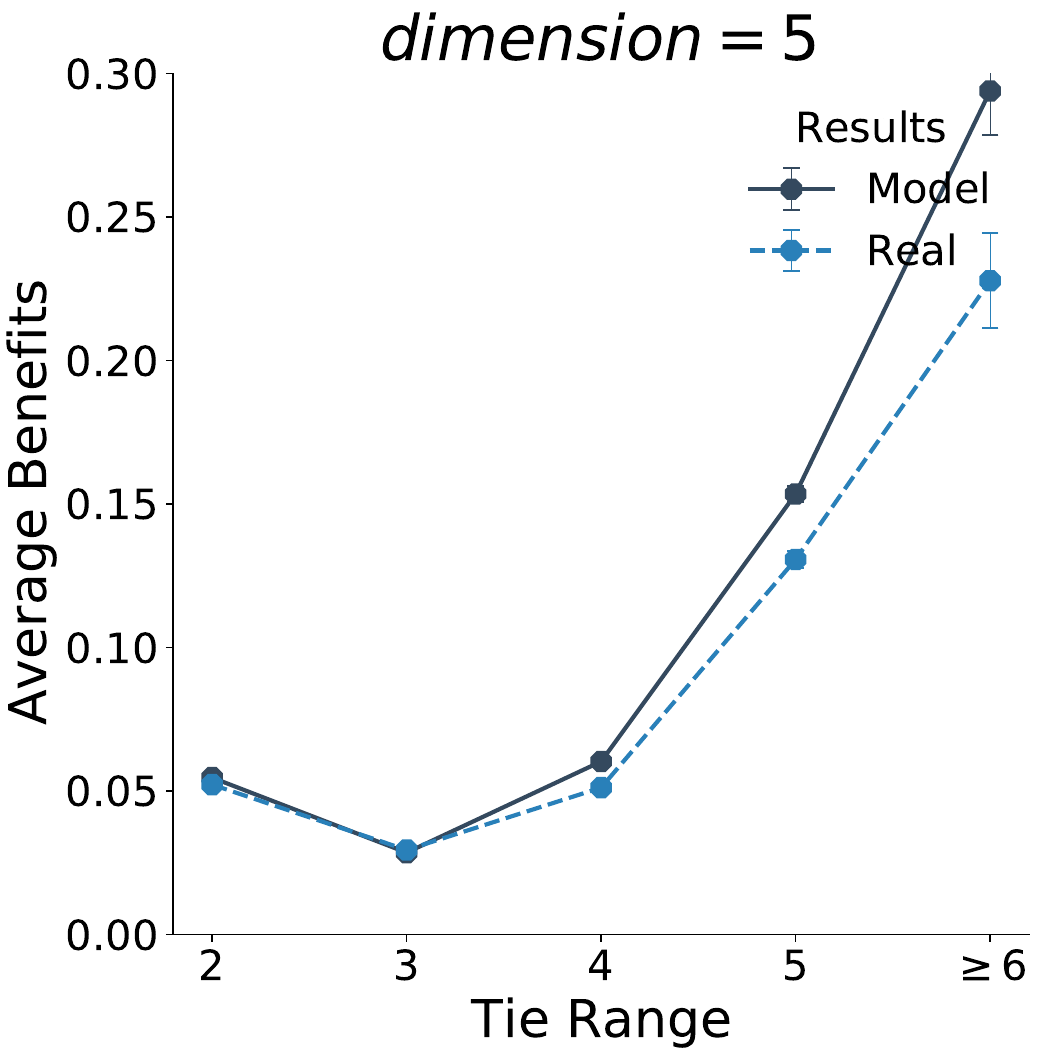}}
    \caption{\textbf{Results of choosing different dimensionality.} Error bars are 95\% confidence intervals for the benefits (assuming normal distribution). Note that error bars are sometimes smaller than the data point markers.}
    \label{fig:Fig.S15}
\end{figure*}

\begin{figure*}
    \captionsetup[subfigure]{labelformat=simple, font={small}}
    \centering
    \subfloat[]{
    \includegraphics[width=0.3\linewidth]{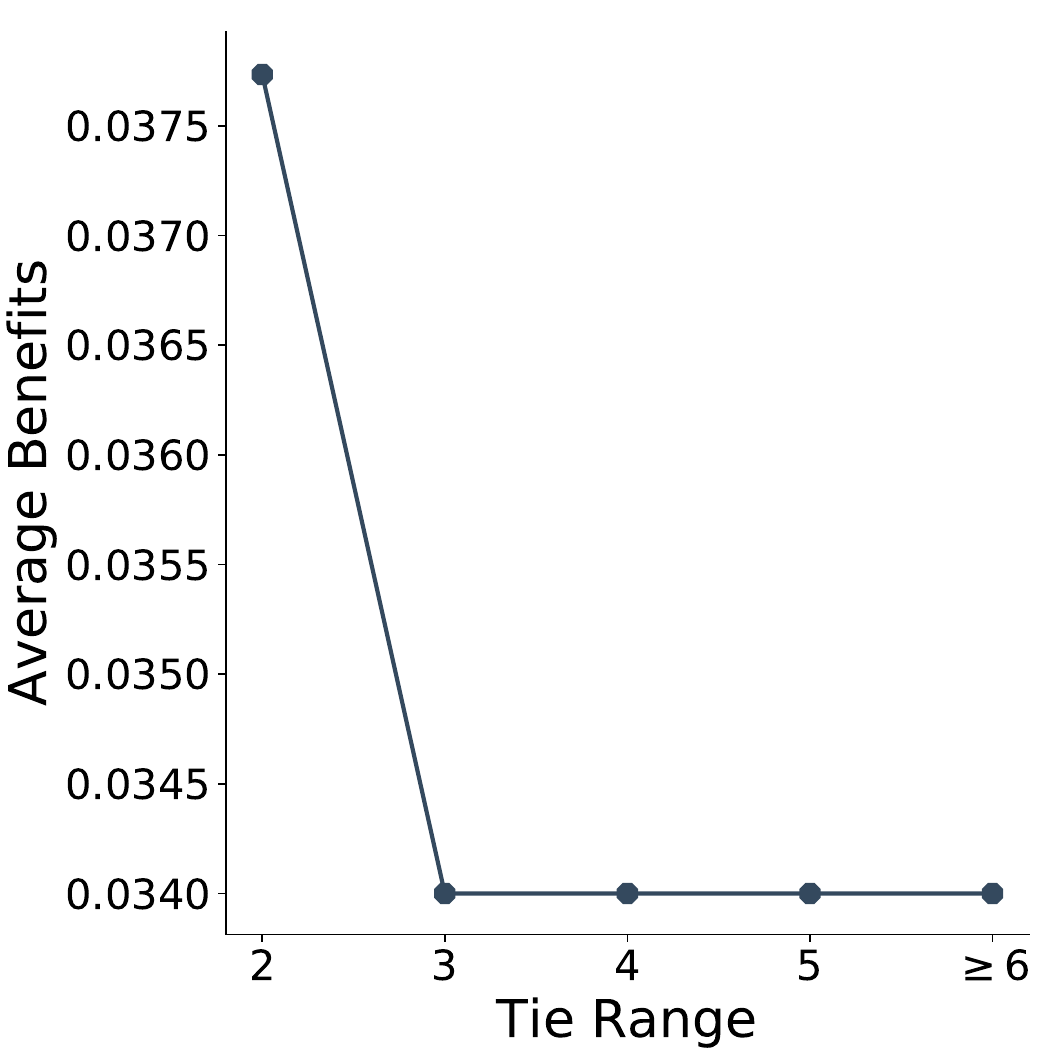}}~
    \subfloat[]{
    \includegraphics[width=0.3\linewidth]{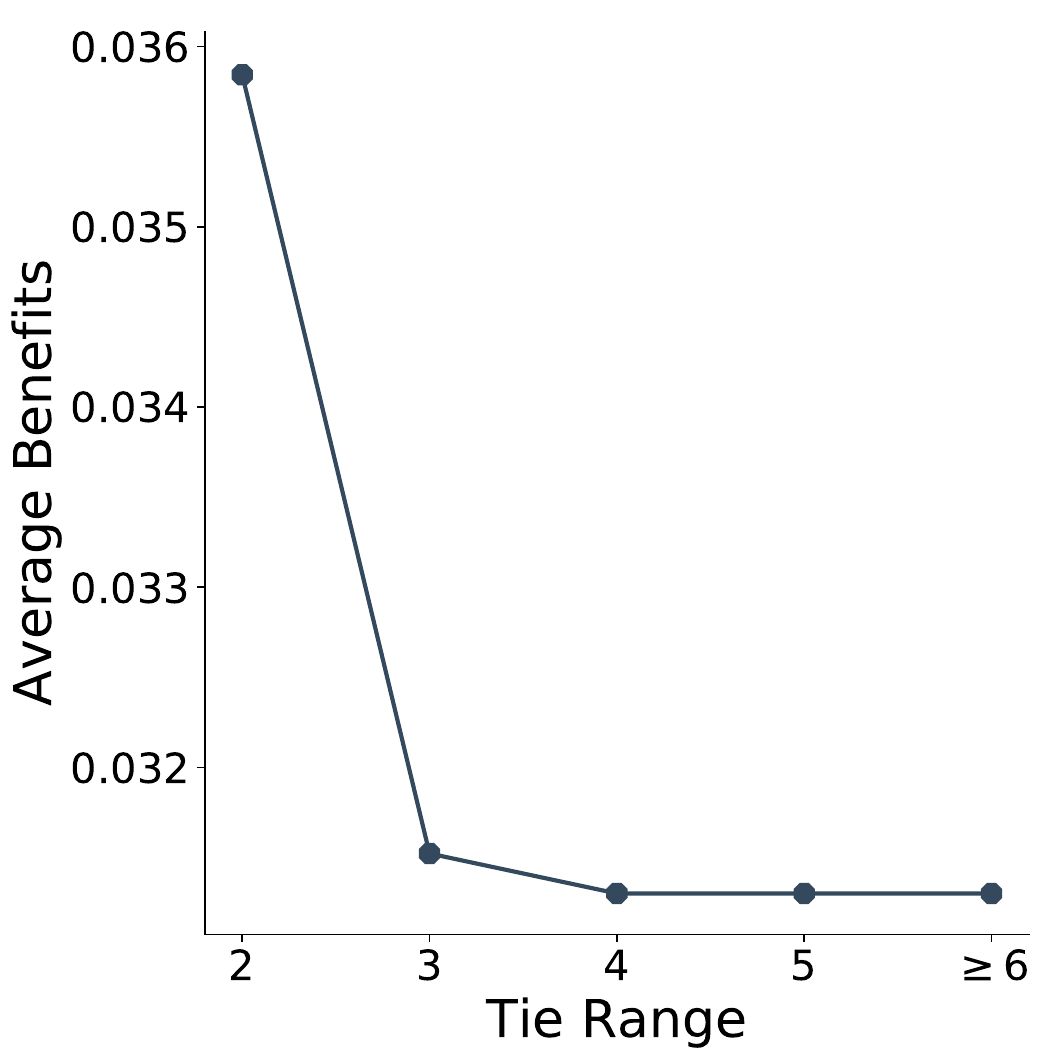}}~
    \subfloat[]{
    \includegraphics[width=0.3\linewidth]{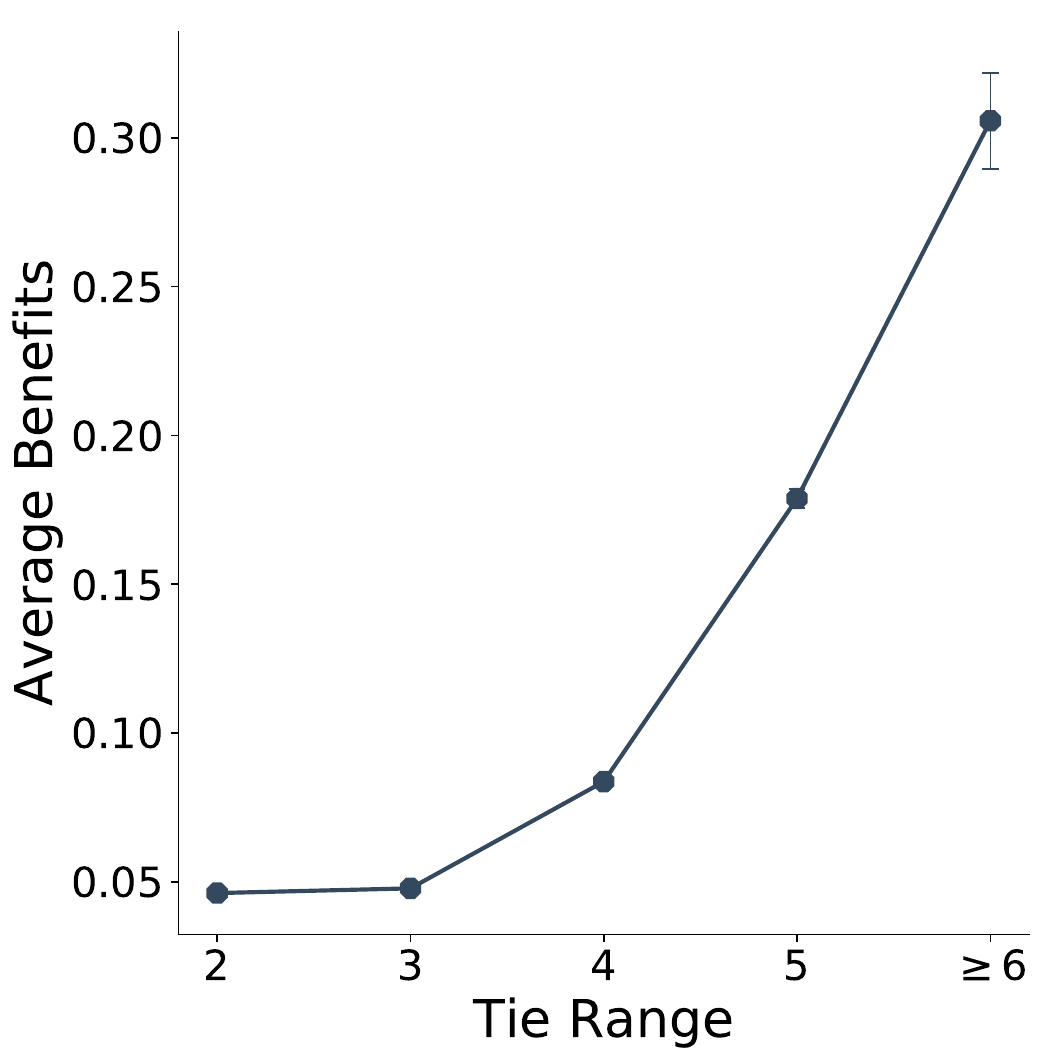}}
    \caption{\textbf{The results implied by the baseline models.}
     They are the connections model of second-degree indirect effects (\textbf{a}), the connections model of three-degree indirect effects (\textbf{b}), and the simplified version of our model (\textbf{c}). None of the baselines can replicate the ``U-shape'' found in empirical data. Note that error bars are sometimes smaller than the data point markers.}
    \label{fig:Fig.S16}
\end{figure*}

\end{document}